\documentclass{article}
 
\usepackage{jheppub}
\usepackage[utf8]{inputenc}
\usepackage{amsmath}
\usepackage{amssymb}
\usepackage{amsfonts}
\usepackage{amsthm}
\usepackage{appendix}
\usepackage{physics}
\usepackage{upgreek}
\usepackage{tensor}
\usepackage{subcaption}
\usepackage{booktabs}
 \makeatletter
\def\moverlay{\mathpalette\mov@rlay}
\def\mov@rlay#1#2{\leavevmode\vtop{%
   \baselineskip\z@skip \lineskiplimit-\maxdimen
   \ialign{\hfil$\m@th#1##$\hfil\cr#2\crcr}}}
\newcommand{\charfusion}[3][\mathord]{
    #1{\ifx#1\mathop\vphantom{#2}\fi
        \mathpalette\mov@rlay{#2\cr#3}
      }
    \ifx#1\mathop\expandafter\displaylimits\fi}
\makeatother
 
\newcommand{\cupdot}{\charfusion[\mathbin]{\cup}{\cdot}}

\def\b1{\bar 1}

\def\Lc{\mathcal{L}}
\def\Hc{\mathcal{H}}
\def\bC{\mathbb{C}}
\def\mK{k}

\def \om {\omega}
\def \Dg {\Delta}
\def \dg {\delta}
\def \ds {\partial}
\def \ag {\alpha}
\def \bg {\beta}
\def \cg {\gamma}
\def \tg {\theta}

\def \La {\Lambda}
\def \la {\lambda}

\def \Zs {\mathbb{Z}}
\def \Rs {\mathbb{R}}

\def \Cs {\mathbb{C}}

\def \oo {\mathcal{O}}

\def \dep{\mathfrak{d}}

\newcommand\vct[1]{\mathbf{#1}}

\title{Gaudin Models and Multipoint Conformal Blocks II: 
Comb Channel Vertices in 3D and 4D}

\subheader{\normalsize
        \vspace*{-3.7em}
        \begin{flushright}
	DESY 21-105\\
        SAGEX-21-14-E\\
        ZMP-HH/21-15
        \end{flushright}
}

\author{Ilija Buri\'c$^1$,}
\author{Sylvain Lacroix$^2$,} 
\author{Jeremy Mann$^1$,} 
\author{Lorenzo Quintavalle$^1$,} 
\author{Volker Schomerus$^{1,2}$}
\affiliation{$^1$ DESY Theory Group, DESY Hamburg, Notkestrasse 85, D-22603 Hamburg,}
\affiliation{$^2$ II. Institut f\"ur Theoretische Physik, Universit\"at Hamburg, Luruper Chaussee 149, D-22761 Hamburg}
\affiliation{Zentrum f\"ur Mathematische Physik, Universit\"at Hamburg, Bundesstrasse 55, D-20146 Hamburg }

\emailAdd{ilija.buric@desy.de}
\emailAdd{sylvain.lacroix@desy.de}
\emailAdd{jeremy.mann@desy.de}
\emailAdd{lorenzo.quintavalle@desy.de}
\emailAdd{volker.schomerus@desy.de}

\date{June 2021}

\abstract{It was recently shown that multi-point conformal blocks in higher dimensional 
conformal field theory can be considered as joint eigenfunctions for a system of commuting 
differential operators. The latter arise as Hamiltonians of a Gaudin integrable system. In 
this work we address the reduced fourth order differential operators that measure the choice 
of 3-point tensor structures for all vertices of 3- and 4-dimensional comb channel conformal 
blocks. These vertices come associated with a single cross ratio. Remarkably, we identify the vertex operators as Hamiltonians of a crystallographic elliptic Calogero-Moser-Sutherland model 
that was discovered originally by Etingof, Felder, Ma and Veselov. Our construction is based on 
a further development of the embedding space formalism for mixed-symmetry tensor fields. The 
results thereby also apply to comb channel vertices of 5- and 6-point functions in arbitrary dimension.}

\begin{document}
\addtolength{\baselineskip}{2mm}
\maketitle

\section{Introduction}

The $N$-point function of a fundamental field in a conformal field theory 
contains a tremendous amount of dynamical information about the theory, more 
than any finite set of 4-point functions could ever capture. Unfortunately, 
this information is difficult to extract, as is well known from the study of 
4-point functions. In general, the extraction is based on conformal partial 
wave expansions wherein the constant coefficients are products of the dynamical data 
that characterizes all operator products. For $N=4$, the relevant partial waves 
(or conformal blocks) are by now well understood through the work of Dolan and 
Osborn and others, see in particular \cite{Dolan:2000ut,Dolan:2003hv,Dolan:2011dv,
Costa:2011dw,SimmonsDuffin:2012uy,Hogervorst:2013sma,Penedones:2015aga,CastedoEcheverri:2015mkz,
Echeverri:2016dun,Costa:2016hju,Isachenkov:2017qgn,Karateev:2017jgd,Erramilli:2019njx,
Fortin:2020ncr} and many references therein. For $N > 4$, similar powerful results on 
conformal blocks do not yet exist, though there is some significant recent activity in this 
area, see for example 
\cite{Rosenhaus:2018zqn,Parikh:2019ygo,Fortin:2019dnq,Parikh:2019dvm,Fortin:2019zkm,
Irges:2020lgp,Fortin:2020yjz,Pal:2020dqf,Fortin:2020bfq,Hoback:2020pgj,
Goncalves:2019znr,Anous:2020vtw,Fortin:2020zxw,Poland:2021xjs}.

In \cite{Buric:2020dyz} we initiated a novel, integrability based approach to multi-point 
conformal blocks. It extends an idea that was 
advanced initially by Dolan and Osborn, namely to characterize conformal blocks
through a set of differential equations they satisfy. For 4-point blocks, these 
differential equations are eigenvalue equations for the set of commuting Casimir 
differential operators that measure the conformal weight and spin of the 
intermediate field. For a higher number $N$ of insertion points, the 
operators that measure the quantum numbers of the $N-3$ intermediate fields 
are still mutually commuting, but they do not suffice to characterize the associated 
blocks. In fact, as one can easily see, the number of such Casimir-like 
differential operators is strictly smaller than the number of cross ratios as 
soon as $N > 4$ and the dimension $d > 2$. The challenge to complete the 
Casimir-like operators into a full set of commuting differential operator was 
solved in \cite{Buric:2020dyz,Buric:2021ywo}. In these papers we explained how 
to obtain the missing differential operators by taking limits of an $N$-site 
Gaudin integrable system~\cite{Gaudin_76a,Gaudin_book83,Feigin:1994in}. 

\begin{figure}[htp]
    \centering
    \includegraphics{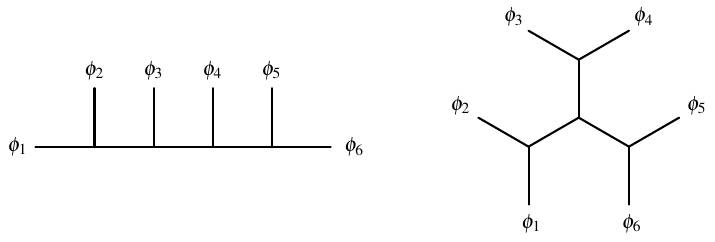}
    \caption{OPE diagrams in the comb channel (left) and snowflake channel (right) for 6-point correlators.}
    \label{fig:Six-points_Comb_Snowflake}
\end{figure}

Let us note that the set of Casimir differential operators is defined by the choice 
of the so-called OPE channel. We have 
displayed two such channels for the case of a scalar 6-point function in 
\autoref{fig:Six-points_Comb_Snowflake}. The latter are referred to as comb and snowflake channel, 
respectively. For $d>3$, the comb channel admits $7$ independent Casimir operators, 
while the snowflake channel gives rise to $6$. The different number is related to 
the fact that in the snowflake channel all the three intermediate fields are 
Symmetric Traceless Tensors (STTs) while in the comb channel one of the intermediate 
fields is a more general tensor. In both cases, the number of independent Casimir 
operators is strictly smaller than the number $n_\textit{cr}(N=6,d\geq 4) = 9$ of 
cross ratios, so the Casimir operators need to be supplemented by two or three 
additional commuting operators. These can be constructed from the Gaudin integrable 
model through a limit that is adapted to the OPE channel under consideration, see 
\cite{Buric:2021ywo} for details. 

The relation between higher-dimensional conformal blocks and integrability first 
surfaced in \cite{Isachenkov:2016gim}. In that work it was pointed out that the 
Casimir operators which Dolan and Osborn used to characterize and construct scalar 
4-point blocks could be reinterpreted as eigenvalue equations for the Hamiltonians
of an integrable 2-particle hyperbolic Calogero-Moser-Sutherland (CMS) model associated with 
the root system $BC_2$. Later this observation was derived more conceptually from the 
very definition of these blocks within the framework of harmonic analysis on the 
conformal group, see \cite{Schomerus:2016epl,Schomerus:2017eny,Isachenkov:2018pef}.
Combining these previous results on the relation between 4-point scalar conformal 
blocks and CMS integrable models with the more recent insight on 
the relation with Gaudin integrable models implies that one can obtain the $BC_2$ hyperbolic 
CMS model through a reduction from a special case of the 4-site 
Gaudin integrable model, a fact that has not been appreciated even among experts 
until recently. 

The work we are about to describe extends this relation between Gaudin and 
CMS integrable models to several $3$-site Gaudin integrable models. 
More specifically, we consider a 3-site Gaudin model for the conformal group. The 
$3$-site model contains no non-trivial parameters, except for the choice of 
representation at each puncture. Here we will consider the Hamiltonian reduction of 
the full Gaudin model by the action of the conformal group and we will select the 
representations at the  
punctures such that the reduced model has a 2-dimensional phase space. Our main 
claim is that for all choices of punctures for which this is the case, the reduced 
model is an elliptic CMS model, namely the lemniscatic model that 
was discovered in \cite{etingof2011107}. Translated back into the context of 
conformal field theory, this implies that the conformal blocks for 3-point functions 
of spinning fields can be regarded as wave functions of the lemniscatic 
CMS model. The statement holds whenever the spins are chosen such 
that the 3-point function is characterized by a single cross ratio. This is the 
case, for example, for 3-point functions of a scalar and two STTs in $d \geq 3$, 
as well as two other cases to be described below.

The vertex operator we study in this work may be considered as one of the most important 
new elements that our approach needs in order to analyze and eventually construct 
multi-point comb channel blocks in 3 and 4D. Such blocks are joint eigenfunctions of 
$N-4$ mutually commuting operators of fourth order. In a full $N$-point function, 
they act on all cross ratios. But upon taking OPE limits and reducing the $N$-point 
function to any one of the $N-4$ non-trivial vertices, one does recover the operator 
we study, i.e. the lemniscatic CMS model. There is another way to 
think about this operator that was advocated in \cite{Buric:2020dyz}. Conformal 
blocks can be constructed through shadow integrals. Their integrands require to 
pick the quantum numbers of the intermediate fields as well as some tensor 
structure at each of the non-trivial vertices. There are many ways to select a 
basis of such 3-point tensor structures. But for generic choices the resulting
integral cannot be characterized as a common eigenfunction of a complete set of 
commuting differential operators. There is only one prescription that guarantees 
such a remarkable feature of the integral, namely to work with a basis whose
elements are eigenfunctions of the lemniscatic CMS model. Therefore, solving the 
eigenvalue problem for this Hamiltonian, or rather $N-4$ copies thereof, 
determines a distinguished set of integrands for the shadow integrals of comb 
channel blocks in 3D and 4D. Once this special choice is adopted, 
we can control the corresponding integrals through the 
powerful methods of integrable systems, just like in the case of 4-point 
blocks. The reduction from both OPE limits and shadow integrals are explained in more detail 
in subsection~\ref{sect:shadowOPE}. They highlight the key role that is played 
by the lemniscatic CMS model in developing a theory of multi-point blocks, 
at least in the comb channel of 3- and 4-dimensional conformal field 
theories. 
\medskip 

Before we end this introduction, let us briefly outline the content of this 
paper. In the next section we provide a more technical review of our 
previous work along with a detailed summary of results. 
Section~\ref{sect:threepointfunctions} is devoted to a construction of spinning 
3-point functions in the embedding space formalism. Such embedding space constructions go back to the 1970s, see e.g.\ \cite{Dobrev:1977qv}. While they have been extended 
in recent years, in particular to go beyond STTs \cite{Costa:2011mg,Costa_2015,Lauria:2018klo}, our treatment of tensor structures in 4-dimensional 
theories appears to be new. For the readers' convenience we have included a detailed
comparison with the more conventional twistor constructions in Appendix \ref{app:twistors_from_emb_space}. With a 
universal treatment of 3-point structures at hand, we can work out the fourth order 
vertex operators for all vertices that admit a single cross ratio in one go, see 
section~\ref{sect:ConstructionOperator}. Our results cover the vertex for one scalar 
and two STT fields in $d \geq 3$ that was already announced in our letter \cite{Buric:2020dyz}, along with two other types of vertices. For the example of the STT-STT-scalar 
vertex, we also explain the precise relation between the (reduced) single variable 
vertex operators and the vertex 
differential operators in multi-point functions constructed in \cite{Buric:2021ywo}, 
both through shadow integrals and OPE limits. In section~\ref{sect:ConstructionOperator},
it will take about two pages to spell out all the coefficients of the vertex differential 
operator. A drastic simplification is then achieved in section~\ref{sect:GeneralizedWeyl}, 
where we rewrite the vertex operator as an element of some appropriate deformation of a 
generalized Weyl algebra \cite{bavula1992generalized} associated with a Kleinian singularity 
\cite{hodges1993noncommutative,crawley1998noncommutative,holland1999quantization}. 
This algebra depends on the dimension $d$ and the spins of the fields, but not on their 
conformal weights. The latter only enter the expression for the vertex operator. A 
crucial step in this discussion is to realize that the 1-dimensional vertex systems 
come equipped with a scalar product that happens to coincide with the scalar product that 
makes Gegenbauer polynomials orthogonal. To make this paper somewhat self-contained, we 
include a detailed derivation of the scalar product in Appendix \ref{app:scalar_product}.
Section~\ref{sect:MappingElliptic} then contains the map of the vertex operators to 
Hamiltonians of the lemniscatic CMS model. The paper concludes with a discussion and 
overview of subsequent steps, along with a list of open problems. 

\section{Review and Summary of Results} 
\label{sect:reviewsummary}

The purpose of this section is to provide a technical review of our earlier 
papers \cite{Buric:2020dyz,Buric:2021ywo}. Once equipped with the relevant notations 
background, we will then be able to spell out the main new results of this paper. 
We proceed in three steps. In the first subsection we review the counting of cross 
ratios for scalar correlation functions in general, and for spinning vertex systems in 
particular cases. Next we sketch a group theoretic interpretation of the cross 
ratios for the spinning vertex system. This will provide the link to the Gaudin 
integrable model and the construction of vertex operators which we outline in 
the third subsection.  

\subsection{Cross ratios and single parameter vertices}

In order to count cross ratios and identify those spinning vertex systems that 
possess one single cross ratio we need a bit of notation. Given an OPE channel 
$\mathcal{C}^N_\textit{OPE}$ for some scalar $N$-point function in $d$-dimensional 
conformal field theory, we enumerate internal lines by Latin indices $r = 1 , \dots, 
N-3$ and vertices by Greek indices $\rho = 1, \dots, N-2$. External legs are 
enumerated by $i = 1, \dots, N$. OPE diagrams are (plane) trees and hence by 
cutting any internal line with label $r$ we separate the diagram into two 
disconnected pieces. Therefore, any choice $r$ of an internal line defines 
a partition of the external fields into two disjoint sets, 
\begin{equation}
\underline{N} = \{ 1, \dots, N\} = I_{r,1} \cupdot I_{r,2}\ .       
\end{equation} 
Similarly, any vertex $\rho$ gives rise to a partition of $\underline N$ into 
three disjoint sets 
\begin{equation} \label{eq:NIrho}
\underline{N} = I_{\rho,1} \cupdot I_{\rho,2} \cupdot I_{\rho,3}\ . 
\end{equation} 
The number of degrees of freedom a single vertex contributes depends on the spin 
of the three fields involved, i.e. whether they are scalars, symmetric traceless 
tensors etc. While scalars possess no vector indices, STTs 
carry a single set of such indices that are totally symmetric. In  dimensions higher than $d=3$, 
one can have more complicated tensor fields that contain several groups of symmetrized 
indices. We shall refer to the number of such groups as the \textit{(spin) depth} $L$ 
of the tensor.\footnote{Throughout this work we shall simply refer to $L$ as the depth,
not as the spin depth. In \cite{Buric:2020dyz} we had already defined the concept of depth 
$\dep$ as $\dep = L+1$. The spin depth we use here is the same concept, but measured 
with respect to the $\textrm{SO}(d)$ subgroup of the conformal group. In the following 
discussion, using $L$ rather than $\dep$ often avoids shifts by one unit in formulas.} 
The depth of the intermediate fields grows with the number of operator products 
that are required to construct them from scalars. More precisely, with the notation introduced above, the depth of a link $r$ in an OPE diagram is given by
\begin{equation} \label{eq:depth} 
L_r(\mathcal{C}^N_\textit{OPE},d) = L(I_{r,1},d), \quad \textit{where} 
\quad L(I,d) = \textit{min}(|I|,N-|I|,\rank_d)-1\ . 
\end{equation} 
Here, $\rank_d$ denotes the rank of the $d$-dimensional conformal algebra, i.e. the 
dimension of its Cartan subalgebra. Let us now look at a particular vertex $\rho$ 
in an OPE diagram in a $d$-dimensional conformal field theory. We call the ordered set 
$(L_{\rho,1},L_{\rho,2},L_{\rho,3})$ of depths $L_{\rho,k}$ of the three adjacent 
legs with $L_{\rho,1}\geq L_{\rho,2} \geq L_{\rho,3}$ the \textit{type} of 
the vertex. This type determines the number of degrees of freedom that are 
associated with $\rho$ according to the formula 
\begin{equation} \label{eq:novertrest} 
n_{\textit{vdo},\rho}(\mathcal{C}^N_\textit{OPE},d) = 
n_\textit{cr}(\sum_{k=1}^3 L_{\rho,k}+3,d) - 
\sum_{k=1}^3 L_{\rho,k} (L_{\rho,k}+1)  \ . 
\end{equation}
Here, $n_\textit{cr}(M,d)$ counts the total number of independent cross ratios of a 
scalar $M$-point function in $d$ dimensions,  
\begin{equation}
n_\textit{cr} (M,d) = \left\{ \begin{array}{ll} \frac12 M (M-3) \quad & M \leq d + 2 \\[3mm] 
Md - \frac12(d+2)(d+1) \quad &  M > d + 2 \end{array} \right. \ .  
\end{equation} 
As is well known, vertices $\rho$ with two scalar legs do not contribute any degree of 
freedom, i.e. $n_{\textit{vdo},\rho} = 0$ for $L_{\rho,1} = 0 = L_{\rho,2}$. Vertices 
for which none of the legs are scalar are easily seen to have at least two degrees of 
freedom. Hence, vertices that possess a single degree of freedom must necessarily have 
one scalar leg. Namely 
\begin{equation} \label{eq:list}  
(L_{\rho,1},L_{\rho,2},L_{\rho,3}) = \left\{ 
\begin{array}{rll}  \text{I}: & (1,1,0) \quad \textit{for} & d \geq 3 \\[1mm]
                   \text{II}: & (2,1,0) \quad \textit{for} & d \geq 4 \\[1mm]  
                   \text{III}:& (2,2,0) \quad \textit{for} & d = 4 
                   \end{array} \right. \ . 
\end{equation} 
\begin{figure}[thb]
\begin{subfigure}[b]{.33\textwidth}
\centering
\includegraphics{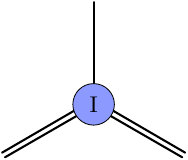}
\caption{STT-STT-scalar}
\label{fig:type1vertex}
\end{subfigure}%
\begin{subfigure}[b]{.33\textwidth}
\centering
\includegraphics{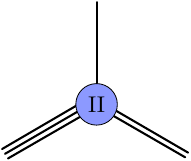}
\caption{MST$_2$-STT-scalar}
\label{fig:type2vertex}
\end{subfigure}%
\begin{subfigure}[b]{.33\textwidth}
\centering
\includegraphics{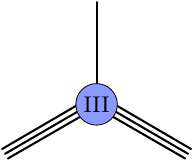}
\caption{MST$_2$-MST$_2$-scalar}
\label{fig:type3vertex}
\end{subfigure}
\caption{Vertices with an associated one-dimensional space of tensor structures. Single-lined legs are scalars; double or triple lines correspond respectively to STT and MST$_2$ representations. For the type III vertex in Figure~\ref{fig:type3vertex}, the space is two-dimensional and reduces to one dimension only in $d=4$.}
\label{fig:typevertices}
\end{figure}
Let us note that in $d > 4$ the vertex of type III possesses two degrees of 
freedom. The reduction to a single degree of freedom in $d=4$ is exceptional. 
In standard terms, type I vertices involve one scalar and two STTs, type II 
occur for one scalar, one STT and one Mixed-Symmetry Tensor (MST) of depth $L=2$ 
while type III contain one scalar and two MSTs of depth $L=2$. These three different 
types, depicted in Figure~\ref{fig:typevertices}, exhaust all those vertices that can appear in the comb channel of scalar $N$-point 
functions in $d=3$ and $d=4$ dimensions. By definition, all vertices in the comb 
channel have at least one external leg which is scalar, i.e. has $L =0$. Let us also 
note that for 5-point functions in any $d$ the only non-trivial vertex is of type I, 
which is included in our list. Similarly, for 6-point functions in the comb channel 
one only needs vertices of type II. In this sense, the theory we are about to describe 
addresses some of the vertices that are most relevant for applications. 

The construction of the single conformal invariant $\mathcal{X}$ that describes the 
vertices in our list \eqref{eq:list} from the coordinates and polarizations of the 
three individual fields, as well as the parameterization of 3-point functions of the 
three types in terms of this cross ratio, will be reviewed in the next section~\ref{sect:threepointfunctions}, see eqs.\ 
\eqref{eq:Vdef}, \eqref{eq:Hdef} and \eqref{eq:threepointcrossratios}. The precise embedding 
of individual (single variable) vertex systems into a multi-point function is illustrated in 
subsection~\ref{sect:shadowOPE} for the example of the type I vertex system and its relation 
with the vertex in a scalar 5-point function.

\subsection{Group theoretic reformulation of the vertex system}  

The main goal of our work is to characterize the three different types of 
vertices listed in eq.\ \eqref{eq:list} through some differential equation of 
fourth order. As we reviewed in the introduction, scalar $N$-point blocks can 
be characterized as joint eigenfunctions of $n_{\textit{cr}}(N,d)$ commuting 
Gaudin Hamiltonians. In an appropriate limit of parameters of the Gaudin model, 
the Hamiltonians were shown to include all the Dolan-Osborn Casimir operators
that measure the weight and spin of intermediate fields. The embedding of these 
operators into an $N$-site Gaudin model guarantees that the Casimir operators 
can be complemented into a full set commuting differential operators, one for 
each cross ratio, and it provides explicit expressions for the additional 
differential operators which can be associated with the vertices of the 
OPE diagram and are thereby referred to as vertex differential operators.  
The Gaudin model allows us to prepare the individual vertex systems, see 
\cite{Buric:2021ywo}. In the context of the present work, this is most 
easily described for the vertex of type III. The 4-dimensional conformal group 
$G=\mathrm{SO}(1,5)$ possesses $15$ generators. It has a number of interesting 
subgroups. In our discussion, two of them play a particular role. The first one 
is the parabolic subgroup $P = (\mathrm{SO}(1,1) \times \mathrm{SO}(4)) \ltimes 
\mathbb{R}^4$ that is generated by dilations, rotations and special conformal 
transformations. The quotient $G/P$ admits a transitive action of translations 
and is therefore 4-dimensional. The representations that are associated with 
scalar fields, i.e.\ representations of depth $L=0$, can be realized on the 
sections of line bundles over $G/P$. Another closely related realization of 
this representation is obtained on the space of holomorphic sections in a 
bundle over the complexification $G_\bC/P_\bC$. It is the latter version we 
shall adopt here. The second subgroup we need is the $9$-dimensional 
Borel subgroup $B_\bC \subset G_\bC$. In this case, the quotient $G_\bC/B_\bC$ 
is a flag manifold and we can realize any representation (of depth $L = 2$) on a 
space of holomorphic sections in a line bundle over it. Given a vertex of 
type I, it is now natural to assign the following coset space, 
\begin{equation} \label{eq:triplecoset}
 \mathcal{M} (2,2,0;d=4) = \left(G_\bC/B_\bC \times G_\bC/B_\bC \times G_\bC/P_\bC
 \right)/G_\bC\ .
\end{equation} 
Here, the complexified conformal group $G_\bC$ in the denominator acts diagonally 
from the left on the three factors in the numerator. Note that the numerator has 
dimension $6+6+4 = 16$. So, once we divide by the $15$-dimensional conformal group we 
end up with a $1$-dimensional quotient space. The coordinate $\mathcal{X}$ of 
this space is the unique degree of freedom that the vertex of type III 
contributes. A triple product of coset spaces, such as the one in equation 
\eqref{eq:triplecoset}, may be regarded as the configuration space of a 
$3$-site Gaudin integrable system. 

To treat other vertices we introduce the following family of subgroups 
$P_{d,L}$, $L = 0, \dots, \rank_d-1$, of the complexified $d$-dimensional 
conformal group $G_\bC$, 
\begin{equation} 
P_{d,L} = \mathcal{S}_{1,1}^{(d)}\left(\mathcal{S}_2^{(d-2)}\left(\cdots 
\mathcal{S}_2^{(d+2-2L)}\left(\mathcal{S}^{(d-2L)}_2(\mathrm{SO}_\bC(d-2L))
\right)\cdots \right) \right)  \subset G_\bC = \mathrm{SO}_\bC(1,d+1) \ .  
\end{equation} 
Here, $\mathcal{S}^{(M)}_2(H) \subset \mathrm{SO}_\bC(M+2)$ denotes a subgroup that is 
defined for any positive integer $M$ and any subgroup $H \subset \mathrm{SO}_\bC(M)$
as 
$$ \mathcal{S}^{(M)}_2 (H) = (\mathrm{SO}_\bC(2) \times H) \ltimes \mathbb{C}^M 
\subset \mathrm{SO}_\bC(M+2) $$
where the carrier space $\mathbb{C}^M$ of the fundamental representation 
of $H \subset \mathrm{SO}_\bC(M)$ is extended to a representation of 
$\mathrm{SO}_\bC(2) \times H$ by requiring that the elements of $\mathbb{C}^M$ carry one unit of $\mathfrak{so}(2)$ charge. This also ensures that 
$\mathcal{S}_2^{(M)}(H)$ becomes a subgroup of $\mathrm{SO}_\bC(M+2)$. We 
use a very similar construction to build 
$$ \mathcal{S}^{(d)}_{1,1}(H) = (\mathrm{SO}_\bC(1,1) \times H) \ltimes \mathbb{\bC}^d $$ 
for any subgroup $H \subset \mathrm{SO}_\bC(d)$. With $P_{d,L}$ fully defined we note 
that the first member $P_{d,0}$ of this family is the parabolic subgroup 
$P_{d,0} = P_\bC$ while the last one with $L = \rank_d-1$ coincides with the Borel 
subgroup $P_{d,\rank_d-1} = B_\bC$. One can thus realize the representation of the conformal group that is associated to a tensor field of depth $L$ on a line bundle over the quotient $G_\bC/P_{d,L}$. The choice of the line bundle is 
determined by the weight and spin of the field. With this notation, we can 
now define 
\begin{equation}
\mathcal{M}(L_1,L_2,L_3;d) = 
\left(G_\bC/P_{d,L_1} \times G_\bC/P_{d,L_2} \times G_\bC/P_{d,L_3}\right)/G_\bC\ .
\end{equation} 
Is is easy to see that the dimension of this space coincides with the number 
of independent conformal invariants that can be constructed from the insertion 
points and polarizations of three fields of depth $L_k$, i.e.\ 
\begin{equation}
 \textit{dim}_\bC\/ \left( \mathcal{M}(L_1,L_2,L_3;d) \right) = 
 n_\textit{cr}(\sum_{k=1}^3 L_{k}+3,d) - 
\sum_{k=1}^3 L_{k} (L_{k}+1)\ . 
\end{equation}
The space $\mathcal{M}$ is the configuration space of the integrable Gaudin 
model on the 3-punctured sphere with punctures of depth $L_k$.  

\subsection{From Gaudin Hamiltonians to Lemniscatic CMS models} 

The Gaudin Hamiltonians~\cite{Gaudin_76a,Gaudin_book83,Feigin:1994in} provide a complete set of commuting higher order 
differential operators on $\mathcal{M}$. The construction of these operators 
has been reviewed in \cite{Buric:2020dyz,Buric:2021ywo}. Here we shall content 
ourselves with a very brief review of the vertex system, see section 2 of 
\cite{Buric:2021ywo}. A key ingredient in the construction of the Gaudin 
model is its so-called Lax matrix, whose components in the basis 
$T^\alpha$ of the conformal Lie algebra are defined as
\begin{equation}\label{eq:Lax3pt}
\Lc^\rho_\alpha(w) = \sum_{k=1}^3 \frac{\mathcal{T}_\alpha^{(k)}}{w-w_k}\,,
\end{equation}
where $w$ is an auxiliary complex variable called the spectral parameter and 
we can fix the three complex parameters $w_k$ to be $w_1 = 0$, $w_2 = 1$ and 
$w_3 = \infty$. The symbols $\mathcal{T}_\alpha^{(k)}$ denote the first order differential 
operators that describe the action of the conformal algebra on the three spinning
primaries at the vertex or, equivalently, on the flag manifolds $G_{\bC}/P_{d,L_k}$
we introduced above. We have placed a superscript $\rho$ on the Lax matrix 
to emphasize that this is the matrix corresponding to the vertex $\rho$. 

For any elementary symmetric invariant tensor $\kappa_p$ of degree $p$ on the 
conformal Lie algebra, there is a corresponding $w$-dependent Gaudin Hamiltonian~\cite{Gaudin_76a,Gaudin_book83,Feigin:1994in}. 
Here we choose $\kappa_p$ such that the Hamiltonian takes the form 
\begin{eqnarray}\label{eq:GaudinHam}
\Hc^{(p)}_\rho(w) & = & \textit{str} \left( \Lc_{\alpha_1}^\rho(w) \cdots 
\Lc_{\alpha_p}^\rho(z) \right) + \dots,
\end{eqnarray}
where $\dots$ represent quantum corrections, involving a smaller number of 
components of the Lax matrix. The construction involves a symmetrized trace 
prescription in some appropriate representation, see \cite{Buric:2021ywo}
for details. The analysis in subsection 2.3 of \cite{Buric:2021ywo} shows that 
for the vertex systems in our list \eqref{eq:list} there is only one such independent Hamiltonian and it is of order $p=4$. Indeed we have argued there that the 
lower order operators are trivial while the higher order ones can be 
rewritten in terms of lower order operators. A non-trivial operator can be 
extracted from the family \eqref{eq:GaudinHam} with $p=4$ as 
\begin{equation} 
\mathcal{D}_\rho\equiv\mathcal{D}_{\rho,13}^{4,3}=
\text{str}\left(\mathcal{T}^{(1)}\mathcal{T}^{(1)}\mathcal{T}^{(1)}
\mathcal{T}^{(3)}\right)\,.
\label{eq:vertexop} 
\end{equation} 
For the single variable vertices listed in \eqref{eq:list} the Gaudin model 
provides the single differential operator of order four which depend on the 
conformal weights and spins of the three fields. We will work it out 
explicitly for all three cases, see section~\ref{sect:ConstructionOperator}. 
Our results extend the formulas 
given in our earlier announcement \cite{Buric:2020dyz} by including also the 
vertices of type II and III which we had not calculated before. The results 
are a bit cumbersome to spell out at first. 

In section~\ref{sect:GeneralizedWeyl} we will massage the answer and thereby pass to a much 
more compact algebraic formulation where we construct the Hamiltonian from 
the generators
of a deformation of some generalized Weyl algebra. The commutation relations of 
its three generators $A, A^\dagger$ and $N$ depend on the spins of the fields, 
see eqs.\ \eqref{NA*} - \eqref{A*A_alpha}. In the limit of $d=3$ this algebra 
is actually well known in the literature on quiver varieties where it appears 
as a generalized Weyl algebra or deformed/quantized Kleinian singularity of 
affine type $\tilde A_3$. Our deformation to $d \neq 3$ can be seen to possess 
finite dimensional representations whenever the spin quantum numbers are 
integers, and the dimension of these representations coincides with the 
number of 3-point tensor structures. Once the algebra generated by $A, A^\dagger$ and $N$ is 
introduced, the expression for the Hamiltonian can be stated in a single 
line, see eq. \eqref{HamFac}. Obviously, this Hamiltonian does depend on 
the choice of conformal weights, unlike the algebra it is a part of. In some 
sense, the formulas of section~\ref{sect:GeneralizedWeyl} provide the most 
compact formulation of our vertex operators and we believe that similar 
formulations are likely to exist for higher dimensional vertex systems. 
Nevertheless, for the main focus of the current paper, the material of 
section~\ref{sect:GeneralizedWeyl} may be considered supplementary. 

Section~\ref{sect:MappingElliptic} contains the main new result of this paper: there we 
show that the vertex operators for all three vertex systems listed in eq. \eqref{eq:list} 
can be mapped to a CMS Hamiltonian, namely the Hamiltonian for a crystallographic 
elliptic model that was originally discovered by Etingof, Felder, Ma and Veselov 
about a decade ago, see \cite{etingof2011107}. This lemniscatic CMS
Hamiltonian is spelled out in equation \eqref{LEFMV}. It is a fourth-order differential
operator in a single variable $z$. The relation between the cross ratio $\mathcal{X}$
of the vertex system and the new elliptic variable $z$ is stated in eq.\ \eqref{X_z_CoV}. 
The map involves Weierstrass' elliptic function $\wp(z)$. The lemniscatic Hamiltonian 
contains three non-trivial coefficient functions $g_p(z)$ which are defined in eqs.\ 
\eqref{gFp2}-\eqref{gFp0}. These coefficient functions depend on 12 multiplicities $m_{i,\nu}$ with $i=1, \dots, 4,$ and $\nu =0,1,2$, subject to the 
five constraints given in eq.\ \eqref{eq:mconstraint}, such that there are only seven remaining independent parameters. These determine the 
coupling constants in the coefficient functions $g_p(z)$ through eq.\ \eqref{eq:kMd}
and eqs.\ \eqref{level2}-\eqref{level4}. For each of the three single variable
vertex systems we determine the parameters $m_{i,\nu}$ in equations 
\eqref{eq:kMd} and \eqref{mMd10}-\eqref{mMd41} (type I, II; for type II one sets 
$\ell_2=0$) and \eqref{mMM410}-\eqref{mMM441} (type III). We note that in 
cases I and II, the vertex Hamiltonians do not exhaust the entire seven parameter family 
of lemniscatic models. In fact, for these two cases the multiplicities satisfy the additional 
constraint \eqref{eq:addmconstraint} that reduces the number of independent parameters to 
six.  Only for vertices of type III are the parameters of the lemniscatic model unrestricted.

\section{Three-Point Functions in Embedding Space}
\label{sect:threepointfunctions}

In our explicit construction of cross ratios and the calculation of the differential
operators we employ the embedding space formalism. Our presentation follows mostly \cite{Costa:2011mg}, which has advocated the usefulness of this formalism in the
context of spinning correlators, though restricted to STTs. More recently, this analysis 
was extended to mixed symmetry tensors, see \cite{Costa_2015,Lauria:2018klo}. 
The first subsection briefly reviews the construction of irreducible representations of 
the conformal group in embedding space, including spinning representations of arbitrary 
depth $L$. In the remaining two subsections we discuss the structure of 3-point functions 
for the three cases listed in eq. \eqref{eq:list}. Vertices of type I and II which exist 
for all sufficiently high dimensions, are treated together in the second subsection. The 
case of type III which is restricted to $d=4$ dimensions requires special treatment in the 
third subsection. Our construction of tensor structures and cross ratios for this case  
seems to be new even though the construction of 3-point tensor structures has a long 
history, see e.g. \cite{Mack:1976pa,Osborn:1993cr}. The use of embedding space formalism 
and polarization variables gives rise to an elegant reformulation that allows us to construct 
3-point correlators easily, up to a function $t$ of conformal invariant variables 
\cite{Buric:2020dyz} that is not determined by conformal symmetry. 

\subsection{Tensor representations in embedding space}
\label{subsec:MSTinembspace}

\begin{figure}[thb]
\centering
\includegraphics{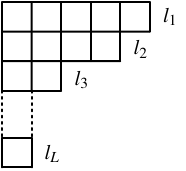}
\caption{\small Young diagram associated with the representation labeled by integers $l_1\ge l_2\ge 
\dots \ge l_L$ where $l_\nu$ represents the length of the $\nu$-th row of the diagram.}
\label{fig:Youngdiagram}
\end{figure}

Tensor fields in $d$-dimensional CFT are irreducible representations of the 
${\mathfrak{so}(1,d+1)}$ algebra labeled by an $\mathfrak{so}(1,1)$ weight $\Delta$, 
the conformal dimension, and an ordered set of $\mathfrak{so}(d)$ weights $l_1\ge l_2
\ge\dots\ge l_L$ which we refer to as the \emph{spins} of the representation. Focusing 
on bosonic representations, we associate representations of $\mathfrak{so}(d)$ with 
Young diagrams, where the integers $l_\nu$ represent the length of the $\nu$-th row 
of the diagram as in~\autoref{fig:Youngdiagram}.

More explicitly, these representations correspond to traceless tensors whose indices can be 
grouped following the rows of the Young diagram, thereby making the symmetry of permuting indices within those 
groups manifest
\begin{equation}
    \mathbf{F}^{\Delta}_{l_1,\dots,l_L}(x)\equiv \left(  F_{\left(a^{(1)}_1 \dots a^{(1)}_{l_1}\right)\dots 
    \left(a^{(L)}_1\dots a^{(L)}_{l_L}\right)}(x) \right)\,.
    \label{MSTindices}
    \end{equation}
Note that one could equivalently focus, instead, on the columns of the Young 
diagram and therefore make the antisymmetries of such indices manifest, as 
detailed in~\cite{Costa_2015}. 

We denote a Mixed-Symmetry Tensor like~\eqref{MSTindices} with $L\ge 2$ groups of 
symmetric indices as MST$_L$, while for the single spin case $L=1$ we use the 
standard terminology of Symmetric Traceless Tensor (STT). For the computation of 
the vertex differential operators and their map to elliptic CMS models we need only to consider the STT and MST$_2$ cases. Finally, with this notation, it is important to note that tensors of the form \eqref{MSTindices} do \emph{not} form irreducible representations when the dimension is even and the depth is maximal, i.e. $d=2L$. The first non-trivial case, $d=2L=4$ and $l_1=l_2=1$,  corresponds to a field strength $F_{ab}=-F_{ba}$ with dual field strength $\star F_{ab}:= \frac{1}{2} \epsilon_{abcd} F^{cd}$. The latter are known to decompose into two irreducible representations of $\mathrm{SO}(4)$: the self-dual part, $\star F = F$, and the anti-self-dual part, $\star F = -F$. The duality map $\star$ that distinguishes irreducible representations at $L=d/2$ can be generalized to any spins $l_1,\dots, l_L$, and we will explain below how we project to the irreducible self-dual and anti-self-dual subrepresentations in our formalism.
\smallskip 

The main purpose of this subsection is to re-express the rather conventional 
description of tensor fields and the associated representations in a way that 
simplifies explicit computations and makes results more compact. This is achieved  
using the so-called embedding space formalism and a generalized index-free notation 
using polarization variables. 

The \textit{embedding space formalism} realizes points on $\mathbb{R}^d$ with a 
non-linear action of the conformal group in terms of projective null rays that 
live in a $(d+2)$-dimensional Minkowski space and admit a linear action of 
$\mathrm{SO}(1,d+1)$. Given any field of conformal weight $\Delta$, one can use 
this relation to define the uplift
\begin{equation}
    \mathbf{F}^\Delta_{\mathbf{l}}(x)\rightarrow \mathbf{F}^\Delta_{\mathbf{l}}(X)\,, 
    \qquad \qquad  \left\{X\in \mathbb{R}^{1,d+1}\,|\, X^2=0\right\}\,,
\end{equation} 
to a function on light-like vectors that is homogeneous of degree $-\Delta$ with respect 
to rescalings of $X$,
\begin{equation} \label{eq:confweight}
    \mathbf{F}_{\mathbf{l}}^{\Delta}(\lambda X)=\lambda^{-\Delta} 
    \mathbf{F}_{\mathbf{l}}^{\Delta}(X)\,. 
\end{equation}
For fields with spin, all physical space indices $a$ get promoted to embedding space 
indices $A$. The tensor \eqref{MSTindices} in physical space can be recovered from 
the tensor in embedding space by contracting each index with a copy of the Jacobian 
$\pdv{X^A}{x^a}$. In order to transform irreducibly under the action of the conformal 
group, the uplifted fields possess a number of additional properties. In particular, 
they are required to be transverse with respect to any index of the tensor in embedding space, 
\begin{equation}
    X^{\,A_i^{(\nu)}}F^{\Delta}_{\left(A^{(1)}_1 \dots A^{(1)}_{l_1}\right)\dots 
    A_i^{(\nu)}\dots \left(A^{(L)}_1\dots A^{(L)}_{l_L}\right)}(X)=0\,,
\end{equation}
and traceless with respect to any pair of indices 
\begin{equation}
    \eta^{A_i^{(\mu)}A_j^{(\nu)}}
    F^{\Delta}_{A^{(1)}_1 \dots A^{(\mu)}_{i}\dots A_j^{(\nu)}\dots A^{(L)}_{l_L}}(X)=0\,.
\end{equation}
Here, the capital letters $A$ are vector indices in embedding space and $(\eta_{AB})$ is the $(d+2)$-dimensional Minkowski metric. The embedding space formalism relates conformal 
transformations of $d$-dimensional space to Lorentz transformations of the embedding 
space coordinates,
\begin{equation}\label{generators-spacetime-part}
 \mathcal{T}_{AB}=X_A 
    \pdv{X^B}-X_{B}\pdv{X^A}\,.
\end{equation}
When transforming tensor fields one needs to add additional terms that act on 
the tensor indices. Instead of detailing these terms, however, we now want to explain 
how to get rid of all the tensor indices. The idea is to encode the 
tensor components of the field as coefficients of some polynomial in several 
variables. This has been known for a long time, at least for STTs , see e.g. 
\cite{Bargmann_1977,Dobrev:1977qv}. The extension to more general mixed-symmetry 
tensors of depth $L > 1$ comes in two variants, one that encodes the 
antisymmetrization between different rows of a Young diagram with fermionic coordinates, see \cite{Costa_2015}, and another that instead encodes the 
symmetrization between different columns of a Young diagram with bosonic coordinates, see \cite{Lauria:2018klo}. 
Here we shall adopt the bosonic approach and introduce one auxiliary polarization 
vector $Z_\nu\in \mathbb{C}^{d+2}$ for every spin quantum number $l_\nu$ of the 
MST; these polarization vectors are contracted with the MST to form a polynomial 
in all of the polarizations
\begin{equation}
    F^{\Delta}_{l_1,\dots,l_L}\left(X,Z_1,\dots,Z_L\right)\equiv 
    F^{\Delta}_{\left(A^{(1)}_1 \dots A^{(1)}_{l_1}\right)\dots \left(A^{(L)}_1\dots A^{(L)}_{l_L}\right)}(X)\,\left(Z_1^{A_1^{(1)}}\cdots 
    Z_1^{A_{l_1}^{(1)}}\right)\cdots \left(Z_L^{A_1^{(L)}}\cdots 
    Z_L^{A_{l_L}^{(L)}}\right)\,.
\end{equation}
The properties of  tracelessness and transversality of 
the tensor $F_{\{A_i^{(\nu)}\}}(X)$ are translated into the conditions 
\begin{equation} \label{eq:XXXZZZ}
    X^2=X\cdot Z_\nu=Z_\nu \cdot Z_\mu=0
\end{equation}
for the coordinates. In addition, these new objects obey the following 
multiple homogeneity condition for a field of conformal weight $\Delta$ and 
with spin labels $l_\nu$, 
\begin{equation}
    F^{\Delta}_{\{l_\nu\}}\!\left(\lambda_0 X, \{ \lambda_\nu
    Z_{\nu}\}\right)=\lambda_0^{-\Delta}\lambda_1^{l_1}\cdots \lambda_L^{l_L}
    F^\Delta_{\{l_\nu\}}(X,\{Z_\nu\})\,.
    \label{homogeneityconditions}
\end{equation}
This extends the condition \eqref{eq:confweight} and rephrases that fields 
with spin $l_\nu$ have a polynomial dependence on $Z_\nu$ with homogeneous 
degree $l_\nu$. Finally, the dependence on the polarizations respects
the following set of gauge invariance conditions
\begin{equation}
    F^\Delta_{\{l_\nu\}}\!\left(X,\left\{Z_\nu+\beta_{\nu,0} X +\sum_{\mu<\nu} 
    \beta_{\nu,\mu} Z_\mu\right\}\right)=F^\Delta_{\{l_\nu\}}(X,\{Z_\nu\})\,, \qquad 
    \forall \beta_{\nu,\mu}\in \mathbb{C}\,.
    \label{gaugeinvariance}
\end{equation}
In this formulation, the generators act on fields as derivations that include terms involving polarizations next to the spacetime part \eqref{generators-spacetime-part}, 
\begin{equation}
    \mathcal{T}_{AB}=X_A\pdv{X^B}+\sum_\nu Z_{\nu\,A}\pdv{Z^{B}_\nu} - 
    (A\leftrightarrow B)\,.
    \label{eq:spinningdiffgenerators}
\end{equation}
Before we conclude this brief presentation of embedding space for tensor fields,
we want to add a couple of comments. First, note that functions in the variables
$X, Z_\nu$ can be assigned a multi-degree that has $L+1$ components, one for 
the variable $X$ and then one for each of the $L$ polarizations $Z_\nu$. The assignment 
is such that a field $F$ with weight $\Delta$ and spins $l_\nu$ has degree $[-\Delta,
l_1, \dots, l_L]$. This degree is measured by the independent rescalings of the 
variables that we have introduced. 
\smallskip 

At this point, we have rephrased the concept of a tensor field of weight $\Delta$ and 
spin $l_\nu$ in terms of functions $F$ of the variables $(X,Z_\nu)$ subject to the 
conditions \eqref{eq:XXXZZZ}. These functions must satisfy the homogeneity conditions
\eqref{homogeneityconditions}, as well as the gauge invariance conditions 
\eqref{gaugeinvariance}. These two conditions ensure that the differential operators 
\eqref{eq:spinningdiffgenerators} give rise to an irreducible representation of the 
conformal algebra. From now on, we will think of tensor fields in terms of functions 
$F(X,Z_\nu)$. Let us note in passing that the homogeneity conditions 
\eqref{homogeneityconditions} can be continued to non-integer values of $l_\nu$. 
\smallskip 

It is also important to notice how the gauge invariance conditions~
\eqref{gaugeinvariance} constrain the way in which the variables $Z_\nu$ can appear 
in expressions that involve the field $F_{\Delta,\{l_\nu\}}$. In fact, the only gauge invariant 
tensors that can be formed from $(X,Z_{\nu})$ are linear combinations or contractions of the wedge products (see also \cite[eq.~27]{Lauria:2018klo}):
\begin{equation}
    C^{(0)}_A=X_A\,, \qquad C^{(\nu)}_{A_1\dots A_{\nu+1}}=
    \left(X\wedge\bigwedge_{\mu=1}^{\nu}Z_\mu\right)_{A_1\dots A_{\nu+1}}\,.
    \label{buildingblocks}
\end{equation}
Let us point out that the projective light ray contains $d$ degrees of freedom. 
After imposing transversality $X \cdot Z_1 = 0$ and the gauge invariance 
\eqref{gaugeinvariance}, there remain $d-2$ degrees of freedom in the polarization 
$Z_1$. Similarly, $Z_2$ contains $d-4$ degrees of freedom, etc. This implies that the variable $Z_{L}$ for tensor fields of maximal depth 
$L=\rank_d-1=d/2$ in even dimensions has no continuous physical degrees of freedom. 
In the reduction from embedding space variables to gauge-invariant tensors, all $C^{(0)},\dots, C^{(L-1)}$ are fixed by $X,\dots, Z_{L-1}$, while $C^{(L)} = C^{(L-1)} \wedge Z_L$. Up to gauge equivalence, this implies that $\mathrm{Span}(Z_L)$ is fixed to be one of two unique null directions in the complex plane orthogonal to $\mathrm{Span}(X,\dots,Z_{L-1})$. To distinguish these two null directions, we can use the fact that $C^{(L)}$ is a $(L+1)$-form in $\Cs^{2(L+1)}$ given by the wedge product of $L+1$ mutually orthogonal null vectors, and must therefore be either self-dual or anti-self-dual with respect to the Hodge star,
\begin{equation}
    \star \tensor{C}{^{(L)}_{A_1\dots A_{L+1}}} = \frac{1}{(L+1)!} \epsilon_{A_1\dots A_{2L+2}} C^{(L) A_{L+2}\dots A_{2L+2}} = \pm\, \mathrm{i}^{L} \,C^{(L)}_{A_1\dots A_{L+1}}.
    \label{hodge}
\end{equation}
The above condition separates the space of gauge equivalence classes of $(X,Z_1,\dots,Z_L)$ into two distinct $\mathrm{SO}(1,d+1)$ orbits: the \emph{self-dual}  and the \emph{anti-self-dual} one, according to the eigenvalue in \eqref{hodge}. Contracting a tensor with null vectors in the (anti-)self-dual orbit projects said tensor to its (anti-)self-dual part, such that the restriction of $F^{\Delta}_{l_1,\dots,l_L}\left(X,Z_1,\dots,Z_L\right)$ to one of these orbits defines an irreducible representation of $\mathfrak{so}(1,d+1)$. In subsection~\ref{sect:four_dimensions_3pt}, we find a concrete parameterization of the two orbits for $d=4$ in a gauge given by two iterated Poincaré patches. To our knowledge, this is the first time that such irreducible representations are constructed directly in $d=4$ embedding space.

All of these remarks on degrees of freedom match nicely with the construction of 
flag manifolds that was outlined in 
the previous section. In particular, we see that physical degrees of freedom that 
reside in $X, Z_1, \dots, Z_L$ encode positions on the quotients\footnote{The 
real form of $\mathrm{SO}(d+2)$ corresponding to the Euclidean conformal group 
allows us to remove the complexification for the first quotient, i.e. one can 
work with real $X$. This is because the parabolic subgroups $P_{d,L}$ admit a 
real form where the translation, special conformal and dilation generators are 
all real. On the other hand, $\mathrm{SO}_\mathbb{R}(d) \cap P_{d,L} = 
\mathrm{SO}_\mathbb{R}(d-2L)$, so we cannot relate representations of 
$\mathrm{SO}_\mathbb{R}(d)$ with reality conditions on the $Z_{\nu}$. The remaining quotients are associated instead with the complex polarizations $Z_\nu$, and the tensor fields are constrained to depend on them holomorphically. This can be understood as a concrete realization of the Borel-Weil theorem for 
unitary representations of $\mathrm{SO}_{\mathbb{R}}(d)$.}  
\begin{equation} 
\mathrm{SO}(1,d+1)/(\mathrm{SO}(1,1) \times \mathrm{SO}(d))\ltimes \mathbb{R}^d 
\quad , \quad 
\mathrm{SO}_\bC(d+2-2\nu)/(\mathrm{SO}_\bC(2) \times \mathrm{SO}_\bC(d-2\nu))\ltimes 
\mathbb{C}^{d-2\nu} 
\end{equation} 
for $\nu = 1, \dots, L$, respectively, that appear as part of the flag manifold 
$G_{\bC}/P_{d,L}$ (see Appendix \ref{ssec:nested_poincare}). The degrees we 
introduced above correspond to the transformation behavior of the quotients under 
the action of $\mathrm{SO}(1,1)$ and the $L$ subgroups $\mathrm{SO}_\bC(2)$ that 
split off when we construct $P_{d,L}$. 

\subsection{Spinning 3-point functions in embedding space}
\label{ssec:3ptI-II} 

We are now interested in those 3-point functions for which conformal symmetry 
leaves one free parameter, i.e. the three configurations of spinning fields 
listed in eq.\ \eqref{eq:list}. These correspond to the vertices for STT-STT-scalar 
in $d\ge 3$, MST$_2$-STT-scalar in $d\ge 4$, and MST$_2$-MST$_2$-scalar in $d=4$, 
respectively. In the section~\ref{sect:ConstructionOperator} we will actually address the computation of the 
vertex operators for these three cases through a single computation by passing 
through the 3-point function for MST$_2$-MST$_2$-scalar in $d > 4$. From there, we can then descend to the three cases we are interested in. As one can easily see, 
the vertex of type $(2,2,0)$ in $d > 4$ comes with two cross ratios and carries seven quantum numbers: three conformal weights and four spin labels. 
In order to descend to the three types in the list \eqref{eq:list}, we need to 
specialize the quantum numbers and restrict to a single cross ratio, see 
below. 

To simplify notation and avoid multiple indices, we will rename our 
variables from here on as
\begin{equation}
    Z\equiv Z_1\,, \quad W\equiv Z_2\,, \quad l\equiv l_1\,, \quad \ell \equiv l_2\,,
\end{equation}
and use Latin indices $i,j,k=1,2,3$ to run over the three points. For instance, the 
first field $\phi_{\Delta_1,l_1,\ell_1}(X_1,Z_1,W_1)$ has spins $l_1 \geq \ell_1 
\geq 0$ and depends on the coordinate $X_1$ and the two polarization vectors $Z_1$ and 
$W_1$, and similarly for the fields $\phi_{\Delta_2,l_2,\ell_2}(X_2,Z_2,W_2)$ 
and $\phi_{\Delta_3}(X_3)$.

The first task now is to find which non-vanishing independent tensor structures 
can be constructed from the gauge invariant quantities~\eqref{buildingblocks}. This means building a set of conformal invariants from the position variables $X_i$ and the 
polarizations $Z_i,W_i$ that generate functions of any degree in all 
of the seven variables, along with the two cross ratios. The latter are conformal invariants 
of vanishing degree. To begin with, we have the scalar products of position vectors
\begin{equation} 
    X_{12}=X_1\cdot X_2\,, \qquad X_{23}=X_2\cdot X_3\,, \qquad X_{31}=X_3\cdot X_1\,.
    \label{scalarproducts}
\end{equation}
If we denote the multi-degree of a MST$_2$-MST$_2$-scalar vertex by 
$[-\Delta_1,l_1,\ell_1;-\Delta_2,l_2,\ell_2;-\Delta_3]$, these scalar products have degree 
$\textit{deg\/}X_{12}=[1,0,0;1,0,0;0]$ etc. Next, it is customary to introduce the following 
contractions of the two-forms with two position vectors
\begin{equation} \label{eq:Vdef} 
    V_1=V_{1,32}=\frac{X_3\cdot\left(X_1\wedge Z_1\right)\cdot X_2}{X_{23}}\,, 
    \qquad V_2=V_{2,13}=\frac{X_1\cdot\left(X_2\wedge Z_2\right)\cdot X_3}{X_{31}}\,. 
\end{equation}
The two objects $V_1$ and $V_2$ have degree $\textit{deg\/}\,V_1 = [1,1,0;0,0,0;0]$ and 
$\textit{deg\/}\,V_2 = [0,0,0;1,1,0;0]$. Another simple tensor structure is given by the 
the contractions of the two-forms
\begin{equation} \label{eq:Hdef} 
    H_{12}=\frac{1}{2}(X_1\wedge Z_1)\cdot(X_2\wedge Z_2)\,. 
\end{equation}
It has degree $\textit{deg\/}\, H_{12} = [1,1,0;1,1,0;0]$. The tensor structures we have introduced 
so far do not depend on the polarizations $W_i$, in contrast to the remaining three variables
that we will introduce now. These include the following contractions of a three-form with a two-form 
and a vector,
\begin{equation} \label{eq:Udef} 
    U_{123}=\frac{1}{2}(X_1\wedge Z_1\wedge W_1)_{ABC}(X_2\wedge Z_2)^{AB}X_3^{C}\,, \qquad U_{213}=\frac{1}{2}(X_2\wedge Z_2\wedge W_2)_{ABC}(X_1\wedge Z_1)^{AB}X_3^{C}\,,
\end{equation}
and, finally, the contraction of the two three-forms
\begin{equation}
    K_{12}=\frac{1}{3!}(X_1\wedge Z_1\wedge W_1)\cdot (X_2\wedge Z_2\wedge W_2)\,.
    \label{Kappaonetwo}
\end{equation}
In the notation of \cite{Lauria:2018klo}, these MST$_2$ tensor structures correspond to $U_{ijk} = T_{i,jk}^{3,21}$ and $K_{ij} = T_{i,j}^{3,3}$. The degrees of these three tensor structures are $\textit{deg\/}\,U_{123} = [1,1,1;1,1,0;1]$, 
$\textit{deg\/}\, U_{213} = [1,1,0;1,1,1;1]$ and $\textit{deg\/}\,K_{12} = [1,1,1;1,1,1;0]$.
This concludes our list of building blocks of tensor structures for the MST$_2$-MST$_2$-scalar vertex in $d>4$. For the reader's convenience we listed the tensor structures and their degrees in Tab. \ref{tab:TS_d>4}.
\smallskip 

As one can easily count, we have written nine independent tensor structures, the degrees of which span the entire $7$-dimensional space of multi-degrees. Since we know that the 
MST$_2$-MST$_2$-scalar vertex admits two cross ratios, the tensor structures we introduced 
indeed suffice to decompose the 3-point function in the following way 
\begin{equation}
    \Phi_{123}(X_i;Z_i,W_i) := \expval{\phi_{\Delta_1,l_1,\ell_1}(X_1,Z_1,W_1)
    \phi_{\Delta_2,l_2,\ell_2}(X_2,Z_2,W_2)\phi_{\Delta_3}(X_3)}=
    \Omega^{\Delta_1,\Delta_2,\Delta_3}_{l_1,l_2;\ell_1,\ell_2} \, t(\mathcal{X},\mathcal{Y})\,,
    \label{eq:threepointtwoMST}
\end{equation}
where $\Omega^{\Delta_1,\Delta_2,\Delta_3}_{l_1,l_2;\ell_1,\ell_2}$ is a prefactor that 
takes care of all homogeneity conditions~\eqref{homogeneityconditions}, i.e. it is a
product of powers of tensor structures that matches the degree of the correlation function 
on the left hand side. The function $t(\mathcal{X},\mathcal{Y})$ is a conformal invariant 
that depends on two variables of vanishing degree, 
\begin{equation}
    \mathcal{X}=\frac{H_{12}}{V_1 V_2}\,, \qquad \mathcal{Y}=\frac{X_{13} X_{23} 
    V_1 V_2 K_{12}}{X_{12}U_{123}U_{213}}\,.
    \label{eq:threepointcrossratios}
\end{equation}
Of course, the prefactor $\Omega$ is not uniquely fixed by the homogeneity condition simply 
because it is possible to form objects of vanishing degree from the nine tensor structures. 
The remaining freedom can be fixed by choosing not to employ $H_{12}$ and 
$K_{12}$ in the construction of $\Omega$. This leaves us with a unique prefactor satisfying 
all of the required homogeneities in 
$(X_i,Z_i,W_i)$, 
\begin{equation}
    \Omega^{\Delta_1,\Delta_2,\Delta_3}_{l_1,l_2;\ell_1,\ell_2}=
    \frac{V_1^{l_1-\ell_1-\ell_2}V_2^{l_2-\ell_1-\ell_2}U_{123}^{\ell_1}U_{213}^{\ell_2}}
    {X_{12}^{\frac{\Delta_1+\Delta_2-\Delta_3+l_1+l_2-\ell_1-\ell_2}{2}}
    X_{23}^{\frac{\Delta_2+\Delta_3-\Delta_1-l_1+l_2+\ell_1+\ell_2}{2}}
    X_{31}^{\frac{\Delta_3+\Delta_1-\Delta_2+l_1-l_2+\ell_1+\ell_2}{2}}}\,.
    \label{prefactor2MST}
\end{equation}
After one has fixed $\Omega$, the remaining freedom in the 3-point function is normally 
taken into account by expanding in a discrete set of 3-point tensor structures with 
vanishing multi-degree. These are then combined with OPE coefficients to make up the full 
correlator. The standard tensor structures, once homogenized, represent a basis for the 
space of $t(\mathcal{X},\mathcal{Y})$, but this basis is of course not unique. As 
we have recalled in the introduction, it is one of the key observations in 
\cite{Buric:2020dyz} that a distinguished basis arises naturally from the study of 
higher-point conformal blocks, as eigenfunctions of our new vertex differential 
operators.%
\medskip%

Now that we have parametrized the MST$_2$-MST$_2$-scalar vertex in terms 
of the function $t(\mathcal{X},\mathcal{Y})$ of two cross ratios, we need to explain how 
to descend to the three types of vertices we are interested in. We will postpone the reduction to type III to subsection~\ref{sect:four_dimensions_3pt}, and we will address here types I and II that are the simplest to discuss. 
In these two cases, the variable $W_2$ does not appear because the spinning field $\phi_{\Delta_2,
l_2,\ell_2}$ is in an STT representation. Therefore,
we only have seven tensor structures whose degrees span a 6-dimensional space of degrees. 
Consequently, there can be only one non-trivial cross ratio which is $\mathcal{X}$. Since 
it is impossible to construct the cross ratio $\mathcal{Y}$, the function $t(\mathcal{X},
\mathcal{Y})$ cannot depend on it, and therefore reduces to $t(\mathcal{X})$.

Going to type I does not impose any further constraint on the remaining variable 
$\mathcal{X}$. Indeed, without a variable $W_1$ it is not possible to construct the
tensor structure $U_{123}$ so that we remain with six tensor structures whose degrees
span a 5-dimensional space of degrees. The construction of $\mathcal{X}$ is not affected. 

The upshot of all this discussion is very simple: for vertices of type I and II in our 
list \eqref{eq:list} the 3-point function assumes the form spelled out in eq.\
\eqref{eq:threepointtwoMST}, but with a function $t$ that depends only on 
$\mathcal{X}$ and not on $\mathcal{Y}$, 
\begin{equation}
    \expval{\phi_{\Delta_1,l_1,\ell_1}(X_1,Z_1,W_1)\phi_{\Delta_2,l_2,\ell_2=0}
    (X_2,Z_2)\phi_{\Delta_3}(X_3)}=\Omega^{\Delta_1,\Delta_2,\Delta_3}_{l_1,l_2;
    \ell_1,\ell_2=0} t(\mathcal{X})\,,
    \label{eq:threepointtwoMSTonecrossratio}
\end{equation}
where the prefactor $\Omega$ is given in eq.\ \eqref{prefactor2MST} and
for vertices of type I one imposes $\ell_1 = 0$. 
\medskip 

Before we conclude this section, we want to carry our discussion of the function $t(\mathcal{X})$
one step further. So far we have not enforced spin labels to be integers so that our general form \eqref{eq:threepointtwoMSTonecrossratio} still applies to 3-point functions of objects with continuous spin. Now we would like to explore the additional conditions that arise from the 
restriction to spins with integer values. We already saw in subsection~\ref{subsec:MSTinembspace} 
that MSTs depend polynomially on the auxiliary variables $Z_i$. This rather basic fact constrains $t(\mathcal{X})$ to live in a finite-dimensional space, as one can infer from the definition \eqref{eq:threepointcrossratios} of the cross ratio $\mathcal{X}$. The tensor structures $V_i$ 
that appear in its denominator each contain factors of $Z_i$. Therefore, the highest power of $V_i$ from 
the denominators of $t(\mathcal{X})$ must not exceed the power of $V_i$ that appears in the numerator 
of the prefactor \eqref{prefactor2MST} in order to ensure polynomial dependence on the $Z$ variables.
This provides an upper bound on the exponent $M$ of $\mathcal{X}^M$ in a series expansion 
of $t(\mathcal{X})$. Negative powers of $\mathcal{X}$ are not possible either, as these would produce 
the tensor structure $H_{12}$ in the denominator, which itself contains both $Z_1$ and $Z_2$
but cannot be compensated by the prefactor $\Omega$ that does not contain $H_{12}$. In conclusion, 
$t(\mathcal{X})$ must be a polynomial of order up to $n_t=\text{min}(l_1-\ell_1,
l_2-\ell_1)$ if $n_t\ge 0$, and it must vanish if $n_t<0$. The set of all allowed functions $t(\mathcal{X})$ therefore spans an $(n_t+1)$-dimensional space of tensor structures. For type III vertices, we will be able to write 3-point functions in the same form as~\eqref{eq:threepointtwoMSTonecrossratio} with $\ell_2\ne 0$. However, this last discussion on the polynomiality of $t(\mathcal{X})$ and the space of tensor structures will be substantially different.

There is a slight twist to this story that is relevant for the STT-STT-scalar vertex in $d=3$. 
Note that all the tensor structures we have introduced so far are even under parity. So all 
of the 3-point tensor structures that they generate are also parity even. But for the type I vertex in $d=3$, it is also possible to construct a parity-odd tensor structure given 
by 
\begin{equation} 
O^{(3)}_{123} = \epsilon_{ABCDE}X_1^A X_2^B X_3^C Z_1^D Z_2^E\ . 
\end{equation}
Its degree in the 5-dimensional space of degrees for type I vertices is $\textit{deg\/}\, O_{123}=
[1,1;1,1;1]$. The square of this parity-odd tensor structure must be parity-even, and it can be
expressed in terms of tensor structures constructed above as 
\begin{equation}
    \left(O^{(3)}_{123}\right)^2 \propto 
    (1-\mathcal{X})\mathcal{X} V_1^2 V_2^2\frac{ X_{13}X_{23}}{X_{12}}\,.
    \label{eq:parityoddsquared}
\end{equation}
We infer from this equation that parity-odd 3-point tensor structures contain factors of
$\sqrt{(1-\mathcal{X})\mathcal{X}}$, generalizing the polynomial space of $t(\mathcal{X})$ to 
also include these half-integer powers. A similar analysis can be done for the type II vertex
in $d=4$. However one finds no extension of the space of polynomials; we will describe this in the following subsection.

\begin{table}[t]
    \centering
\begin{tabular}{l|c c c c c c c}
    &  $-\Delta_1$ & $-\Delta_2$ & $-\Delta_3$ & $l_1$ & $l_2$ & $\ell_1$ & $\ell_2$ \\
    \midrule
$H_{12}$ & $1$ & $1$ & $0$ & $1$ & $1$ & $0$ & $0$  \\
$K_{12}$ & $1$ & $1$ & $0$ & $1$ & $1$ & $1$ & $1$  \\
$V_1$ & $1$ & $0$ & $0$ & $1$ & $0$& $0$ & $0$  \\
$V_2$ & $0$ & $1$ & $0$ & $0$ & $1$& $0$ & $0$ \\
$U_{123}$ & $1$ & $1$ & $1$ & $1$ & $1$& $1$ & $0$ \\
$U_{213}$ & $1$ & $1$ & $1$ & $1$ & $1$& $0$ & $1$
\end{tabular}
     \caption{Degrees of tensor structures of the MST$_2$-MST$_2$-scalar 3-point function in $d>4$.}
   \label{tab:TS_d>4}
\end{table}

\subsection{Embedding space construction in \textit{d} = 4 dimensions}
\label{sect:four_dimensions_3pt}

We now address the restriction of the 3-point function
\eqref{eq:threepointtwoMST} to $d=4$, thereby describing vertices of type III. 
In going from $d>4$ to $d=4$, the number of independent cross-ratios reduces 
from two to just one. One may think of this reduction in terms of a constraint 
that is imposed on the variable $\mathcal{Y}$. The simplest way to understand 
the need of a reduction in cross-ratio space is by observing that the embedding 
space for a theory in $d=4$ dimensions is 6-dimensional, while the MST$_2$-MST$_2$-scalar
3-point function described in the previous subsection depends on seven vectors
$(X_1,Z_1,W_1,X_2,Z_2,W_2,X_3)$. Seven vectors in a six-dimensional space must be 
linearly dependent, which is equivalent to the vanishing of the determinant of their Gram matrix (matrix of scalar products). For the case at hand, the Gram determinant is easily computed in terms of our tensor structures,
and the vanishing condition becomes 
\begin{equation}
    \frac{X_{12}}{X_{13} X_{23}}\frac{ U_{213} U_{123}}{V_1 V_2}
    \frac{\mathcal{Y}^2(-1+\mathcal{X})+\mathcal{Y}}{\mathcal{X}}=0\,,
    \label{GramDeterminant}
\end{equation}
with two solutions in cross-ratio space: $\mathcal{Y}=0$ and
$\mathcal{Y}=1/(1-\mathcal{X})$. To understand the reason why two different
solutions appear and what each one means, we first need to explain in more detail some properties of the embedding space representation of mixed-symmetry 
tensors in $d=4$.

As we already anticipated at the end of section~\ref{subsec:MSTinembspace}, the
representations labeled by Young diagrams that we considered so far are reducible in 
even dimensions when $L=d/2$, and further decompose into irreducible self-dual 
and anti-self-dual representations. To see this more concretely, let us write 
out the $X$ and $Z$ vectors in the following Poincar\'e patches,
\begin{equation}
    \begin{gathered}
    X=\left(1,x^2,x^\mu\right)\,,\\
    Z=\left(0,2x\cdot z, z^{\mu}\right) \qquad z^\mu =
    \left(1,-\zeta_+\zeta_-,\zeta_+,\zeta_-\right)\,,
\end{gathered}
\end{equation}
where we use three pairs of coordinates $(X_{+\,1},X_{-\,1},X_{+\,2},X_{-\,2},
X_{+\,3},X_{-\,3})$ in which metric given by  $ds^2 = -dX_{+\,1}dX_{-\,1}+dX_{+\,2}
dX_{-\,2}+dX_{+\,3}dX_{-\,3}$. The variable $W$ has been introduced to parameterize 
the space of solutions to $X\cdot W=Z\cdot W= W^2=0$, quotiented by the gauge and 
projective equivalence $W \sim \lambda W+\alpha Z +\beta X$. These conditions allow for two independent solutions in a six-dimensional 
embedding space, namely 
\begin{align}
    W=\left(0,2x\cdot w, w^\mu\right) \qquad &w^\mu = \left(0,-\zeta_+,0,1\right) 
    \quad \mathit{or} \label{PoincarepatchW}\\
    \overline{W}=\left(0,2x\cdot \overline{w}, \overline{w}^\mu\right) 
    \qquad &\overline{w}^\mu =
    \left(0,-\zeta_-,1,0\right)\,.\label{PoincarepatchWbar}
\end{align}
It is easy to check that $\star (X\wedge Z \wedge W) = X \wedge Z\wedge W$, and $\star (X\wedge Z \wedge W) = - X \wedge Z\wedge \overline{W}$. As a result, $W$ and $\overline{W}$ define two distinct orbits of the conformal group in embedding space, and the restriction to the $W$ or $\overline{W}$ orbit projects a spinning representation to its self-dual or anti-self-dual part. To make the choice of orbit and duality explicit, we now slightly modify the homogeneity conditions of \eqref{homogeneityconditions}: we redefine $W$ and $\overline{W}$ to have opposite homogeneity degree, such that $\ell > 0$ denotes self-dual representations encoded by polynomials of order $\abs{\ell}$ in $W$, and $\ell < 0$ denotes anti-self-dual representations encoded by polynomials of order $\abs{\ell}$ in $\overline{W}$. To motivate this prescription, recall that the double cover of the Lorentzian conformal group $\mathrm{SO}(2,4)$ by $\mathrm{SU}(2,2)$ defines a map from $\Cs^4$ twistor fields to $\Rs^{2,4}$ embedding space fields. In the twistor formalism, it is customary to label representations by the positive integers $(j,\bar{\jmath})$ that respectively count the number of indices transforming in the chiral and anti-chiral representation of the $\mathrm{SL}_{\Cs}(2)$ Lorentz subgroup\footnote{More specifically the double cover of the Lorentz subgroup $\mathrm{SO}(1,3)\subset \mathrm{SO}(2,4)$.}. Using the explicit map from gauge invariant embedding space tensors to twistor space variables constructed in Appendix \ref{app:twistors_from_emb_space}, our prescription to label self-dual and anti-self-dual representations is then equivalent to the identification
\begin{equation}
    l=\frac{j+\bar{\jmath}}{2}\,, \qquad \ell=\frac{j-\bar{\jmath}}{2}\,,
\end{equation}
which is standard in the CFT$_4$ literature. 
\smallskip 

With the introduction of these two vectors $W$ and $\overline{W}$, the space of tensor 
structures that one can construct changes dramatically. To see this, we begin by 
evaluating our expression for the tensor structure $K$ with $W_i$, respectively $\overline{W}_i, \,
i=1,2$, taken from the same Poincar\'e patch \eqref{PoincarepatchW}, respectively 
\eqref{PoincarepatchWbar}, 
\begin{equation} \label{eq:Kvanish} 
    K_{12}=\frac{1}{3!}\left(X_1\wedge Z_1\wedge W_1\right)\cdot \left(X_2\wedge Z_2 \wedge
    W_2\right)=0 = \frac{1}{3!}\left(X_1\wedge Z_1\wedge \overline{W}_1\right)\cdot 
    \left(X_2\wedge Z_2 \wedge \overline{W}_2\right) = K_{\bar{1}\bar{2}}\ . 
\end{equation}
On the right hand side, we introduced the notation that barred indices $\bar{\imath}$ 
in tensors correspond to occurrences of the variable $\overline{W}_i$, as opposed to 
$W_i$. The vanishing of the tensor structures $K_{12}$ and $K_{\bar{1}\bar{2}}$ in $d=4$ 
forces us to introduce new non-vanishing structures\footnote{if for no other reason than to write the 2-point function of fields with $\ell \neq 0$.}. 

It turns out that it is possible to construct two non-vanishing tensor structures 
by mixing the self-dual and anti-self-dual Poincar\'e patches in the expression of $K_{12}$,
\begin{equation}
    \mK_{1\bar{2}}=\sqrt{\frac{1}{3!}\left(X_1\wedge Z_1\wedge W_1\right)\cdot \left(X_2\wedge 
    Z_2 \wedge \overline{W}_2\right)}\,, 
    \qquad \mK_{\bar{1}2}=\sqrt{\frac{1}{3!}\left(X_1\wedge Z_1\wedge \overline{W}_1\right)
    \cdot \left(X_2\wedge Z_2 \wedge W_2\right)}\,.
\end{equation}
We introduced square roots because it turns out 
that the arguments are perfect squares. Therefore, 
even with inclusion of the square roots in the definition, $\mK_{1\bar{2}}$ and
$\mK_{\bar{1}2}$ are both polynomials in the $d$-dimensional variables $x_i$, 
$z_i$. These two structures satisfy the following relation
\begin{equation}
    H_{12}= 2 \mK_{1\bar{2}} \mK_{\bar{1}2}\,. 
    \label{relationHKKbar}
\end{equation}
Let us also spell out the degrees of the two new tensor structures, 
\begin{equation}
    \textit{deg}\, \mK_{1\bar{2}}=\left[\frac{1}{2},\frac{1}{2},\frac{1}{2};\frac{1}{2},
    \frac{1}{2},-\frac{1}{2};0\right], \qquad \textit{deg}\,
    \mK_{\bar{1}2}=\left[\frac{1}{2},\frac{1}{2},-\frac{1}{2};\frac{1}{2},\frac{1}{2},
    \frac{1}{2};0\right]. 
\end{equation}
In conclusion, we have now replaced the two tensor structures $K_{12}$ and $H_{12}$ of the previous subsection by the two tensor structures $\mK_{1\bar{2}}$
and $\mK_{\bar{1}2}$. Furthermore, one can show that the objects $U_{ij3}$ can be 
decomposed into
\begin{equation}
    U_{123}=\mho\, \mK_{1\bar{2}}\,, \qquad U_{213}=\mho\, \mK_{\bar{1}2}\,,
    \label{Utensorstructure_mho}
\end{equation}
with a new tensor structure $\mho$ defined as 
\begin{equation}
    \mho = \sqrt{X_3^A\left(X_1\wedge Z_1\wedge W_1\right)_{ABC}
    \left(X_2\wedge Z_2 \wedge W_2\right)^{BCD} X_{3\,D}}\,.
\end{equation}
The degree of $\mho$ is given by 
\begin{equation}
    \textit{deg}\, \mho =\left[\frac{1}{2},\frac{1}{2},\frac{1}{2};\frac{1}{2},
    \frac{1}{2},\frac{1}{2};1 \right]\ ,
\end{equation}
The tensor structure $\mho$ uses only $W_i$. Of course, it is also possible to 
construct a similar tensor structure $\overline{\mho}$ in terms of $\overline{W}_i$
as 
\begin{equation}
    \overline{\mho} = \sqrt{X_3^A\left(X_1\wedge Z_1\wedge
    \overline{W}_1\right)_{ABC}\left(X_2\wedge Z_2 \wedge 
    \overline{W}_2\right)^{BCD} X_{3\,D}}\,, 
\end{equation}
with 
\begin{equation} 
\textit{deg}\, \overline{\mho}
=\left[\frac{1}{2},\frac{1}{2},-\frac{1}{2};\frac{1}{2},
\frac{1}{2},-\frac{1}{2};1 \right]\ .
\end{equation} 
In direct analogy with eq.\ \eqref{Utensorstructure_mho} we also find that
\begin{equation}
    U_{\bar{1}23}=\overline{\mho}\, \mK_{\bar{1}2}\,, \qquad
    U_{\bar{2}13}=\overline{\mho}\, \mK_{1\bar{2}}\, . 
    \label{Ubartensorstructure_mhobar}
\end{equation}
At this point we have nine basic tensor structures at our disposal, namely 
$\mK_{\bar{1}2}, \mK_{1\bar{2}}, \mho$ and $\overline{\mho}$ in addition to 
$X_{12}, X_{23}, X_{13}, V_1$ and $V_2$. Their degrees certainly span the 
7-dimensional space and in addition, we can construct the unique cross ratio 
$\mathcal{X}$ as 
\begin{equation}\label{eq:X4d} 
\mathcal{X} = 2 \frac{\mK_{\bar{1}2} \mK_{1\bar{2}}}{V_1 V_2}\ . 
\end{equation} 
Finally, the nine fundamental tensor structures that we introduced satisfy one 
relation,
\begin{equation}
    (X_{23} V_1) (X_{13}V_2) = X_{12}\mho\, \overline{\mho}+
    2\mK_{1\bar{2}}\mK_{\bar{1}2} X_{13} X_{23}\,.
    \label{relationmhobarmho}
\end{equation}
For the reader's convenience we listed these additional tensor structures and their degrees in Tab. \ref{tab:TS_d=4}.

\begin{table}
    \centering
\begin{tabular}{l|c c c c c c c}
    &  $-\Delta_1$ & $-\Delta_2$ & $-\Delta_3$ & $l_1$ & $l_2$ & $\ell_1$ & $\ell_2$ \\
    \midrule
$k_{1\bar{2}}$ & $1/2$ & $1/2$ & $0$ & $1/2$ & $1/2$ & $1/2$ & $-1/2$  \\
$k_{\bar{1}2}$ & $1/2$ & $1/2$ & $0$ & $1/2$ & $1/2$ & $-1/2$ & $1/2$  \\
$\mho$ & $1/2$ & $1/2$ & $1$ & $1/2$ & $1/2$& $1/2$ & $1/2$ \\
$\bar{\mho}$ & $1/2$ & $1/2$ & $1$ & $1/2$ & $1/2$& $-1/2$ & $-1/2$
\end{tabular}
     \caption{Degrees of additional tensor structures of the MST$_2$-MST$_2$-scalar 3-point function in $d=4$.}
    \label{tab:TS_d=4}
\end{table}

Having introduced this new set of tensor structures for vertices of type III, we immediately see that the two solutions to the vanishing of the Gram determinant 
\eqref{GramDeterminant} in $d=4$ arise very naturally when trying to construct a 
second $\mathcal{Y}$-like cross ratio. First, note that the cross ratio 
$\mathcal{Y}$ introduced in eq.\ \eqref{eq:threepointcrossratios} vanishes
when $\ell_1$ and $\ell_2$ have the same sign,  
\begin{equation}
 \mathcal{Y}_{++} = \frac{X_{13} X_{23} 
    V_1 V_2 K_{12}}{X_{12}U_{123}U_{213}} = 0 = 
     \frac{X_{13} X_{23} 
    V_1 V_2 K_{\bar{1}\bar{2}}}{X_{12}U_{\bar{1}23}U_{\bar{2}13}}
    = \mathcal{Y}_{--}\,,  
    \label{VanishingYCases}
\end{equation} 
because of the property \eqref{eq:Kvanish}. On the other hand, when the fields 
have opposite duality, one can only construct a cross ratio with the help of the 
non-vanishing tensor structures $\mK_{\bar{1}2}$ or $\mK_{1\bar{2}}$,  
\begin{equation}
\mathcal{Y}_{-+} = \frac{X_{13} X_{23} 
    V_1 V_2 \mK^2_{\bar{1}2}}{X_{12}U_{\bar{1}23}U_{213}} = 
    \frac{1}{1-\mathcal{X}} 
    =  \frac{X_{13} X_{23} 
    V_1 V_2 \mK^2_{1\bar{2}}}{X_{12}U_{123}U_{\bar{2}13}}
    = \mathcal{Y}_{+-}\ .  
\end{equation} 
To compare with eq.\ \eqref{VanishingYCases}, note that $K_{12} = k_{12}^2$.
In evaluating the expressions for $\mathcal{Y}_{-+}$ and $\mathcal{Y}_{+-}$, 
we have used the relation \eqref{relationmhobarmho} before inserting the 
definition \eqref{eq:X4d} of the cross ratio $\mathcal{X}$. In this sense, 
the second zero of the Gram determinant can be associated with 3-point 
functions in which the spins $\ell_i$ of the MSTs have opposite sign.

Keeping in mind that we are also allowed to have negative values of the 
spin $\ell_i$ in $d=4$, we can now write a generic 3-point function as in equation 
\eqref{eq:threepointtwoMSTonecrossratio}, but with the prefactor $\Omega$ given 
by 
\begin{equation}
    \Omega^{\Delta_1,\Delta_2,\Delta_3}_{l_1,l_2;\ell_1,\ell_2} =
    \frac{V_1^{l_1-\abs{\ell_1}-\abs{\ell_2}}V_2^{l_2-\abs{\ell_1}-\abs{\ell_2}}
    U_{s_123}^{\abs{\ell_1}}U_{s_213}^{\abs{\ell_2}}}
    {X_{12}^{\frac{\Delta_1+\Delta_2-\Delta_3+l_1+l_2-\abs{\ell_1}-\abs{\ell_2}}{2}}
    X_{23}^{\frac{\Delta_2+\Delta_3-\Delta_1-l_1+l_2+\abs{\ell_1}+\abs{\ell_2}}{2}}
    X_{31}^{\frac{\Delta_3+\Delta_1-\Delta_2+l_1-l_2+\abs{\ell_1}+\abs{\ell_2}}{2}}}
    \label{generic4dthree-point}
\end{equation}
instead of \eqref{prefactor2MST}. In spelling out the new prefactor that is 
defined for arbitrary integer values of $\ell_i$, we have introduced the notation
\begin{equation}
    s_i=\begin{cases}
    \,i \qquad \ell_i\ge 0\\
    \,\bar{\imath} \qquad \ell_i<0
    \end{cases}.
\end{equation}
Note that, despite the presence of absolute values in \eqref{generic4dthree-point},
representations with any sign of $\ell_i$ are allowed, as the possible presence of
$\overline{W}_i$ takes full care of negative homogeneity degrees. Formula 
\eqref{generic4dthree-point} is the main result of this subsection. 
\smallskip 

As in the previous subsection, we can use our expression for the 3-point function to count the number of tensor structures when we impose spins to acquire 
integer values. In order to do so, we need to expand \eqref{generic4dthree-point} in terms of the tensor structures specific to $d=4$, in a form that depends specifically on the duality of the two MST$_2$ involved. To distinguish between those cases, we introduce the notation $\Omega_{\sigma_1 \sigma_2}$, where $\sigma_i=+,-$ depending whether the field $i$ is in a self-dual or anti-self-dual representation respectively. Those prefactors satisfy the relations $\Omega_{++}=\overline{\Omega}_{--}$ and $\Omega_{+-}=\overline{\Omega}_{-+}$, where the bar operation exchanges $W_i\leftrightarrow \overline{W}_i$ for both $i=1,2$; we can therefore focus on only the $\Omega_{++}$ and $\Omega_{+-}$ cases, as the results in these cases can easily be translated to the other two cases.

In analyzing the first 
case, involving $\Omega_{++}$, we can express the prefactor in terms of 
$\mho$, $\mK_{1\bar{2}}$ and $\mK_{\bar{1}2}$, leading to the 3-point function
\begin{equation}
    \Omega_{++}t(\mathcal{X})=
    \frac{V_1^{l_1-\abs{\ell_1}-\abs{\ell_2}}V_2^{l_2-\abs{\ell_1}-\abs{\ell_2}}
    \mho^{\abs{\ell_1}+\abs{\ell_2}}\mK_{1\bar{2}}^{\abs{\ell_1}}
    \mK_{\bar{1}2}^{\abs{\ell_2}}}
    {X_{12}^{\frac{\Delta_1+\Delta_2-\Delta_3+l_1+l_2-\abs{\ell_1}-\abs{\ell_2}}{2}}
    X_{23}^{\frac{\Delta_2+\Delta_3-\Delta_1-l_1+l_2+\abs{\ell_1}+\abs{\ell_2}}{2}}
    X_{31}^{\frac{\Delta_3+\Delta_1-\Delta_2+l_1-l_2+\abs{\ell_1}+\abs{\ell_2}}{2}}}
    t(\mathcal{X})\, . 
    \label{prefactor_self-self}
\end{equation}
By requiring polynomial dependence on the variables $Z_i$, $W_i$, $\overline{W}_i$ in this expression, it is easy to see that $t(\mathcal{X})$ must contain integer 
powers of the cross ratio \eqref{eq:X4d}, with exponents that are bounded from above 
by the minimum exponent of the $V_i$ in the prefactor, and bounded from below 
by the minimum exponent of the $\mK_{ij}$. As a result, the function $t(\mathcal{X})$ must 
take the form 
\begin{equation}
    t(\mathcal{X})=\sum_n c_n \mathcal{X}^n\,,
\end{equation}
with the sum over exponents restricted by the inequalities
\begin{equation} \label{eq:nconstraint}
-\min(\abs{\ell_1},\abs{\ell_2})\le n\le \min(l_1,l_2)-\abs{\ell_1}-\abs{\ell_2}\ . 
\end{equation} 
In cases where $\ell_1$ and $\ell_2$ have opposite sign, e.g. $\ell_1>0$, $\ell_2<0$ and the prefactor $\Omega_{+-}$ is used, the discussion is a bit 
different. Here we can use the relation
\eqref{relationmhobarmho} to eliminate one of the tensor structures $\mho$ or 
$\overline \mho$ from the prefactor and write the 3-point function as 
$$
    \frac{V_1^{l_1-\abs{\ell_1}-\abs{\ell_2}}V_2^{l_2-\abs{\ell_1}-\abs{\ell_2}}
    \mho^{\abs{\ell_1}-\min(\abs{\ell_1},\abs{\ell_2})}\overline{\mho}^{\abs{\ell_2}-
    \min(\abs{\ell_1},\abs{\ell_2})}\mK_{1\bar{2}}^{\abs{\ell_1}+\abs{\ell_2}}
    \left( V_1 V_2 -2\mK_{1\bar{2}}\mK_{\bar{1}2}\right)^{\min(\abs{\ell_1},\abs{\ell_2})}}
    {X_{12}^{\frac{\Delta_1+\Delta_2-\Delta_3+l_1+l_2-\abs{\ell_1}-\abs{\ell_2}}{2}}
    X_{23}^{\frac{\Delta_2+\Delta_3-\Delta_1-l_1+l_2+\abs{\ell_1}+\abs{\ell_2}}{2}}
    X_{31}^{\frac{\Delta_3+\Delta_1-\Delta_2+l_1-l_2+\abs{\ell_1}+\abs{\ell_2}}{2}}
    \left(\frac{X_{12}}{X_{13}X_{23}}\right)^{\min(\abs{\ell_1},\abs{\ell_2})}}
    t(\mathcal{X})\, . 
$$
In order to analyze the resulting constraints on the function $t(\mathcal{X})$, we 
shall think of $t$ as a function of 
\begin{equation} 
1-\mathcal{X}=\frac{V_1 V_2-2\mK_{1\bar{2}}\mK_{\bar{1}2}}{V_1 V_2} .   
\end{equation} 
Requiring polynomial dependence on the polarization vectors, $t$ is constrained to 
contain integer powers of $(1-\mathcal{X})$ that are bounded from above by the 
minimum power of the $V_i$ as in the previous case, and are bounded from below by
the power of the factor $(V_1V_2-2\mK_{1\bar{2}}\mK_{\bar{1}2})$. This can be written concretely as
\begin{equation}
    t(\mathcal{X})=\sum_n c_n^{\prime} (1-\mathcal{X})^n\,,
\end{equation}
with the same constraint on the sum over exponents $n$ that was spelled out 
in eq.\ \eqref{eq:nconstraint}. Thus, in both cases, we have determined that 
the space of tensor structures $t(\mathcal{X})$ has dimension
\begin{equation}
    \min(l_1,l_2)-\max(\abs{\ell_1},\abs{\ell_2})+1\,,
\end{equation}
which matches exactly the expected number from the representation theory of the
conformal group. 

For the computation of the vertex operator in section~\ref{sect:ConstructionOperator}, it will prove useful to spell out an explicit relation between the prefactors $\Omega_{++}$ and $\Omega_{+-}$, 
\begin{equation}
   \Omega_{+-}\Big|_{\abs{\ell_2}\rightarrow -\abs{\ell_2}} = \left(\frac{\mathcal{X}(1-\mathcal{X})}{2}\right)^{-\abs{\ell_2}}
   \Omega_{++}\,.
    \label{relation_prefactor_sd_asd}
\end{equation}
In other words, the prefactor for one self-dual and one anti-self-dual field is
related to the one for two self-dual MSTs through a simple function of the cross 
ratio $\mathcal{X}$, plus a change of sign for the homogeneity of the field for which we are changing duality. This equation will allow us to relate the vertex operator computed in one case to the other one, see eq. \eqref{Hpm_to_Hpp}.

We now address the tensor structures 
that can be constructed with the use of a six-dimensional Levi-Civita symbol for 
vertices of type II and III. Using only one $W_i$ or $\overline{W}_i$ vector, it is possible to 
construct the structures
\begin{equation} 
O^{(4)}_{ijk} = \epsilon_{ABCDEF}X_i^A X_j^B 
X_k^C Z_i^D Z_j^E W_i^F\,, \qquad O^{(4)}_{\bar{\imath}jk} = \epsilon_{ABCDEF}X_i^A X_j^B 
X_k^C Z_i^D Z_j^E \overline{W}_i^F\,.
\label{LeviCivita4D_oneW}
\end{equation} 
These are however easily seen to be proportional to the $U_{ijk}$ tensor 
structure
\begin{equation}
    \left(O^{(4)}_{s_i jk}\right)^2\propto \left(U_{s_i jk}\right)^2\,,
\end{equation}
so that the vertex function $t(\mathcal{X})$ is unaffected by the introduction
of parity-odd tensor structures for vertices of type II in $d=4$. For vertices 
of type III we can also construct parity odd tensors of the form
\begin{equation}
    \widetilde{O}^{(4)}_{12} = \epsilon_{ABCDEF}X_1^A X_2^B 
Z_1^C Z_2^D W_1^E W_2^F\,, \qquad \widetilde{O}^{(4)}_{\bar{1}2} = 
\epsilon_{ABCDEF}X_1^A X_2^B 
Z_1^C Z_2^D \overline{W}_1^E W_2^F\,,
\label{LeviCivita4D_twoMST}
\end{equation}
as well as their images under $1\leftrightarrow \bar{1}$ 
and $2\leftrightarrow \bar{2}$. However, these structures are once again proportional 
to tensors that we have already introduced:
\begin{equation}
    \left(\widetilde{O}^{(4)}_{12}\right)^2\propto
    \left(K_{12}\right)^2\stackrel{d=4}{=}0\,, \qquad \qquad
    \left(\widetilde{O}^{(4)}_{\bar{1}2}\right)^2\propto \left(K_{\bar{1}2}\right)^2\,.
\end{equation}
We can therefore conclude that structures of the type
\eqref{LeviCivita4D_oneW} or 
\eqref{LeviCivita4D_twoMST} do not extend the space of $t(\mathcal{X})$ for vertices of type II and III.

Before ending this section, we would like to point out that the construction of $d=4$ 3-point tensor structures in this section is the embedding space version of the twistor based construction of tensor structures in \cite{SimmonsDuffin:2012uy,
Elkhidir:2014woa}. We describe in more detail
in Appendix~\ref{app:twistors_from_emb_space} the 
dictionary from embedding space variables to twistor variables 
and vice versa.

\section{The Single Variable Vertex Operator} 
\label{sect:ConstructionOperator}

Having assembled all of the required background, including in particular 
a detailed discussion of 
single variable 3-point functions of the form \eqref{eq:threepointtwoMSTonecrossratio}, we now move on to our central goal. In this 
section, we work out the explicit expression for the action of our vertex 
differential operators on the function $t(\mathcal{X})$ that multiplies the 
prefactors~\eqref{prefactor2MST} or \eqref{generic4dthree-point}. Our strategy 
is to obtain the results for all three sub-cases listed in eq.\ \eqref{eq:list}
by studying the MST$_2$-MST$_2$-scalar vertex in $d \geq 4$. Note that passing 
through this 2-variable vertex is just a trick that allows us to shorten the 
discussion and avoid displaying multiple long expressions for all different 
cases; using the same procedure described in this section, one can compute 
the vertex operator in each individual case and easily verify that the 
answer is the same as what is obtained by reduction of the more general 
vertex. The results of this section should be seen as 
providing raw data that we will process in the subsequent sections. We 
will also comment on the relation between our formulas and the vertex 
differential operator of a 5-point function in $d\geq 3$, see 
\cite{Buric:2021ywo}. To this end, we shall look at both shadow 
integrals and OPE limits in the second subsection. 

\subsection{Construction of the reduced vertex operator}

As we had argued in~\cite{Buric:2020dyz,Buric:2021ywo}, see also the 
review in section \ref{sect:reviewsummary}, there is a distinguished basis for the 
vertex functions $t$ that is selected by solving the eigenvalue equations of some 
commuting set of vertex differential operators. A full prescription of how to construct 
these operators for 3-point vertices with sites of arbitrary depth $L_i$, 
$i=1,2,3$ was given in \cite{Buric:2021ywo}. For the MST$_2$-MST$_2$-scalar
vertex there are two such operators, one of order four and the other of order 
six. When we descend from there to the single variable vertices in the list 
\eqref{eq:list} via the constraint $\mathcal{Y}=0$, the sixth order operator 
becomes dependent. Hence to achieve our goals it is sufficient to work out 
the fourth order operator. The operator starts its existence as a differential 
operator on the space of coordinates and polarizations of three fields. We use the embedding space constructions that were reviewed in the previous 
section to write the operator 
\begin{equation}
    \mathcal{D}_\rho\equiv\mathcal{D}_{\rho,13}^{4,3}=\text{str}\left(\mathcal{T}^{(1)}\mathcal{T}^{(1)}\mathcal{T}^{(1)}\mathcal{T}^{(3)}\right)
    \label{eq:vertexoperatorabstract}
\end{equation}
in terms of the simple first order differential operators \eqref{eq:spinningdiffgenerators} encoding the action of the conformal 
generators on the variables $(X_i,Z_i,W_i)$. Let us also recall that 
$\text{str}$ stands for symmetrized trace. The action of the differential 
operator \eqref{eq:vertexoperatorabstract} can be reduced to the cross-ratio 
space of $t(\mathcal{X})$ by conjugation with the prefactor $\Omega^{\Delta_1,
\Delta_2,\Delta_3}_{l_1,l_2;\ell_1,\ell_2}$, as is the case for 
Casimir differential operators, 
\begin{equation}
    H^{(d,\Delta_i,l_i,\ell_i)} t(\mathcal{X})=
    \frac{1}{\Omega^{\Delta_1,\Delta_2,\Delta_3}_{l_1,l_2;\ell_1,\ell_2}}
    \mathcal{D}_\rho \left(\Omega^{\Delta_1,\Delta_2,\Delta_3}_{l_1,l_2;
    \ell_1,\ell_2} t(\mathcal{X})\right)\,.
    \label{eq:conjugationvertexoperator}
\end{equation}
By plugging eqs.\ \eqref{eq:spinningdiffgenerators} and \eqref{eq:vertexoperatorabstract} in eq.\ \eqref{eq:conjugationvertexoperator}, it 
is then possible to compute the action of $\mathcal{D}_\rho^{\mathcal{X}}$ in cross ratio space. To do this, we implemented the action of generators \eqref{eq:spinningdiffgenerators} in \texttt{Mathematica}\footnote{for an implementation of the code used to compute the fourth-order operator, see the supplementary material attached to this publication.} and first obtained the conjugation with the prefactor expressed in terms of scalar products of $X_i$, $Z_i$, $W_i$. We then solved the expressions of the cross ratios \eqref{eq:threepointcrossratios} for two scalar products, set $\mathcal{Y}=0$, and plugged these expressions in the conjugated differential operator; due to conformal invariance all of the remaining scalar products drop out, and one is left with a differential operator in one cross ratio of the form 
\begin{equation}
    H^{(d,\Delta_i,l_i,\ell_i)}=h_0(\mathcal{X})+\sum_{q=1}^4h_q(\mathcal{X})\mathcal{X}^{q-1}(1-\mathcal{X})^{q-1}\partial_{\mathcal{X}}^q\,.
    \label{H_basis}
\end{equation}
Apart from a constant piece in $h_0(\mathcal{X})$, all of the coefficients $h_q(\mathcal{X})$ are symmetric under exchange of fields $1\leftrightarrow 2$, and we can therefore represent them as
\begin{equation}
    h^{(d,\Delta_i,l_i,\ell_i)}_q(\mathcal{X})=\chi^{(d,\Delta_i,l_i,\ell_i)}_q(\mathcal{X})+(1\leftrightarrow 2)\,.
\end{equation}
Finally, we write $h_0$ as
\begin{equation}
    h^{(d,\Delta_i,l_i,\ell_i)}_0(\mathcal{X})=\left[\chi^{(d,\Delta_i,l_i,\ell_i)}_0(\mathcal{X})+(1\leftrightarrow 2)\right]+\tilde{\chi}_0^{(d,\Delta_i,l_i,\ell_i)}\,.
\end{equation}
These coefficients take the following form:
\begin{eqnarray*}
    \chi_4(\mathcal{X})& = &-2\,, \\[0.5em]
    \chi_3(\mathcal{X})& = &-8 \mathcal{X} \left(l_1-2 \left(\ell _1+1\right)\right)+2 \Delta _3+4 l_1-8 \ell _1-d-8\,,\\[0.5em]
    \chi_2(\mathcal{X})& = &-4 \mathcal{X}^2 \left(l_1 \left(2 l_2-6 \ell _1-9\right)+6 \ell _1 \left(-l_2+\ell _1+\ell _2+3\right)+l_1^2+7\right)\\
    & &\begin{aligned}+\mathcal{X} \bigl[&-2 l_1 \left(d-2 \Delta _3-2 l_1-4 l_2+12 \ell _1+18\right)+2 \ell _1 \left(3 d-6 \Delta _3-12 l_2+12 \ell _1+13 \ell _2+36\right)\\
    &d^2-2 d (\Delta _1-1)-\Delta _3(d+4)+2 \Delta _1 \Delta _2+28\bigr]
    \end{aligned}\\
    & &+l_1 \left(-2 \Delta _3-l_1-l_2+6 \ell _1+2 d+4\right)+\ell _1 \left(-2 d+6 \Delta _3+6 l_2-5 \ell _1-7 \ell _2-16\right)-3 d+\Delta _1^2\\
    & &-\Delta _1 \Delta _2+\frac{\Delta _3}{2}(-\Delta _3+2d+4)-2\,,\\[0.5em]
\chi_1(\mathcal{X})& = &8 \mathcal{X}^3 \bigl[-3 l_1 \ell _1^2-3 l_2 \ell _1^2+l_1^2 \ell _1+l_2^2 \ell _1-6 l_1 \ell _1+4 l_1 l_2 \ell _1-6 l_2 \ell _1-6 l_1 \ell _2 \ell _1+l_1^2-3 l_1-l_1^2 l_2+2 l_1 l_2\\
&&\qquad \, +2 \ell _1^3+6 \ell _2 \ell _1^2+6 \ell _1^2+6 \ell _2 \ell _1+6 \ell _1+1\bigr]\\
    &&+\mathcal{X}^2 \bigl[l_1 \left(2 d(d-\Delta _3-\Delta_1)+4 \Delta _1 \Delta _2-l_2 \left(d-2 \Delta _3+24\right)+4 \ell _1 \left(d-2 \Delta _3+19 \ell _2+18\right)+36\right)\\
        &&\qquad-2 \ell _1\!\left(2 d^2\!-\!2 l_2 \left(d+9 \ell _1+18\right)\!+\!(3 d+37) \ell _2+\ell _1 \!\left(3 d+40 \ell _2+36\right)+d+6 l_2^2+12 \ell _1^2+36\right)\\
        &&\qquad-d^2+2 d \Delta _1 \left(-l_2+2 \ell _1+2 \ell _2+1\right)+\Delta _3 \left(4 \ell _1 \left(d-2 l_2+3 \ell _1+3 \ell _2+1\right)+d\right)-48 l_1 l_2\ell _1\\
        &&\qquad+36 l_1 \ell _1^2+12 l_1^2 \left(l_2-\ell _1-1\right)-2 \Delta _1 \Delta _2 \left(4 \ell _1+1\right)-12\bigr]\\
    &&+\frac{1}{2} \mathcal{X} \bigl[-2 l_1\! \left(-2 \Delta _1\! \left(d-2 \Delta _2\right)-2 l_2 \!\left(d+10 \ell _1+3\right)+4 \ell _1\! \left(d+10 \ell _2+7\right)+d (d+6)+16 \ell _1^2+4\right)\\
        &&\qquad\quad+2 \ell _1 \left(-2 \Delta _1 \left(d+\Delta _1-4 \Delta _2\right)-4 (d+7) l_2+(5 d+34) \ell _2+d (3 d+4)+6 l_2^2+24\right)\\
        &&\qquad\quad+4 \Delta _3 l_1 \left(d-l_2+4 \ell _1\right)+2\Delta _1 \left(-(d-2) \Delta _1+2 d l_2-2 \ell _2 \left(d+\Delta _1\right)+(d-4) d+2 \Delta _2\right)\\
        &&\qquad\quad+2\ell _1^2 \left(5 (d+6)-16 l_2+42 \ell _2\right)\!+\!8 d\!+20 \ell _1^3+2 l_1^2 \left(d-4 l_2+6 \ell _1+2\right)+\!\Delta _3^2 \left(d+4 \ell _1\!-\!2\right)\\
        &&\qquad\quad-\Delta _3 \left(d^2+4 \ell _1 \left(3 d-4 l_2+6 \ell _1+7 \ell _2+2\right)\right)\bigr]\\
    &&\frac{1}{4} \bigl[4 \Delta _3 \left(-l_1 \left(d+2 \ell _1-2\right)+\ell _1 \left(d-2 l_2+2 \ell _1+4 \ell _2+2\right)+d-2\right)-2 \ell _1^2 \left(d-4 l_1-4 l_2+6\right)\\
        &&\quad+4 \left(l_1+l_2\!-\!2\right) \ell _1 \left(d-\!l_1\!-\!l_2\right)-2 \ell _2 \ell _1 \left(d\!-\!16 l_1\!+\!14 \ell _1\!+\!10\right)-2 (d-2) \left(l_1 \left(l_1+l_2-4\right)+2\right)\\
        &&\quad+2 \Delta _1^2 \left(d+2 \ell _1+2 \ell _2-2\right)-\Delta _3^2 \left(d+4 \ell _1-2\right)-2 \Delta _1 \Delta _2 \left(d+4 \ell _1-2\right)-4 \ell _1^3\bigr]\,,\\[0.5em]
\tilde{\chi}_0&=&\frac{1}{6} \left(\Delta _1-\Delta _2\right) \left(d-\Delta _1-\Delta _2\right) \left(d^2-3 d \left(\Delta _1+\Delta _2+\Delta _3+1\right)+3 \left(\Delta _1^2+\Delta _2^2+\Delta _3^2\right)\right)\\
&&-\frac{1}{6} \left(l_1-l_2\right) \left(d+l_1+l_2-2\right) \left(d^2+3 d \left(-\Delta _3+l_1+l_2\!-\!3\right)+3 \left(\Delta _3^2+l_1^2-2 l_1+l_2^2-2 l_2+2\right)\right)\\
&&-\frac{1}{6} \!\left(\ell _1\!-\!\ell _2\right) \left(d+\ell _1+\ell _2\!-\!4\right) \left(d^2+3 d \left(-\Delta _3+\ell _1+\ell _2\!-\!5\right)+3 \left(\Delta _3^2+\ell _1^2-4 \ell _1+\ell _2^2-4 \ell _2\!+\!8\right)\right)\!,
\end{eqnarray*}
\begin{eqnarray*}
\chi_0(\mathcal{X})&=&2 \mathcal{X}^2 \bigl[l_1 \left(-4 l_2^2 \ell _1+l_2 \left(8 \ell _1 \left(\ell _1+\ell _2+1\right)+1\right)-2 \ell _1 \left(\ell _1+1\right) \left(2 \ell _1+6 \ell _2+1\right)\right)\\
    &&\qquad\,+2 \ell _1 \left(\ell _2 \left(2 \ell _1 \left(-3 l_2+2 \ell _1+3\right)+1\right)+\left(\ell _1+1\right) \left(\ell _1-l_2\right) \left(-l_2+\ell _1+1\right)+3 \ell _1 \ell _2^2\right)\\
    &&\qquad\,+l_1^2 \left(-2 l_2 \left(2 \ell _1+1\right)+l_2^2+2 \ell _1 \left(\ell _1+2 \ell _2+1\right)\right)\bigr]\\
&&+\mathcal{X} \bigl[2 \ell_1 l_2 \left(d^2-d \left(\Delta _1+\Delta _3+1\right)+2 \left(\Delta _1 \Delta _2+\Delta _3+1\right)+\ell _1 \left(d-2 \Delta _3+14 \ell _2+6\right)+4 \ell _1^2\right)\\
&&\qquad-d \Delta _1 l_2 \ell _2 +2\ell _1^2 \left(-d^2+\Delta _1 \left(d-2 \Delta _2\right)+(d-2) \Delta _3-3 \ell _2 \left(d-2 \Delta _3+4\right)+d-8 \ell _2^2-2\right)\\
&&\qquad+2\ell _1 \ell _2 \left(-d^2+2 \Delta _1 \left(d-\Delta _2\right)+(d-2) \Delta _3+d-2\right)-2\ell _1^3 \left(d-2 \Delta _3+10 \ell _2+4\right)\\
&&\qquad+l_1 \bigl(-l_2 \left(2 \Delta _1 \left(\Delta _2-d\right)+2 \ell _1 \left(d+9 \ell _2+8\right)+(d-1) d+16 \ell _1^2+2\right)\\
&&\qquad \qquad +2\ell _1 \left(-\Delta _1 \left(d-2 \Delta _2\right)-(d-2) \Delta _3+2 \ell _2 \left(d-2 \Delta _3+6\right)+(d-1) d+2\right)\\
&& \qquad \qquad -2d \Delta _1 \ell _2+8 \ell _1^3+2\ell _1^2 \left(d-2 \Delta _3+14 \ell _2+6\right)+\Delta _3 l_2 \left(d+4 \ell _1-2\right)+8 l_2^2 \ell _1\bigr)\\
&&\qquad+2d \Delta _1 \ell _2^2-4 l_2^2 \left(\ell _1+1\right) \ell _1-4 \ell _1^4-2 l_1^2 \left(-2 l_2 \left(2 \ell _1+1\right)+l_2^2+2 \ell _1 \left(\ell _1+2 \ell _2+1\right)\right)\bigr]\\
&&-\frac{\ell _1 \ell _2}{2 \mathcal{X}} \left(\Delta _3^2+4 \Delta _3 \left(l_1-\ell _1\right)+2 \left(\Delta _1 \left(\Delta _2-\Delta _1\right)+l_1 \left(l_2-2 \ell _1\right)+\ell _1 \left(-2 l_2+\ell _1+\ell _2\right)+l_1^2\right)\right)\\
&&+\frac{1}{12} \bigl[6 l_1 \bigl(l_2 \left(d^2+2 \!\Delta _1 \left(\Delta _2\!-d\right)-(d-2) \Delta _3+2 \ell _1 \left(d-2 \Delta _3+7 \ell _2+2\right)+8 \ell _1^2-2\right)\!-\!(d\!-\!2) \Delta _3^2\\
&&\qquad \qquad \,\,-2 \left(\ell _1 \left(d \left(d-\Delta _1-1\right)+(d+6) \ell _2+2 \Delta _1 \Delta _2\right)-d \Delta _1 \ell _2+\ell _1^2 \left(d+12 \ell _2+2\right)+2 \ell _1^3\right)\\
&&\qquad \qquad\,\,+\Delta _3 \left(2 \ell _1 \left(d+2 \ell _1+6 \ell _2-2\right)+(d-2) d\right)+2 (d-2) (d-1)-4 l_2^2 \ell _1\bigr)\\
&&\qquad \,\, -\Delta _3^2 \left(6 \Delta _1 \left(\Delta _1-d\right)+(d-3) d\right)+d \Delta _3 \left(6 \Delta _1 \left(\Delta _1-d\right)+(d-3) d\right)\\
&&\qquad\,\, -6 l_1^2 \left(-d \Delta _3+(d-6) d+\Delta _3^2+4 l_2 \ell _1-l_2^2-2 \ell _1 \left(\ell _1+4 \ell _2+1\right)+6\right)\\
&&\qquad\,\,-6 \Delta _1 \bigl(\left(\Delta _1-\Delta _2\right) \left(\left(d-\Delta _1\right){}^2+\Delta _1 \Delta _2\right)+2 l_2^2 \left(d-\Delta _1\right)+2 l_2 \left((d-2) \left(d-\Delta _1\right)-d \ell _2\right)\\
    &&\qquad \qquad\quad\,\,\,\, +2 \ell _2 \left(\Delta _1-d\right)+2 d \ell _2^2\bigr)\\
&&\qquad\,\, +6 \ell _1^2 \bigl(\Delta _3 \left(-3 d+\Delta _3+4\right)+2 \ell _2 \left(3 d-10 \Delta _3+5\right)+(d-2) (d+8)-2 \Delta _1 \left(\Delta _1-2 \Delta _2\right)\\
    &&\qquad\qquad \quad\,\, +4 l_2 \left(\Delta _3-6 \ell _2-2\right)+4 l_2^2+15 \ell _2^2\bigr)\\
&&\qquad\,\, +6 \ell _1 \bigl(2 l_2 \left(\Delta _1 \left(d-2 \Delta _2\right)+(d-2) \Delta _3-5 d+8\right)+(d-2) \Delta _3^2+2 (d-3) \Delta _1 \left(d-\Delta _1\right)\\
&&\qquad \qquad \quad  +\ell _2 \left(-2 \Delta _1 \left(d-3 \Delta _2\right)+\Delta _3 \left(-5 d+2 \Delta _3+10\right)+3 (d-2) d-4 \Delta _1^2\right)\\
&& \qquad \qquad \quad -(d-2) d \Delta _3+2 (d-3) l_2^2-4 d+8\bigr)\\
&&\qquad\,\, -12 (d-2) l_1^3+12 \ell _1^3 \left(-2 \Delta _3-2 l_2+8 \ell _2+5\right)-6 l_1^4+6 \ell _1^4\bigr]\,,
\end{eqnarray*}
From the operator above, it is easy to reduce to the vertex operators of type I and II: 
one has simply to impose the corresponding $\ell_i=0$. For vertices of type III, where
representations labeled by Young diagrams are reducible, the reduction requires some
further comments. Let us note that the prefactor \eqref{generic4dthree-point} for two
self-dual fields (respectively two anti-self-dual fields) acquires precisely the same form as 
the $d>4$ one, modulo the replacement of both second polarizations with their $d=4$
counterparts $W_i$ (respectively $\overline{W}_i$), and the replacement of the $\ell_i$ with 
their absolute values $\abs{\ell_i}$. This means that the computation of the vertex
operator for these cases will proceed in exactly the same way as in $d>4$ up to the
replacements described above. Furthermore, we observed with equation
\eqref{VanishingYCases} that the cross ratio $\mathcal{Y}$ vanishes in $d=4$ when 
expressed only in terms of (anti-)self-dual variables. We can therefore conclude 
that the vertex operator for two (anti-)self-dual fields corresponds to
\eqref{eq:vertexoperatorabstract} with $\ell_i\rightarrow \abs{\ell_i}$, as we have 
already imposed $\mathcal{Y}=0$ in the computation of the $d>4$ vertex operator. If 
instead we wish to describe the type III vertex operator with one self-dual and one
anti-self-dual field, we can use our observation \eqref{relation_prefactor_sd_asd} relating prefactors in the $\ell_1 \ell_2>0$ case to the $\ell_1 \ell_2<0$ case. 
In particular, labeling the operator with $\ell_1, \ell_2>0$ as $H=H_{++}$ and 
the operator for $\ell_1>0$, $\ell_2<0$ as $H_{+-}$, we find that
\begin{equation}
H_{+-}^{(d=4;\Dg_i;l_i;\abs{\ell_1},-\abs{\ell_2})} = \left(\frac{\mathcal{X}(1-\mathcal{X})}{2}\right)^{\abs{\ell_2}} H_{++}^{(d=4;\Dg_i;l_i;\abs{\ell_1},\abs{\ell_2})}\left(\frac{\mathcal{X}(1-\mathcal{X})}{2}\right)^{-\abs{\ell_2}} \,.
\label{Hpm_to_Hpp}
\end{equation}
and analogously for $H_{-+}$ with $\ell_1\leftrightarrow \ell_2$. This concludes our 
construction of the vertex differential operators for all three single variable
cases listed in eq.\ \eqref{eq:list}

Having written out the results of our computations, let us add a few 
quick remarks and observations. First of all, it is important to note that
almost all terms have a polynomial dependence on the cross ratio $\mathcal{X}$. 
The only exception appears in our expression for $\chi_0(\mathcal{X})$, which 
contains one term proportional to $\ell_1 \ell_2 \mathcal{X}^{-1}$. 
For vertices of type I and II, where $\ell_2 = 0$, this non-polynomial 
term is absent, while it remains present for vertices of type III.  For vertices 
of type I, the expression is equivalent to the one introduced in \cite{Buric:2020dyz}, up to normalization and a constant shift. 
Let us stress again that our derivation is valid for $\ell_2 \neq 0$ and for 
arbitrary dimension $d \geq 4$. As we shall show in 
section~\ref{sect:MappingElliptic}, the mapping of our operator \eqref{H_basis}
to the elliptic CMS model of \cite{etingof2011107} 
also works for all cases, including MST$_2$-MST$_2$-scalar vertices $d > 4$ with 
kinematics reduced to $\mathcal{Y}=0$. Nevertheless, it turns out that the 
map has significantly different features when it is applied beyond the list 
\eqref{eq:list} of single variable vertex systems, c.f. section~\ref{sect:MappingElliptic} and 
appendix~\ref{app:d_def} for a discussion. Our analysis of the results 
in the next section will be restricted to the cases with $\ell_2 = 0$ 
which possess polynomial coefficients.

\subsection{Relation with vertex operator for 5-point functions}
\label{sect:shadowOPE}

It is worth to pause our analysis of the single variable vertex operators for a moment 
and to explain how this differential operator is related to the vertex operator for a 
5-point function in $d\geq 3$ that we worked out in \cite{Buric:2021ywo}. As usual, 
we split the scalar 5-point function 
\begin{equation}   \label{five_point_correlator}
    \expval{\phi_1(X_1)\phi_2(X_2)\phi_3(X_3)\phi_4(X_4)\phi_5(X_5)}=
    \Omega_5^{\Delta_i}(X_i)\, G(u_i)
    \end{equation} 
   
into a function $G$ of cross ratios and a prefactor $\Omega$  that accounts for the 
nontrivial covariance law of the scalar fields under conformal transformations. 
The former can be further decomposed into conformal blocks, 
\begin{equation} 
 G(u_i) = \sum_{\Delta_a,l_a,\Delta_b,l_b,t}\! \lambda_{12a}\lambda_{a 3 b;t}\lambda_{b45} \psi^{(\Delta_{12},\Delta_3,\Delta_{45})}_{(\Delta_a,\Delta_b;l_a,l_b;t)}
    (u_i)\,,
\end{equation}
while the latter is given by 
\begin{equation}
		\Omega^{(\Delta_i)}_5(X_i)=\frac{\left(\frac{X_2\cdot X_3}{X_1\cdot X_3}\right)^{\frac{\Delta_1-\Delta_2}{2}} \left(\frac{X_2\cdot X_4}{X_2\cdot X_3}\right)^{\frac{\Delta_3}{2}} 		\left(\frac{X_3\cdot X_5}{X_3\cdot X_4}\right)^{\frac{\Delta_4-\Delta_5}{2}}}{\left(X_1\cdot X_2\right)^{\frac{\Delta_1+\Delta_2}{2}}\left( X_3\cdot X_4\right)^{\frac{\Delta_3}{2}}
		\left(X_4\cdot X_5\right)^{\frac{\Delta_4+\Delta_5}{2}}} \,. 
		\label{Omegafivepts}
	\end{equation}
For $N=5$ points in $d\geq 3$, one can construct five cross ratios. Two common sets of 
such conformal invariant coordinates are denoted by $z_1,\bar{z}_1,z_2,\bar{z}_2,w$ and 
$u_i$, respectively. These are obtained from the embedding space variables by the 
following relations
\begin{equation}
        \begin{gathered}
        u_1=\frac{\left(X_1\cdot X_2\right) \left(X_3\cdot X_4\right)}{\left(X_1\cdot X_3\right) \left(X_2 \cdot X_4\right)}=z_1 \bar{z_1}\,,\qquad u_2=\frac{\left(X_1\cdot X_4\right) \left(X_2\cdot X_3\right)}{\left(X_1\cdot X_3\right) \left(X_2 \cdot X_4\right)}=(1-z_1)(1-\bar{z}_1)\,,\\
    u_3=\frac{\left(X_2\cdot X_3\right) \left(X_4\cdot X_5\right)}{\left(X_2\cdot X_4\right) \left(X_3 \cdot X_5\right)}=z_2 \bar{z}_2\,, \qquad u_4=\frac{\left(X_2\cdot X_5\right) \left(X_3\cdot X_4\right)}{\left(X_2\cdot X_4\right) \left(X_3 \cdot X_5\right)}=(1-z_2)(1-\bar{z}_2)\,,\\
    u_5=\frac{\left(X_1\cdot X_5\right) \left(X_2\cdot X_3\right)\left(X_3\cdot X_4\right)}{\left(X_2\cdot X_4\right) \left(X_1 \cdot X_3\right)\left(X_3\cdot X_5\right)}=w(z_1-\bar{z}_1)(z_2-\bar{z}_2)+(1-z_1-z_2)(1-\bar{z}_1-\bar{z}_2)\,.
        \end{gathered}
\end{equation} 
Since the OPE diagram for a  5-point function contains two internal fields of depth $L=1$, 
i.e.\ two STTs, its blocks are characterized by four Casimir and one 
vertex operator. The latter is of type I and it was constructed in \cite{Buric:2021ywo} as 
a fourth order operator acting on the five cross ratios.
\medskip 

One way to express the relation between this full vertex operator and the reduced 3-point 
vertex operator we have of the previous subsection makes use of the shadow formalism \cite{Ferrara:1972uq}. Shadow integrals turn the graphical 
representation of a conformal block, such as that of Fig.~\ref{fig:Six-points_Comb_Snowflake}, into an integral formula. 
Just as in the case of Feynman integrals, the `shadow integrand' is built from 
relatively simple building blocks that are assigned to the links and 3-point vertices 
of the associated OPE diagram. For a scalar 5-point function, the only non-trivial 
vertex is of type I. Within this subsection we label the two internal STT 
lines that are attached to this vertex by $a$ and $b$ rather than $1$ and $2$, to 
distinguish them from the external lines. The basic building block for the integrand 
of the shadow integral is the 3-point function $\Phi$ that was introduced in eq.\ 
\eqref{eq:threepointtwoMST}. In the context of the 5-point function, only two 
special cases of this formula appear. On the one hand, there are two 1-STT-2 scalar 
vertices $\Phi_{1a2}$ and $\Phi_{b54}$ that are completely fixed by conformal 
symmetry, i.e. where $t$ is trivial. On the other hand, there is the central 
vertex  $\Phi_{ab3}$ of type I. With these notations, the shadow integral for 
scalar 5-point blocks of weight $\Delta_i, i=1, \dots,5$ reads
\begin{eqnarray}
\Psi^{(\Delta_1,...,\Delta_5)}_{(\Delta_a,\Delta_b;l_a,l_b;t)} (X_1,...,X_5) & = & 
\label{eq:5pshadow} \\[2mm]
 & & \hspace*{-3.5cm} =\! \prod_{s =a,b} \int \dd \mu(X_a,X_b,Z_a,Z_b) 
  \Phi_{1\tilde{a}2}(X_1,X_a,X_2;\bar{Z}_a) \,  \Phi^t_{ab3}(X_a,X_b,X_3;Z_a,Z_b)
\, \Phi_{\tilde{b}54}(X_b,X_5,X_4;\bar{Z}_b) \ . \nonumber 
\end{eqnarray}
Here the tilde on the indices of the first and third vertex means that we use eq.\
\eqref{eq:threepointtwoMST} for two scalar legs but with $\Delta_a$ and $\Delta_b$ replaced by $d-\Delta_a$ and $d-\Delta_b$, respectively. We have placed a superscript 
$t$ on the vertex function of the central vertex to remind the reader that this depends on a function $t$ of the 3-point cross ratio. Integration is performed with the conformal 
invariant measure $\dd \mu$ of the embedding space variables \eqref{full_isospin_integral}. After splitting off the 
prefactor \eqref{Omegafivepts},
\begin{equation}
\Psi^{(\Delta_i)}_{(\Delta_a,\Delta_b;l_a,l_b;t)} (X_i)  =
\Omega^{(\Delta_i)} (X_i)
\psi^{(\Delta_{12},\Delta_3,\Delta_{45})}_{(\Delta_a,\Delta_b;l_a,l_b;t)}
(u_1,...,u_5)
\end{equation} 
the shadow integral \eqref{eq:5pshadow} gives rise to a finite conformal integral 
that defines the conformal block $\psi$ as a function of the five conformal 
invariant cross ratios $u_i$. These integrals depend on the choice of 
$(\Delta_a,l_a)$, $(\Delta_b,l_b)$ and the function $t(\mathcal{X})$. 

In \cite{Buric:2021ywo} we constructed five differential equations for these blocks. 
Four of these are given by the eigenvalue equations for the second and fourth order 
Casimir operators for the intermediate channels,
\begin{align}
		&\mathcal{D}^{2}_{(12)}=\left(\mathcal{T}_1+\mathcal{T}_2\right)_{[AB]}
		\left(\mathcal{T}_1+\mathcal{T}_2\right)^{[BA]}\,,\label{fivepointsQuadCasimir12}\\
		&\mathcal{D}^{2}_{(45)}=\left(\mathcal{T}_4+\mathcal{T}_5\right)_{[AB]}
		\left(\mathcal{T}_4+\mathcal{T}_5\right)^{[BA]}\,,\label{fivepointsQuadCasimir45}\\
		&\mathcal{D}^{4}_{(12)}=\left(\mathcal{T}_1+\mathcal{T}_2\right)_{[AB]}
		\left(\mathcal{T}_1+\mathcal{T}_2\right)^{[BC]}
		\left(\mathcal{T}_1+\mathcal{T}_2\right)_{[CD]}
		\left(\mathcal{T}_1+\mathcal{T}_2\right)^{[DA]}\,,\\
		&\mathcal{D}^{4}_{(45)}=\left(\mathcal{T}_4+\mathcal{T}_5\right)_{[AB]}
		\left(\mathcal{T}_4+\mathcal{T}_5\right)^{[BC]}
		\left(\mathcal{T}_4+\mathcal{T}_5\right)_{[CD]}
		\left(\mathcal{T}_4+\mathcal{T}_5\right)^{[DA]}\,. 
\end{align} 		
5-point conformal blocks are eigenfunctions of these four differential operators 
with eigenvalues determined by the conformal weights $\Delta_a, \Delta_b$ and the 
spins $l_a,l_b$ of the two internal fields $a=(12)$ and $b=(45)$ that appear in the 
operator products  $\phi_1 \phi_2$ and $\phi_4\phi_5$, respectively. The shadow 
integrals $\psi$ for conformal 5-point blocks turn out to be eigenfunctions of 
the following fifth differential operator 
\begin{equation} 		
		\mathcal{D}_{\rho,(12)3}^{4,3}=\left(\mathcal{T}_1+\mathcal{T}_2\right)_{[AB]}\left(\mathcal{T}_1+\mathcal{T}_2\right)^{[BC]}\left(\mathcal{T}_1+\mathcal{T}_2\right)_{[CD]}\left(\mathcal{T}_3\right)^{[DA]}\,
	    \label{fivepointsvertexop}
\end{equation} 
\textit{if and only if} the vertex functions $t(\mathcal{X})$ we use in the integrand to 
represent the central vertex of the OPE diagram is an eigenfunction of the reduced 
vertex operator of the previous subsection, specialized to vertices of type I. 
In this sense, the shadow integral intertwines the full 5-point vertex 
operator constructed explicitly in \cite{Buric:2021ywo} with the reduced vertex 
operator above. 
\medskip 

\begin{figure}[htp]
    \centering
    \includegraphics[width=\textwidth]{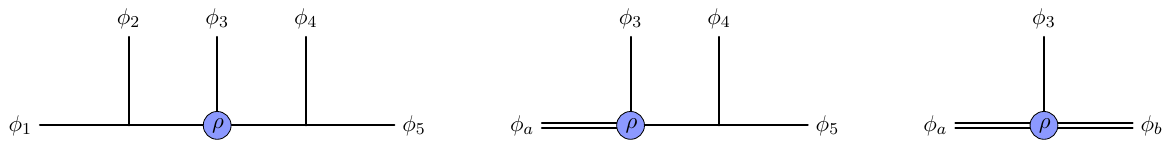}
    \caption{Scalar five point function (left), which in the OPE limit of fields $\phi_1$ and $\phi_2$ gets reduced to the 4-point function with a spinning leg $\phi_a$ (center), and after a second OPE limit for fields $\phi_4$ and $\phi_5$ gets fully reduced to a type I vertex (right).}
    \label{fig:OPElimit}
\end{figure}

There is another way to relate the full 5-point operator with the reduced one for 
type I vertices that employs OPE limits. In order to work out the reduction, we 
make use of the OPE in the limit where fields $(\phi_1,\phi_2)$ and $(\phi_4,\phi_5)$ 
are taken to be colliding, and are replaced with fields $\phi_a$ and 
$\phi_b$ whose conformal dimension and spin belongs to the tensor product 
of their representations. The first step is to reduce the operators to act on a 
spinning 4-point function, as in Figure~\ref{fig:OPElimit}.

The OPE of the fields  $\phi_1$ and $\phi_2$ acquires schematically the form
\begin{equation}
    \phi_1(X_1)\phi_2(X_2)=\sum_{\Delta_a,l_a} \frac{1}{(X_1\cdot X_2)^{\frac{\Delta_1+\Delta_2-\Delta_a+l_a}{2}}}C_{\phi_a}(X_1,X_2,Y,\partial_Y)\phi_a(Y)\,. 
    \label{OPEphi1phi2}
\end{equation}
Plugging this into the left hand side of eq.\ \eqref{five_point_correlator} 
allows us to rewrite the equation as
\begin{multline}
    \sum_{\Delta_a,l_a} \Biggl(\frac{1}{(X_1\cdot X_2)^{\frac{\Delta_1+\Delta_2-\Delta_a+l_a}{2}}}C_{\phi_a}(X_1,X_2,Y,\partial_Y)\expval{\phi_a(Y)\phi_3(X_3)\phi_4(X_4)\phi_5(X_5)}\\
    - \Omega_5^{\Delta_i}(X_i)\!\sum_{\Delta_b,l_b,t}\! \lambda_{12a}\lambda_{a 3 b;t}\lambda_{b45} \psi^{(\Delta_{12},\Delta_3,\Delta_{45})}_{(\Delta_a,\Delta_b;l_a,l_b;t)}(z_1,\bar{z}_1,z_2,\bar{z}_2,w) \Biggr)=0\,. 
    \label{five_point_to_four_point}
\end{multline}
The whole sum over weights $\Delta_a$ and spins $l_a$ on the left hand side can only vanish if 
every term vanishes separately. By considering the $X_1\cdot X_2\rightarrow 0$ limit in this 
expression, we can reproduce the pole on the first term only by imposing a specific leading 
behavior of the conformal blocks in the second term. If by convention we pick $\bar{z}_1$ to 
be the cross ratio that vanishes for $X_1\cdot X_2\rightarrow 0$ (otherwise we can simply 
rename variables $z_1\leftrightarrow \bar{z}_1$) and take into account that the prefactor 
$\Omega_5^{\Delta_i}(X_i)$ contains $(X_1\cdot X_2)^{\frac{\Delta_1+\Delta_2}{2}}$, we can 
then deduce the following behavior of the conformal blocks:
\begin{equation}
    \psi(z_1,\bar{z}_1,z_2,\bar{z}_2,w)\stackrel{\bar{z}_1\rightarrow 0}{\sim}\bar{z}_1^{\frac{\Delta_a-l_a}{2}} \psi(z_1,z_2,\bar{z}_2,w)\,.
\end{equation}
Imposing this leading behavior in the eigenvalue equations for the differential operators \eqref{fivepointsQuadCasimir12}--\eqref{fivepointsvertexop} allows to reduce the action of 
the differential operators to a 4-dimensional subspace of cross ratios in the following way:
\begin{equation}
    \lim_{\bar{z}_1\rightarrow 0}\left[\bar{z}_1^{-\frac{\Delta_a-l_a}{2}}\mathcal{D}\left( \bar{z}_1^{\frac{\Delta_a-l_a}{2}} \psi(z_1,z_2,\bar{z}_2,w)\right)\right]=\mathcal{E}\,\psi(z_1,z_2,\bar{z}_2,w).
\end{equation}
To complete the OPE limit, one needs to impose $X_2\rightarrow X_1$ which is of course 
stronger than the condition $X_1\cdot X_2=0$ we discussed up to now. Doing so requirs a bit of caution. One subtlety is that the limit is quite sensitive to the set 
of cross ratios used to parameterize the 5-point function. For example, if we 
had decided to work with the cross ratios $u_i$, the complete OPE limit $X_1=X_2$ would 
have implied both $u_2=1$ and $u_5=u_4$, thereby overreducing the space of cross ratios. This issue is avoided in the $\{z_i, \bar{z}_i,w\}$ set of cross ratios, provided that the limit is taken in the appropriate way.
The correct way to address this limit in coordinate space is to take $X_1=X_a-\epsilon Z_a$, $X_2=X_a+\epsilon 
Z_a$, and then take the $\epsilon\rightarrow 0$ limit. Note how this step requires either 
Lorentzian signature or an analytic continuation: given that we have already imposed the light-cone 
condition $X_1\cdot X_2=0$, the variable $Z_a$ must satisfy the constraints $Z_a\cdot X_a=Z_a\cdot Z_a=0$. The latter are the same as the orthogonality and null constraints of an STT polarization vector, which explains why we used the symbol $Z_a$ to parameterize the difference between $X_2$ and
$X_1$. Following this prescription, the $\epsilon\rightarrow 0$ limit 
in cross ratio space corresponds to taking the sole limit $z_1\rightarrow 0$, with $z_2, 
\bar{z}_2$ being unaffected and $w$ acquiring a dependence on the newly introduced coordinate $X_a$ and spin 
variable $Z_a$. In order to reproduce the correct eigenvalue for the quadratic Casimir
$\mathcal{D}_{(12)}^2$ in the $z_1\rightarrow 0$ limit, the Euclidean conformal blocks must exhibit leading behavior in $z_1$ of the type
\begin{equation}
     \psi(z_1,z_2,\bar{z}_2,w)\stackrel{z_1\rightarrow 0}{\sim} 
     z_1^{\frac{\Delta_a+l_a}{2}} \psi(z_2,\bar{z}_2,w).
\end{equation}
After these conjugations and limits, both the quadratic and quartic Casimirs associated 
to the internal leg $a=(12)$ are reduced to constants, and the remaining three operators
\begin{equation}
    \mathcal{D}_{(45)}^2\,, \qquad     \mathcal{D}_{(45)}^4 \,, 
    \qquad \mathcal{D}_{\rho,a3}^{4,3}
\label{fourpoint_operators}
\end{equation}
characterize the spinning 4-point function that is shown in  Figure~\ref{fig:OPElimit} 
(center). The latter depends only on three cross ratios $z_2,\bar{z}_2,w$, the spacetime 
dimension $d$, and the external data
\begin{equation}
        \frac{\Delta_{a}-\Delta_3}{2}\,, \qquad 
        \frac{\Delta_5-\Delta_4}{2}\,, \qquad l_a\,.
\end{equation}
It is straightforward to repeat the same procedure we just outlined for the leg $b=(45)$ 
in the remaining correlator, i.e. one can impose leading behaviors of the type
\begin{equation}
    \psi(z_2,\bar{z}_2,w)\stackrel{\bar{z}_2\rightarrow 0}{\sim} 
    \bar{z}_2^{\frac{\Delta_b-l_{b}}{2}}\psi(z_2,w)
    \stackrel{z_2\rightarrow 0}
    {\sim} z_2^{\frac{\Delta_{b}+l_{b}}{2}}
    \bar{z}_2^{\frac{\Delta_{b}-l_{b}}{2}}\psi(w)
\end{equation}
to ensure that the quadratic and quartic Casimir of the internal leg $b=(45)$ assume
constant values that are determined by the weight and spin of the intermediate field. 

At the end of this procedure, one is left with a 3-point block of two STT's 
and one scalar, that is to say a vertex of type I, which is characterized by the sole vertex 
operator $\mathcal{D}_{\rho,a3}^{4,3}$ acting on the remaining cross 
ratio $w$. By replacing 
$X_5=X_b-\epsilon'Z_b$, $X_4=X_b+\epsilon' Z_b$ and taking $\epsilon'\rightarrow 0$, we find the following 
expression for $w$ in terms of the external 3-point data,
\begin{equation}
    w=1-\frac{ (X_3\cdot X_a)(X_3\cdot X_b)\left[(X_a\cdot Z_b)(X_b\cdot Z_a)-(X_a\cdot X_b)(Z_a\cdot Z_b)\right]}{\left[(X_3\cdot Z_b)(X_a\cdot X_b)-(X_3\cdot X_b)(X_a\cdot Z_b)\right] \left[(X_3\cdot Z_a)(X_a\cdot X_b)-(X_3\cdot X_a)(X_b\cdot Z_a)\right]}\,.
\end{equation}
After further inspection, this expression can be identified with
\begin{equation}
    w=1-\frac{H_{ab}}{V_{a,3b}V_{b,a3}}    =1-\mathcal{X}'\,,    
\end{equation}
where the cross ratio $\mathcal{X}'$ is equal to $\mathcal{X}$ with the replacement $(1,2,3)\rightarrow (a,b,3)$. The resulting operator can be easily identified with the $\ell_1=\ell_2=0$ case of
our general expression \eqref{H_basis}.

\section{Vertex Operator and Generalized Weyl algebras} 
\label{sect:GeneralizedWeyl}

The Hamiltonians we constructed in our letter \cite{Buric:2020dyz} and the extension discussed 
in the previous section have nice properties, even though they may look a bit uninviting at 
first. In this section we exhibit some of their underlying algebraic structure. This allows us to recast the vertex operator into a one-line expression, somewhat analogous to the 
harmonic oscillator that possesses a particularly simple representation in terms of creation 
and annihilation operators. Here we define a generalized Weyl algebra with relatively 
simple commutation relations and then build our vertex operators directly in terms of its 
generators. An important role in our discussion is played by the scalar 
product of the vertex system. 

\subsection{Single variable vertices and the Gegenbauer scalar product} 

Functions on the configuration space $\mathcal{M}$ inherit a scalar product 
from the Haar measure on the conformal group. This is the case in general, but in 
particular for the 1-dimensional spaces we are dealing with in this paper. Working 
out this scalar product is straightforward in principle, but a bit cumbersome in 
practice. Since we have not found the answer in the literature, we included the full 
calculation in Appendix \ref{app:scalar_product}. The result is surprisingly simple: 
it turns out that, when written in the variable $s=1-2\mathcal{X}$, the group theoretic 
scalar product on the configuration space $\mathcal{M}$ coincides with 
the Gegenbauer scalar product, 
\begin{equation}
\langle f, g \rangle_{\ag(d;\ell_i)} :=  \int_{-1}^{+1} \dd s \, (1-s^2)^{\ag(d;\ell_i)- 
\frac{1}{2}} \, \overline{f(s)} \, g(s)\, ,  
\label{GegSP}
\end{equation}
with the parameter $\alpha$ given by 
\begin{equation}  \label{eq:alpha}
\ag(d;\ell_i) := \ell_1+\ell_2+\frac{d-3}{2}\,.
\end{equation} 
In the following we shall implicitly assume that the parameters assume only those 
values that appear in the context of our three single parameter vertices, i.e. for $d=3$
we have $\ell_1 = 0 = \ell_2$ while for $d > 3$ only $\ell_2 = 0$. The only case for 
which $\ell_2$ can also be non-zero is in $d=4$. Gegenbauer polynomials $C_n^{(\ag)} 
(s)$ provide an orthogonal basis for $\langle -, - \rangle_{\ag}$:
\begin{equation}
\langle C_m^{(\ag)}, C_n^{(\ag)} \rangle_{\ag} = \frac{\pi\, 2^{1-2\ag} }{\Gamma (\ag)^2} 
\frac{\Gamma (n+2 \ag)  \Gamma (n+\ag)}{\Gamma(n+\ag+1) \Gamma(n+1)} \, \dg_{mn}\,. \label{normGeg}
\end{equation}
As one may check by explicit computation, our vertex differential operators $H^{(d;\Dg_i;l_i)}
(\mathcal{X}, \ds_{\mathcal{X}})$ are hermitian with respect to a Gegenbauer scalar product 
whenever the conformal weights $\Dg_i$ and the STT spins $l_i$ are analytically continued to satisfy,
\begin{equation} \label{conjugation} 
\bar{\Dg}_i = d- \Dg_i\,, \qquad \bar{l}_i = 2-d-l_i\,,
\end{equation}
i.e. $(\Dg_i;l_i) \in \left(\frac{d}{2} + \mathrm{i} \Rs \right) \times \left(\frac{2-d}{2}+\mathrm{i} \Rs \right)$, while $(d;\ell_i)$ are kept as real parameters. Our goal now is to compute the Hamiltonian 
in the basis of Gegenbauer polynomials. In doing so we shall restrict to the case with $d>3$ 
and $\ell_2 = 0$, i.e. we exclude the somewhat special case of $d=4$ with $\ell_1 \neq 0 
\neq \ell_2$ for which our Hamiltonian contains a non-polynomial term, see comments at the 
end of the previous section. The computation of the Hamiltonian in the Gegenbauer basis 
relies on its expression as
\begin{equation}
    H^{(d,\Dg_i,l_i,\ell_i)} = h_0^{(s)}(s) + \sum_{q=1}^4 h_q^{(s)}(s) (1-s^2)^{q-1} \ds_s^q\,,
    \label{H_s_var}
\end{equation}
where all $h_q^{(s)}(s)$ are polynomials of order at most $3$ in $s$ whenever $\ell_2 = 0$. 
It proceeds with the help of three well-known identities:
\begin{itemize}
\item[(S)] the recursion relation of Gegenbauer polynomials
\begin{equation}
s \cdot C_n^{(\ag)} = \frac{(n+1) C_{n+1}^{(\ag)} + (n+ 2\ag -1) C_{n-1}^{(\ag)}}{2(n+\ag)}\,,
\label{recGeg}
\end{equation}
\item[(D)] the Gegenbauer differential equation in self-adjoint form,
\begin{equation}
\mathcal{D}_{\ag} \cdot C_n^{(\ag)} = (1-s^2)^{\frac{1}{2}-\ag} \ds_s \, (1-s^2)^{\frac{1}{2}+\ag} \ds_s \cdot C_n^{(\ag)} = - n(n+2\ag) C_n^{(\ag)}\,,
\label{stlGeg}
\end{equation}
\item[($\Theta$)] and the first order differentiation operator 
\begin{equation}
\ds_{\Theta} \cdot C_n^{(\ag)} = (1-s^2) \ds_s  \cdot C_n^{(\ag)} = \frac{(n+2\ag-1)(n+2\ag) C_{n-1}^{(\ag)} - n(n+1) C_{n+1}^{(\ag)}}{2(n+\ag)}\,.
\label{dphGeg}
\end{equation}
\end{itemize}
Using these building blocks (S),(D),($\Theta$), our Hamiltonians can be recast into the form 
\begin{align}
H^{(d,\Dg_i,l_i,\ell_i)}  = &\, \mathcal{D}_{\ag} \ds_{\Theta}^2 + \left( k_{3,1}s + k_{3,0} \right) \mathcal{D}_{\ag} \ds_{\Theta} + \left( k_{2,2} s^2 + k_{2,1} s + k_{2,0} \right) \mathcal{D}_{\ag} \nonumber \\
&+ \left( k_{1,1} s + k_{1,0} \right) \ds_{\Theta} + k_{0,2} s^2 + k_{0,1} s + k_{0,0}\,, \label{algebraicExpGeg}
\end{align}
where the coefficients $k_{i,j} = k_{i,j}(d,\Dg_i,l_i,\ell_i)$ are given by 
\begin{align}
& k_{3,0} = 2\mathrm{i}\cg_3\,, \qquad k_{3,1} = 2(\nu_1+\nu_2+\ag)+3\,,  \nonumber\\
& k_{2,0} = \cg_1^2+\cg_2^2-\cg_3^2-2\nu_1\nu_2 -2 (\ag-1)(\nu_1+\nu_2) - \ag(1+3\ag) + \frac{13}{4}\,, \nonumber\\
& k_{2,1} = -2\cg_1 \cg_2 + \mathrm{i}\cg_3 \left(  2(\nu_1+\nu_2+\ag)+3 \right), \qquad k_{2,2} = \nu_1^2+\nu_2^2 + 4 l\nu_1\nu_2+(4\ag+1)(\nu_1+\nu_2) + 4\ag^2+3\,, \nonumber\\
& k_{1,0} = 2(\nu_1+\nu_2+\ag) \cg_1\cg_2 + \mathrm{i}\cg_3 \left(-2\nu_1\nu_2 + (2\ag+1)(\nu_1+\nu_2) -4\ag^2 + 2 \ag+2\right), \label{kcoefficients}\\
& k_{1,1} = -2(\nu_1^2\nu_2+\nu_2^2\nu_1)+ (1-2\ag)\left(\nu_1^2+\nu_2^2+4 \nu_1 \nu_2 +(1+4\ag)(\nu_1+\nu_2+\ag+1)-\ag-2 \right),  \nonumber\\
& k_{0,1} =2 \nu_1\nu_2 \left(\cg_1\cg_2+\mathrm{i}\cg_3(\ag+\frac{1}{2})  \right), \qquad k_{0,2} = -\nu_1\nu_2(\nu_1-1)(\nu_2-1)\,.\nonumber
\end{align}
The parameters $\cg_k, \nu_k$ are defined through
\begin{equation}
\Dg_k := \frac{d}{2}+\mathrm{i} \cg_k\,, \qquad k=1,2,3\,, \qquad \qquad \nu_k := l_k-\ell_1\,, \qquad k=1,2\,,
\label{new_params}
\end{equation}
and $\alpha$ was introduced in eq.\ \eqref{eq:alpha}. Let us stress once again that the solution 
we have displayed here applies to $\ell_2 =0$ and $d\geq 3$. When $\ell_2 \neq 0$, the Hamiltonian
contains a non-polynomial term in $h_0^{(s)}(s)$ which is proportional to $(1-s)^{-1}$. So, while 
it is in principle possible to compute the action of the MST$_2$-MST$_2$-scalar Hamiltonian in 
$d=4$ on Gegenbauer polynomials with $\ag = (d-3)/2 + \ell_1+\ell_2$, it does not directly fit our Ansatz \eqref{algebraicExpGeg}. 
 
Plugging the identities \eqref{recGeg}, \eqref{stlGeg}, \eqref{dphGeg} back into 
\eqref{algebraicExpGeg} one can obtain simple explicit formulas for the matrix elements 
\begin{equation}
H_{mn}^{(d;\Dg_i;l_i;\ell_i)} = \frac{\langle C_m^{(\ag)}, H^{(d;\Dg_i;l_i;\ell_i)} \cdot C_n^{(\ag)}\rangle_{\ag}}{\langle C_m^{(\ag)}, C_m^{(\ag)} \rangle_{\ag} }\,. \label{MEstoSPs}
\end{equation}
One observes that these vanish whenever $|n-m| > 2$, so our Hamiltonian in the Gegenbauer basis has non-vanishing matrix elements only close to the diagonal. In terms 
of the matrix elements, the hermiticity property of the vertex differential operators 
reads 
\begin{equation}
    H_{mn}^{(d;\Dg_i;l_i;\ell_i)} = \frac{\langle C_n^{(\ag)}, C_n^{(\ag)} \rangle_{\ag}}{\langle C_m^{(\ag)}, C_m^{(\ag)} \rangle_{\ag}}    H_{nm}^{(d;d-\Dg_i;2-d-l_i;\ell_i)}\,.
    \label{hermicity}
\end{equation}
With our formulas \eqref{algebraicExpGeg} and \eqref{kcoefficients} we have fulfilled our 
first promise, namely to write the Hamiltonian in a much more compact form that fully replaces the two pages of formulas we spelled out in the previous section.

\subsection{A generalized Weyl algebra acting on tensor structures}
\label{sec:Weyl}

We want to go one step further and write the vertex Hamiltonians in terms of the 
generators of some Weyl-like algebraic structure that acts on Gegenbauer polynomials
and hence on 3-point tensor structures. Our algebra contains three generators $A,A^\dagger$
and $N$ and it depends on the parameters $\ag$, $\nu_1$ and $\nu_2$ which we introduced in 
eqs. \eqref{eq:alpha} and \eqref{new_params}. When acting on Gegenbauer polynomials, these 
three operators are given by 
\begin{align}
    N C_n^{(\ag)} &:= n C_n^{(\ag)}, \label{NCn} \\
    A C_n^{(\ag)} &:= (n+ \nu_1+2\ag)(n+\nu_2+2\ag) 
    \frac{n+2\ag-1}{n+\ag} C_{n-1}^{(\ag)}\,,\label{ACn} \\
    A^{\dagger} C_n^{(\ag)} &:= (n-\nu_1)(n-\nu_2) \frac{n+1}{n+\ag} 
    C_{n+1}^{(\ag)}, \label{A*Cn}
\end{align}
where eq.\ \eqref{ACn} applies to all $n>0$, and $A C_0^{(\ag)} = 0$ when $n=0$, i.e.\ the 
state $C_0^{(\ag)}$ is annihilated by the lowering operator $A$. Similarly, 
the action of the raising operator $A^\dagger$ vanishes if $n=\nu_1$ or $n=\nu_2$.
Consequently, one can restrict the action of $A,A^\dagger$ and $N$ to the finite 
dimensional subspace that is spanned by $C_n^{(\ag)}$ for $n = 0,\dots,\mathrm{min}
(\nu_1,\nu_2)$. This should remind us of the space of 3-point tensor structures we discussed 
at the very end of section \ref{sect:threepointfunctions}. There we argued that the space of 
3-point tensor structures has dimension $n_t+1$ with $n_t = \textit{min}\, (l_1 - \ell_1, l_2 - \ell_1)= \textit{min}\, 
(\nu_1, \nu_2)$ in the case of $\ell_2=0$, using the definition \eqref{new_params} of $\nu_k$. Therefore, the truncation of the 
action of $A,A^\dagger$ and $N$ to a finite dimensional subspace of 
Gegenbauer polynomials we observe here is fully consistent with the 
finiteness of the space of 3-point tensor structures, at least for 
$d > 3$. We will discuss the special case of $d=3$ below. 
\smallskip 

From the action on Gegenbauer polynomials it is possible to check that the 
operators $A,A^{\dagger}$ and $N$ obey the following relations  
\begin{align}
    &[N,A^{\dagger}] = A^{\dagger}, \label{NA*}\\
    &[N,A]=-A\,, \label{NA} \\
& A A^{\dagger} = \frac{(N+1)(N+2\ag)}{(N+\ag)(N+\ag+1)} (N-\nu_1)(N+2\ag+\nu_1+1)(N-\nu_2)(N+2\ag+\nu_2+1)\,, \label{AA*_alpha}\\ 
        & A^{\dagger} A = \frac{N(N+2\ag-1)}{(N+\ag-1)(N+\ag)} (N+\nu_1+2\ag)(N-\nu_1-1)(N+\nu_2+2\ag)(N-\nu_2-1)\,. \label{A*A_alpha}
\end{align}
We can use these to define a family of abstract algebras that depends 
parametrically on $\ag = \ell_1 +(d-3)/2$ and $\nu_k = l_k - \ell_1$. This 
family comes equipped with an involutive antiautomorphism $(-)^\ast$ defined by 
$N^\ast = N$ and $A^\ast = A^\dagger$. It coincides with the adjoint whenever 
$d$ is real and the spins $l_i$ satisfy the relation \eqref{conjugation} that 
is needed in order for our vertex operators to be hermitian or, equivalently, 
\begin{equation}
    \bar{\ag} = \ag\,, \qquad \bar{\nu}_1 = -(2 \ag+1+\nu_1)\,, \qquad \bar{\nu}_2 = -(2 \ag+1+\nu_2)\,.
\end{equation}
Having introduced the algebra generated by $A, A^\dagger$ and $N$, the vertex operator $H$ can now be written as a rational combination of the generators of this algebra:
\begin{align}
H^{(d;\Dg_i;l_i;\ell_i)} =& B^{\dagger} B - \Gamma  (N+\ag)^2+ \frac{\ag(\ag-1)\, K}{(N+\ag)^2-1} + E^{(d;\Dg_i;l_i;\ell_i)}.
\label{HamFac}
\end{align}
Here, we defined the operators
\begin{align}
B^{\dagger} :=& \frac{A-A^{\dagger}}{2\mathrm{i} } - \mathrm{i} (2\cg_1\cg_2 +\mathrm{i} \cg_3)(N+\ag)\,,  \\
B :=& \frac{A-A^{\dagger}}{2\mathrm{i} } + \mathrm{i} (2\cg_1\cg_2 -\mathrm{i} \cg_3)(N+\ag)\,,
\end{align}
and the two parameters
\begin{align}
\Gamma :=& \frac{1}{4} (1+4\cg_1^2)(1+4\cg_2^2)\,, \\
K :=&  (\nu_1+\ag)(\nu_2+\ag)(\nu_1+\ag+1)(\nu_2+\ag+1)\,.
\end{align}
The constant term $E^{(d;\Dg_i;l_i;\ell_i)}$ is obtained by relating $H^{(d;\Dg_i;l_i;\ell_i)} \cdot 1 = h_0^{(d;\Dg_i;l_i;\ell_i)}(0)$ to the action of $A,A^{\dagger},N$ on $C_0^{(\ag)}$. In this way we find
\begin{align*}
E^{(d;\Dg_i;l_i;\ell_i)} =& h_0^{(d;\Dg_i;l_i;\ell_i)}(0) - \ag^4 - (1+2\nu_1+2\nu_2)\ag^3 \\
& +\ag^2 \left( \frac{1}{4} + \cg_1^2 + \cg_2^2- \cg_3^2 +\nu_1(\nu_1+1)+\nu_2(\nu_2+1)+4\nu_1\nu_2 \right) \\
&-2 \nu_1\nu_2 \ag -\nu_1\nu_2(1+\nu_1\nu_2).
\end{align*}
This concludes the algebraic reformulation of our vertex Hamiltonians.
It is remarkable that the algebra only depends on the spins and dimension 
$d$, i.e.\ that all the dependence on the conformal weights of the three 
fields resides in the Hamiltonian. 
\medskip

The case of $d=3$, which implies $\ell_1=0$, requires additional consideration since in this case 
one can also have 
odd-parity tensor structures, see our discussion at the end of subsection~\ref{ssec:3ptI-II}. 
As we saw there, the STT-STT-scalar vertex in $d=3$ is unique in that it admits a total 
number
\begin{equation}
  ( \mathrm{min}(\nu_1,\nu_2)+1)+\mathrm{min}(\nu_1,\nu_2) = \mathrm{min}(2l_1+1,2l_2+1)\, 
\end{equation}
of 3-point tensor structures. A complete and orthogonal basis can be obtained from the union of 
Chebyshev polynomials of the first and second kind,
$$\{C_n^{(0)}(s)\}_{n=0,\dots,\mathrm{min}(l_1,l_2)} \quad , \quad 
\{ \sqrt{1-s^2} C_n^{(1)}(s)\}_{n=1,\dots,\mathrm{min}(l_1,l_2)}\ . $$  
The action of $A,A^{\dagger},N$ in $d=3$, however, is more conveniently written in the 
Fourier basis $e^{\mathrm{i}n\tg}$, where $n=-\mathrm{min}(l_1,l_2),\dots,+\mathrm{min}(l_1,l_2)$ 
and the new variable $\theta$ is related to our cross ratio $\mathcal{X}$ by 
$\mathcal{X} = \sin^2 \frac{\tg}{2}$. In this case, the action of 
$A, A^\dagger$ and $N$ on the Fourier basis is 
\begin{equation*}
N \, e^{\mathrm{i} n \tg} := n e^{\mathrm{i} n \tg}, \qquad A^{\dagger}  e^{\mathrm{i} n \tg} := 
(n-l_1)(n-l_2) e^{\mathrm{i} (n+1) \tg}, \qquad A \, e^{\mathrm{i} n\tg} := (n+l_1)(n+l_2) e^{\mathrm{i} (n-1) \tg}. 
\end{equation*}
It is easy to see that these operators satisfy the following polynomial relations
\begin{align}
&[N,A^{\dagger}] = A^{\dagger}, \label{NA*3} \\
&[N,A] = -A\,, \label{NA3} \\
& A A^{\dagger} = (N-l_1)(N-l_2)(N+l_1+1)(N+l_2+1)\,, \label{AA*} \\
& A^{\dagger} A = (N-l_1-1)(N-l_2-1)(N+l_1)(N+l_2)\,. \label{A*A} 
\end{align}
These relations agree with those we found in eqs.\ \eqref{NA*}-\eqref{AA*_alpha} above for the 
special choice $\ag = \ell_1 + (d-3)/2 = 0$ relevant for vertices in $d=3$, where
$\ell_1 = 0$. In other words, we have now shown that for $d=3$, the algebra we have introduced
above possesses a finite dimensional representation on the space of Chebyshev polynomials of 
first and second kind. 

For $\alpha=0$ and $\alpha = 1$, the algebra of $A,A^\dagger$ and $N$ is one special example of a larger family of 
algebras of the form $ A^{\dagger}A = f(N)$, $A A^{\dagger}  = f(N+1)$ that can be associated 
with a polynomial $f(N)$. Such families of algebras have been studied for a long time in the 
mathematics literature, going back at least as early as \cite[\S 3]{joseph1977generalization}. 
The representation theory of these algebras was studied in \cite{smith1990class} and in
\cite{bavula1992generalized}, where the latter author first used the term "generalized Weyl 
algebra". It was then in \cite{hodges1993noncommutative} that these algebras were first reformulated as 
non-commutative deformations of the Kleinian singularity of type $\tilde{A}_{n-1}$ when $f$
is a polynomial of degree $n$. Finally, using quiver theory, the authors of \cite{crawley1998noncommutative} 
generalized this analysis to non-commutative deformations of the Kleinian singularities 
associated to any finite subgroup of $\mathrm{SL}_{\Cs}(2)$. In this 
context, the algebra with relations \eqref{NA*3} --- \eqref{A*A} is thus called a generalized 
Weyl algebra or deformed Kleinian singularity of type $\tilde{A}_3$.

For $\ag \neq 0,1,$ the relations \eqref{AA*_alpha} and \eqref{A*A_alpha} are no longer 
polynomial, at least not in the way we wrote them. Nevertheless, they can be recast as an 
$\ag$-dependent family of generalized Weyl algebras if we are willing to sacrifice the 
property $A^\ast = A^\dagger$. Indeed, any rescaling of the operators $A$ and 
$A^{\dagger}$ by a rational function of $N$ defines a homomorphism of 
algebras\footnote{We thank Pavel Etingof for pointing this out to us.}. In the 
case of eqs.\ \eqref{AA*_alpha} and \eqref{A*A_alpha}, it is natural to take
\begin{equation}
U := \frac{(N+\ag)(N+\ag+1)}{(N+1)(N+2\ag)} A\,, \qquad V := A^{\dagger}\,, \label{A_to_U}
\end{equation}
in which case the modified relations read
\begin{align}
    &[N,V] = V\,, \label{NV}\\
    &[N,U]=-U\,, \label{NU} \\
    & U V =  (N-\nu_1)(N+\nu_1+2\ag+1)(N-\nu_2)(N+\nu_2+2\ag+1)\,, \label{altAA*_alpha}\\ 
    & V U =  (N-\nu_1-1)(N+\nu_1+2\ag)(N-\nu_2-1)(N+\nu_2+2\ag)\,, \label{altA*A_alpha}
\end{align}
and also define a generalized Weyl algebra of type $\tilde{A}_3$, but now with an extra 
deformation parameter $\ag$. In any given representation, the homomorphism $A \longleftrightarrow U$ in \eqref{A_to_U} is bijective if and only if $-1,-\alpha,-(\alpha+1), 
-2\alpha \notin \mathrm{Spec}(N)$. This condition is indeed satisfied in the Gegenbauer polynomial
representations, where $U$ is explicitly represented as  
\begin{equation}
U \cdot C_n^{(\ag)} = (n+\nu_1+2\ag)(n+\nu_2+2\ag) \frac{n+\ag-1}{n} C_{n-1}, \,\,\, \forall n>0, 
\qquad U \cdot C_0^{(\ag)} = 0.
\end{equation}
As a result, $(A,A^{\dagger},N) \mapsto (U,V,N)$ defines an isomorphism, and all vertex systems of type I and II are representations of the generalized $\tilde{A}_3$ Weyl algebra with relations \eqref{NV} --- \eqref{altA*A_alpha}.
\medskip 

Our final comment in this subsection concerns the fact that our expression \eqref{HamFac}
for the Hamiltonian depends on parameters only through the 
combinations $\ag,\nu_i,\cg_i$, 
at least up to the constant term. It follows that (see appendix \ref{app:d_def} for a 
further generalization)
\begin{equation}
    H^{(d;\Dg_i;l_i;\ell_1,0)} = H^{(d+2\ell_1;\Dg_i+\ell_1;l_i-\ell_1;0,0)} + 
    \Dg E^{(d;\Dg_i;l_i;\ell_1)},
    \label{dim_shift}
\end{equation}
where
\begin{align*}
\ell_1^{-1}\Dg E^{(d;\Dg_i;l_i;\ell_1)} =& -\frac{2}{3} \ell_1^3 + \frac{8 \ag + 26}{3}  \ell_1^2 \\
&+ \frac{4}{3}\left(\! -2\ag^2+ 2(\nu_1-\nu_2-41/2)\ag -\cg_1^2 +\cg_2^2-\cg_3^2 +
\nu_1(\nu_1+1) - \nu_2(\nu_2+1)   -33 \! \right)\! \ell_1\\
&+ \frac{16}{3} \ag^2 ( \nu_2-\nu_1 + 17/4)+ \frac{8}{3} \ag
\left(\cg_1^2-\cg_2^2+\cg_3^2-\nu_1(\nu_1+4) + \nu_2(\nu_2+5/2) +9 \right) \\
&+ 2 \left(2\cg_1^2-\cg_2^2+\cg_3^2-2\nu_1(\nu_1+1) +\nu_2(\nu_2+1) +\frac{9}{2} \right).
\end{align*}
Now, the $\ag$-deformed relations \eqref{AA*_alpha} and \eqref{A*A_alpha}  of the generalized 
Weyl algebra coincide with the $d=3$ relations \eqref{AA*}, \eqref{A*A} whenever $\ag = 0$ 
or $\ag =1$. In the former case, $\ag = 0 \iff (d,\ell_1) = (3,0)$. On the other hand, the 
latter case $\ag=1$ can occur in two situations,
\begin{equation}
(d,\ell_1) = (5,0)\,,\qquad \mathrm{or} \qquad (d,\ell_1) = (4,\frac{1}{2})\,, 
\end{equation}
in which case
\begin{equation}
\nu_i = l_i\,,\qquad \mathrm{or} \qquad \nu_i = l_i - \frac{1}{2}\,, 
\end{equation}
and the two operators are the same up to a constant shift,
\begin{equation}
    H^{(d=5;\Dg_i;l_i;0,0)}(\mathcal{X},\ds_{\mathcal{X}}) = 
    H^{(d=4;\Dg_i-\frac{1}{2};l_i+\frac{1}{2};\frac{1}{2},0)}(\mathcal{X},\ds_{\mathcal{X}})+\Dg E\,.
\end{equation}
In both of these cases, the Gegenbauer polynomials become Chebyshev polynomials of the second 
kind $\{ C_n^{(1)}(s)\}_{n=0,\dots, \min(l_1,l_2)}$, and the two vertex operators are related 
to the $d=3$ operator by a similarity transformation,
\begin{equation}
 H^{(d=5;\Dg_i;l_i;0,0)}(\mathcal{X},\ds_{\mathcal{X}})  =
    \frac{1}{\sqrt{\mathcal{X}(1-\mathcal{X})}} H^{(d=3;\Dg_i-1;l_i+1)}(\mathcal{X},
    \ds_{\mathcal{X}})    \sqrt{\mathcal{X}(1-\mathcal{X})} + \Dg E\,.
\end{equation}
In particular, the parity-even 3-point tensor structures of two STTs in $d=5$, are equivalent to the 
parity-odd 3-point tensor structures of two STTs in $d=3$.

\section{Map to the Lemniscatic CMS model}
\label{sect:MappingElliptic}

In the previous section we have found quite an elegant reformulation of our vertex operators that 
makes it seem a bit more tractable than the original formulas we displayed in section~\ref{sect:ConstructionOperator}. All this is somewhat similar to the Casimir operators 
of Dolan and Osborn, which appeared a bit uninviting at first, but were found to possess 
interesting algebraic structure that led to explicit solutions, in particular 
in even dimensions. In \cite{Isachenkov:2016gim}, it was discovered that the usual Casimir operator can be mapped to another well studied operator, 
namely the Hamiltonian of an integrable 2-particle CMS model. Here we establish a very similar statement for the vertex operators. By 
explicit computations, these operators can be mapped to the lemniscatic CMS  
model, a special case of the  crystallographic elliptic CMS models found 
by Etingof, Felder, Ma and Veselov \cite{etingof2011107}. We review this model in the first 
subsection before constructing the map from our vertex differential operator. The 
third subsection contains a complete identification of parameters. 

\subsection{The elliptic \texorpdfstring{$\Zs/4 \Zs$}{Z/4Z} CMS model}

While our vertex system examples have received little attention, there has been much 
discussion in similar cases of the relation between deformations of Kleinian singularities 
on the one hand, and CMS models for the corresponding complex 
reflection group on the other hand --- see e.g. \cite{oblomkov2005deformed} for 
$\tilde{A}_{n-1}$, and \cite{etingof2007harish} for the general case, both of which are 
based on \cite{holland1999quantization}. Thus, the identification of our operator with a 
lemniscatic CMS model is less surprising in light of the $\tilde{A}_3$ singularity of 
\S\ref{sec:Weyl}. That said, apart from $\tilde{A}_{1}$, the only integrable models 
explicitly studied in this particular context have so far always been rational. And while 
the integrable systems in \cite{etingof2008lie} include, amongst others, the compact
$\mathfrak{so}_{\Rs}(6)$ analogue of our MST$_2$-MST$_2$-scalar vertex in $d=4$, the authors do 
not make the connection with the elliptic integrable models of \cite{etingof2011107}. 

The elliptic CMS models associated with the complex reflection groups $\Zs_m$ form a family 
of quantum mechanical integrable systems with (complexified) coordinate on an orbifold 
curve of the form 
\begin{equation}
\mathcal{M} =  \Cs / \left(\Zs_m \ltimes (\Zs \oplus \tau \Zs)\right),
\label{full_quotient}
\end{equation}
where $\Zs \oplus \tau \Zs \subset \Cs$ is a 2-dimensional lattice with elliptic modulus 
$\tau$ in the upper half of the complex plane, and elements of the 
group $\Zs_m \subset \mathrm{SO}(2)$ act on the lattice as a point group, i.e. through 
rotations by angles $\varphi_n = n/2m\pi$ where $n=1,\dots,m$. It is well known that 
the only 2-dimensional lattices with a non-trivial point group $\mathbb{Z}_m$ appear 
for $m=2,3,4,6$. Apart from $m=2$, the elliptic modulus $\tau$ is also fixed so that 
spaces $\mathcal{M}$ of the form 
\eqref{full_quotient} only appear for 
\begin{equation}
(m,\tau) \in \{2\} \times \Cs_+\,, \qquad \mathrm{or} \qquad (m,\tau) \in 
\{3, e^{2 \pi \mathrm{i} /3}\} \cup \{4, \mathrm{i}\} \cup \{6, e^{\pi \mathrm{i} /3}\}\,. 
\end{equation}
In \cite{etingof2011107}, the authors construct new integrable models on each of these 
curves, but only the case $\tau = \mathrm{i}$ with group action of $\mathbb{Z}_4$ 
turns out to be relevant for us. In order to proceed, let us write the associated curve 
\eqref{full_quotient} as the quotient of the so-called lemniscatic elliptic curve 
$E_\mathrm{i}$ by a $\Zs_4$ action, 
\begin{equation}
\mathcal{M} = E_{\mathrm{i}}/\Zs_4\,, \quad \textit{where}\quad  E_{\mathrm{i}} = \Cs/(\Zs \oplus \mathrm{i} \Zs) 
= \{ z \in \Cs \,\vert\, z \sim z+1\sim z+\mathrm{i} \}\,.  
\end{equation}
Here, the $\Zs_4$ action is the obvious one that is given by multiplication of $z
\in E_{\mathrm{i}}$ with any fourth root of unity $\zeta^4 =1$, i.e.\ $z \mapsto \zeta \cdot z$. 
Under this action, the lemniscatic curve $E_\mathrm{i}$ has four fixed points:
\begin{align}
   & \om_0 := 0\,, \qquad \quad \,\,\,\,\, \zeta \cdot 0 = 0\,, \\
   & \om_1 := \frac{1+\mathrm{i}}{2}\,,\qquad \zeta \cdot \frac{1+\mathrm{i}}{2} = \frac{-1+\mathrm{i}}{2} \sim \frac{1+\mathrm{i}}{2}\,, \\
& \om_2 := \frac{\mathrm{i}}{2}\,, \qquad \zeta^2 \cdot \frac{\mathrm{i}}{2} = -\frac{\mathrm{i}}{2} \sim \frac{\mathrm{i}}{2}\, , \\
 \zeta^3 \cdot \om_2 =\, & \om_3 := \frac{1}{2}\,, \qquad \zeta^2 \cdot \frac{1}{2} = -\frac{1}{2} \sim \frac{1}{2}\, ,
\end{align}
where $\zeta \in \mathbb{Z}_4$ denotes the generating element $\zeta = \mathrm{i}$, and the equivalence relation $\sim$ identifies 
points that are obtained from one another by lattice shifts. 
From the short computation in the second column we conclude that $\om_0,\om_1$ are fixed points stabilized by the entire $\mathbb{Z}_4$, i.e.\ they are fixed points of order $4$, while $\om_2,\om_3$ are fixed points of order $2$ with a stabilizer subgroup $\mathbb{Z}_2 \subset 
\mathbb{Z}_4$. These last two fixed points are mapped to each other by the nontrivial 
$\mathbb{Z}_4$ transformation on $E_\mathrm{i}$. They thus give rise to the same point in the 
quotient $\mathcal{M}= E_\mathrm{i}/\Zs_4$. We conclude that $\mathcal{M}$ has three (singular) 
orbifold points which we denote as 
\begin{equation}
z_0:= \om_0,\,\,\, z_1:= \om_1,\,\,\, z_2:= \om_2 \sim \om_3. 
\end{equation}
At these points, the orbifold singularities are of orders $4,4,2$, respectively. The elliptic 
CMS model associates to each of these singular points $z_{\nu}$, $\nu = 0,1,2$, a family of multiplicities $m_{i,\nu}$, $i = 1,\dots,4$ such that
\begin{align} \label{eq:mconstraint} 
\sum_{i=1}^4 m_{i,\nu} := 6 \,, \quad \nu = 0,1,2\,, \qquad m_{1,2}+m_{2,2} =1\,, \qquad   m_{3,2}+m_{4,2} = 5\,.
\end{align}
Note that there is one relation among the four multiplicities we associate with the fixed 
points of order four, while there are three relations among the four multiplicities that are 
associated with the fixed point of order two. Given that there are three relations that 
constrain the four multiplicities $m_{i,2}$, it is often convenient to parametrize the 
solutions in terms of a single parameter $k$ which we define as  
\begin{equation}
    m_{1,2} := k+1\,.
    \label{def_k}
\end{equation}
The Hamiltonian $L_{\mathrm{EFMV}}$ of the lemniscatic CMS model has a relatively 
complicated dependence on the multiplicities. On the other hand, it may be uniquely 
characterized by a rather simple set of conditions: if we require that that the $\mathbb{Z}_4$-invariant operator $L_{\mathrm{EFMV}}(z,\ds_z)$ takes the normalized form
\begin{equation}
    L_{\mathrm{EFMV}} = \ds_z^4 + \oo(\ds_z^2)\,,
    \label{L_canonical_form}
\end{equation} 
then its dependence on the multiplicities is uniquely determined by the following 
set of conditions 
\begin{align}
L_{\mathrm{EFMV}}(z,\ds_z) \cdot (z-z_{0})^r =&  \prod_{i=1}^{4} (r-m_{i,0}) \, (z-z_{0})^{r-4} + \mathcal{O}( (z-z_{0})^{r})\,, \label{L_near_z0}\\
L_{\mathrm{EFMV}}(z,\ds_z) \cdot (z-z_{1})^r =&  \prod_{i=1}^{4} (r-m_{i,1}) \, (z-z_{1})^{r-4} + \mathcal{O}( (z-z_{1})^{r})\,, \label{L_near_z1} \\
L_{\mathrm{EFMV}}(z,\ds_z) \cdot (z-z_{2})^r =&  \prod_{i=1}^{4} (r-m_{i,2}) \, (z-z_{2})^{r-4}  \label{L_near_z2} \\
&+ \lambda (r-m_{1,2})(r-m_{2,2}) (z-z_2)^{r-2} +\mathcal{O}( (z-z_{2})^{r})\,, \nonumber
\end{align}
which we assume to hold for some constant $\la \in \Cs$, in the neighborhood of the singular 
points $z = z_{\nu}$. In order to write the Hamiltonian explicitly over the full orbifold $\mathcal{M}$, we first introduce the Weierstrass elliptic function 
\begin{equation}
\wp(z) := \frac{1}{z^2} + \sum_{w \in (\Zs\oplus i\Zs)\backslash \{0\}} \left( \frac{1}{(z-w)^2}-\frac{1}{w^2} \right),
\label{wp}
\end{equation}
which is double-periodic by construction, $\wp(z)=\wp(z+1)=\wp(z+\mathrm{i})$. Then the Hamiltonian of the 
lemniscatic $\Zs/4 \Zs$ CMS model is given by \cite[Eq.~4.3]{etingof2011107} 
\begin{equation}
   L_{\mathrm{EFMV}}(z,\ds_z) = \ds_z^4 + \sum_{p=0}^2 g^{(z)}_p (z) \ds_z^p\,,
   \label{LEFMV}
\end{equation}
where 
\begin{align}
g_2^{(z)} (z)  =&  \sum_{\nu=0}^3 a_{\nu} \, \wp(z-\om_{\nu})\,, \label{gFp2}  \\
g_1^{(z)} (z) =& \sum_{\nu=0}^3 b_{\nu} \, \wp'(z-\om_{\nu})\,,\label{gFp1}  \\
g_0^{(z)} (z) =& \sum_{\nu=0}^3 c_{\nu} \, \wp^2(z-\om_{\nu}) + \wp(\om_3)(a_0-a_1) k(k+1)  \left( \wp(z-\om_2)- \wp (z-\om_3) \right).  \label{gFp0}
\end{align}
The various coefficients $(a_{\nu},b_{\nu},c_{\nu})$ for $\nu=0,1,2$ are related to 
the multiplicities as \cite[Example~7.7]{etingof2011107}
\begin{align}
    & a_{\nu} :=-11+ \sum_{1\leq i < j \leq 4} m_{i,\nu} m_{j,\nu}\,, \label{level2} \\
&b_{\nu} := \frac{1}{2} \left(- a_{\nu} - 6 + \sum_{1\leq i < j < k \leq 4} m_{i,\nu} m_{j,\nu} m_{k,\nu}  \right),\label{level3} \\
& c_{\nu} := \prod_{i=1}^4 m_{i,\nu}\,,   \label{level4}
\end{align}
$(a_3,b_3,c_3) = (a_2,b_2,c_2)$ and the parameter $k$ is $k = m_{1,2}-1$. Note that in 
eq.\ \eqref{gFp2}-\eqref{gFp0} we are summing over the four fixed points $(\om_0,\om_1,\om_2,\om_3)$ of the elliptic curve $E_\mathrm{i}$, and that we need $(a_3,b_3,c_3) \equiv (a_2,b_2,c_2)$ to ensure the $\Zs_4$-symmetry of the Hamiltonian. 

\subsection{Construction of the map}
In order to recast our vertex operator $H$ in the form \eqref{L_canonical_form} of the lemniscatic CMS Hamiltonian, we need 
to find a change of variables from our cross-ratio $\mathcal{X}$ to a new 
variable $\phi$ and a `gauge transformation' $\Theta(\mathcal{X}(\phi))$ such that 
\begin{equation}
     \Theta^{-1} H \Theta = (\mathrm{const}) \ds_{\phi}^4 + \oo (\ds_{\phi}^2)\,. 
\end{equation}
Looking at the terms of order $\ds_{\mathcal{X}}^4$ and $\ds_{\mathcal{X}}^3$, we see that $\phi$ can be taken to solve 
the differential equation 
\begin{equation}
\frac{\dd \phi}{\dd \mathcal{X}} = \frac{(-)^{\frac{1}{4}}}{4} \mathcal{X}^{-\frac{3}{4}} (1-\mathcal{X})^{-\frac{3}{4}}, \qquad \phi(\mathcal{X}=0) = 0\,,
\label{dphidX}
\end{equation}
and $\Theta$ must be of the form
\begin{equation}
    \Theta = \Theta_0 \, \mathcal{X}^{\frac{l_1+l_2-2(\ell_1+\ell_2) + \Delta_3 + (1-d)/2}{4}} (1-\mathcal{X})^{\frac{l_1+l_2-2(\ell_1+\ell_2)-\Delta_3+(1+d)/2}{4}}\,,
\end{equation}
where $\Theta_0 \in \Cs\backslash \{0\}$ is an arbitrary multiplicative constant.
The solution to eq.\ \eqref{dphidX} is proportional to the incomplete Beta function, which has the known analytic expression (see \cite[Eq.~8.17.7]{dlmf})
\begin{equation}
\phi(\mathcal{X}) =\frac{(-)^{\frac{1}{4}}}{4} \int_0^{\mathcal{X}} \dd \mathcal{X}' \, \mathcal{X}'^{-\frac{3}{4}} (1-\mathcal{X}')^{-\frac{3}{4}} = (-\mathcal{X})^{\frac{1}{4}} \, \tensor[_2]{F}{_1}\left(\frac{1}{4}, \frac{3}{4}; \frac{5}{4}; \mathcal{X} \right).
\end{equation}
If we now apply the Pfaff transformation of the Gauss hypergeometric function, 
\begin{equation}
\tensor[_2]{F}{_1}(a,b;c;\mathcal{X}) = (1-\mathcal{X})^{-a} \tensor[_2]{F}{_1}\left(a,c-b;c; \frac{\mathcal{X}}{\mathcal{X}-1} \right),
\end{equation}
the above function can be expressed in terms of the inverse arc length function for the lemniscate curve (see \cite[Eq.~5]{lemniscate}),
\begin{equation}
\phi(\mathcal{X}) = \left( \frac{\mathcal{X}}{\mathcal{X}-1} \right)^{\frac{1}{4}} \tensor[_2]{F}{_1}\left(\frac{1}{4}, \frac{1}{2}; \frac{5}{4};  \frac{\mathcal{X}}{\mathcal{X}-1}\right) \equiv \, \mathrm{arcsinlemn} \, \left( \frac{\mathcal{X}}{\mathcal{X}-1} \right)^{\frac{1}{4}}.
\label{arcsinlemn}
\end{equation}
Using \cite[Eq.~21]{lemniscate}, the change of variables can be inverted to 
\begin{equation}
\sqrt{\frac{\mathcal{X}}{\mathcal{X}-1}} =  \frac{\mathrm{sd}^2 \left( \phi \sqrt{2}, \frac{1}{\sqrt{2}} \right)}{2}\,, 
\label{Xtosd}
\end{equation}
where $\mathrm{sd}(u,k)$ is one of the Jacobi elliptic functions. This can be equivalently expressed in terms of the Weierstrass function $\wp(z)$ defined in eq.\ \eqref{wp}, 
\begin{equation}
\mathcal{X} = \frac{\wp(\om_3)^2}{\wp(\om_3)^2- \wp(z)^2}\,, \qquad \phi = \wp(\om_3) \, z\,, 
\label{X_z_CoV}
\end{equation}

To re-express the corresponding operator in the form \eqref{LEFMV} and solve the parameters $k,a_{\nu},b_{\nu},c_{\nu}$ in eqs.\  \eqref{gFp2}, \eqref{gFp1}, \eqref{gFp0} for $\Dg_i,l_i,\ell_i,d$, we made a symbolic computation in \texttt{Mathematica}\footnote{The corresponding notebook can be found in the supplementary material of this publication.}. This symbolic computation specifically avoids the use of the special functions \texttt{JacobiSD} and \texttt{WeierstrassP} that appear in eqs.\ \eqref{Xtosd} and \eqref{X_z_CoV}, because \texttt{Mathematica} does not efficiently make use of the derivative and addition formulas of these special functions. Instead, we only use the \texttt{Hypergeometric2F1} in eq.\ \eqref{arcsinlemn} to compute the coefficients $g_0^{(z)},g_1^{(z)},g_2^{(z)}$ as functions of $\mathcal{X}$. More specifically, we start by determining the three functions
\begin{align*}
h_0^{(\phi)}(\phi(\mathcal{X})) &= H(\mathcal{X},\ds_{\mathcal{X}}) \cdot 1\,, \\
h_1^{(\phi)}(\phi(\mathcal{X})) &= H(\mathcal{X},\ds_{\mathcal{X}}) \cdot \phi(\mathcal{X}) - \phi(\mathcal{X}) h^{(\phi)}_0(\phi(\mathcal{X}))\,, \\
h_2^{(\phi)}(\phi(\mathcal{X})) &= H(\mathcal{X},\ds_{\mathcal{X}}) \cdot \frac{\phi(\mathcal{X})^2}{2} - \phi(\mathcal{X}) h_1^{(\phi)}(\mathcal{X}) - \frac{\phi(\mathcal{X})^2}{2} h^{(\phi)}_0(\mathcal{X})\,,
\end{align*}
that are related to the $g$-coefficients by
\begin{equation}
g_p^{(z)}(z(\mathcal{X})) =  4^3 \wp(\om_3)^{2- \frac{p}{2}}  \left( h_p^{(\phi)} \left( \phi \left( \mathcal{X} \right) \right) - \delta_{p,0}\,E_{\mathrm{EFMV}} \right), \qquad p=0,1,2\,,
\label{h_to_g}
\end{equation}
where $E_{\mathrm{EFMV}}^{(d;\Dg_i;l_i;\ell_i)}$ is the constant shift of the Hamiltonian given in appendix \ref{ssec:ECMS}. Following eq.\ \eqref{X_z_CoV}, it is easy to show that the $g_p^{(z)}$ computed from $H(\mathcal{X},\partial_{\mathcal{X}})$ are algebraic functions of $\wp (z)$. Similarly, we can express each term in eqs.\ \eqref{gFp2}, \eqref{gFp1}, \eqref{gFp0} as a rational function of $\wp (z)$ using the addition formulas 
\begin{align}
 &\wp\left( z -\om_1  \right) =- \frac{\wp(\om_3)^2}{\wp(z)}\,,\label{p1+i} \\
 &\wp(z-\om_2) = - \wp(\om_3) \frac{\wp(z)-\wp(\om_3)}{\wp(z)+\wp(\om_3)}\,,  \label{pi}\\
&\wp(z-\om_3) = \wp(\om_3) \frac{\wp(z)+\wp(\om_3)}{\wp(z)-\wp(\om_3)}\,, \label{p1} 
\end{align}
and the derivative formula
\begin{equation}
\wp'(z)^2 = 4 \wp(z) \left( \wp(z)^2-\wp(\om_3)^2 \right),
\label{dp}
\end{equation}
for the lemniscatic Weierstrass elliptic function. Following these identities, each of the
coefficient functions $g_p^{(z)}$ in the Hamiltonian is the product of $\wp^{\frac{p}{2}-2}\left(\wp(\om_3)^2-\wp^2\right)^{\frac{p}{2}-2}$ with a polynomial function in $\wp$. We can then identify each polynomial coefficient expressed as a function of $k,a_{\nu},b_{\nu},c_{\nu}$ with its expression in terms of $\Delta_i,l_i,\ell_i,d$ to obtain the map from spins and conformal dimensions to multiplicities.

\subsection{CMS multiplicities from weights and spins} 
\label{ssec:CMS_multiplicities}

In all cases, the multiplicity associated to $z_2$ (see eq.\ \eqref{def_k}) can be computed from the spin quantum numbers 
$l_1,l_2$ as 
\begin{equation} \label{eq:kMd} 
k = l_1 -l_2 - \frac{1}{2} \qquad \mathrm{or} \qquad k= l_2-l_1-\frac{1}{2}\,.
\end{equation}
Going from one choice to the other in eq.\ \eqref{eq:kMd} is equivalent to the change of parameters $k \rightarrow -(k+1)$, which leaves the CMS Hamiltonian invariant. For the MST$_2$-STT-scalar (type II) vertex in all $d\geq 3$ we have
\begin{align}
&m_{1,0} = 3\, \frac{5-d}{2}-(l_1+l_2)-\Dg_3-2\ell_1\,, \label{mMd10} \\
&m_{2,0} = \frac{d-1}{2}-(l_1+l_2)-\Dg_3+2\ell_1\,, \label{mMd20} \\
&m_{3,0} = \frac{d-1}{2}+(l_1+l_2) + \Dg_3+2(\Dg_1-\Dg_2)\,, \label{mMd30}\\
& m_{4,0} = \frac{d-1}{2}+(l_1+l_2) + \Dg_3-2(\Dg_1-\Dg_2)\,, \label{mMd40}
\end{align}
and 
\begin{align}
&m_{1,1} =-5\, \frac{d-3}{2}-(l_1+l_2)+\Dg_3-2\ell_1 \,, \label{mMd11} \\
&m_{2,1} = -\frac{d+1}{2}-(l_1+l_2)+\Dg_3+2\ell_1\,,  \label{mMd21}\\
&m_{3,1} = -\frac{d+1}{2}+(l_1+l_2)- \Dg_3+2(\Dg_1+\Dg_2)\,,  \label{mMd31}\\
&m_{4,1} = \frac{7d-1}{2}+(l_1+l_2)- \Dg_3-2(\Dg_1+\Dg_2) \,,\label{mMd41} 
\end{align}
which also contains the particular case of type I with two spinning fields in $d\geq 3$ simply by setting $\ell_1=0$. It is easy to 
observe from eqs.\ \eqref{mMd10}---\eqref{mMd41} that 
\begin{equation}
m_{i,\nu}(d; \Dg_i, l_i;\ell_1)  = m_{i,\nu}\left( d+\dg d; \Dg_i+ \frac{\dg d}{2};l_i- \frac{\dg d}{2};\ell_1-\frac{\dg d}{2} \right),
\end{equation}
or equivalently 
\begin{equation}
m_{i,\nu}(2\ag+3-2\ell_1; \ag-\ell_1+\frac{3}{2}+\mathrm{i}\cg_i, l_i;\nu_i+\ell_1;\ell_1)  = \mathrm{function}\left( \ag;\cg_i;\nu_i \right).
\end{equation}
This is a direct consequence of the observation made in eq.\ \eqref{dim_shift}. We conclude that the three weight and three 
spin labels along with the dimension $d$ of the MST$_2$-STT-scalar vertices do not exhaust the full 7-dimensional parameter space of 
the elliptic $\Zs_4$ CMS model. In fact, it is easy to see 
that the parameters are constrained by
\begin{equation}
m_{1,0}-m_{2,0} = m_{1,1}-m_{2,1}\,. \label{eq:addmconstraint}
\end{equation}
When specializing to $d=4$, this last constraint defines the restriction of the generic MST$_2$-MST$_2$-scalar (type III) vertex to the MST$_2$-STT-scalar case. We have determined that the full MST$_2$-MST$_2$-scalar vertex in $d=4$ yields the CMS multiplicities 
\begin{align}
& m_{1,0} = \frac{3}{2} -(l_1+l_2)-\Dg_3 -2(\ell_1-\ell_2)\,, \label{mMM410}\\
& m_{2,0} = \frac{3}{2}-(l_1+l_2)-\Dg_3 + 2(\ell_1-\ell_2)\,,  \label{mMM420}\\
& m_{3,0} = \frac{3}{2} + (l_1+l_2) +  \Dg_3 + 2 (\Dg_1-\Dg_2)\,,  \label{mMM430} \\
& m_{4,0} = \frac{3}{2} + (l_1+l_2)+ \Dg_3 -2(\Dg_1-\Dg_2)\,, \label{mMM440} \\
& m_{1,1} = - \frac{5}{2} - (l_1 + l_2) + \Dg_3 - 2(\ell_1+\ell_2)\,,  \label{mMM411}\\
& m_{2,1} = - \frac{5}{2} -( l_1 + l_2) + \Dg_3 + 2 (\ell_1+\ell_2)\,, \label{mMM421} \\
& m_{3,1} = - \frac{5}{2} + (l_1 +l_2) -  \Dg_3 + 2(\Dg_1+\Dg_2)\,,   \label{mMM431}\\
& m_{4,1} = \frac{27}{2} + (l_1+l_2) - \Dg_3 - 2(\Dg_1+\Dg_2)\,. \label{mMM441}
\end{align} 
Let us note that this set of multiplicities does not satisfy any additional 
constraints. This concludes our description of the precise relation between the 
vertex differential operators for single variable vertices and the 
lemniscatic CMS model of \cite{etingof2011107}. We would like 
to finish this section off with two additional comments.
\medskip 

\textit{Comments on algebraic integrability:} The CMS operator is said to be algebraically integrable if the multiplicities $m_{i,\nu}$ defined by eqs.\ \eqref{L_near_z0}, \eqref{L_near_z1}, \eqref{L_near_z2} are integers (see \cite[Corollary~2.4]{etingof2011algebraically}). In this case, according to \cite[Theorem~2.5]{etingof2011algebraically}, a generic eigenfunction of $L_{\mathrm{EFMV}}$ will take the form
\begin{equation}
    \psi_{\la}(z) = e^{\bg z} \prod_{i=1}^4 \frac{\theta(z-\ag_i)}{\theta(z-\bg_i)},
    \label{eigenfunction}
\end{equation}
where $\theta(z)$ is the first Jacobi theta-function of the lemniscatic elliptic curve, and
$\bg,\bg_1,\ag_1,\dots,\bg_4$, $\ag_4$ are certain parameters that can be solved in terms for the multiplicities and eigenvalue $\lambda$ by writing the eigenvalue equation $L_{\mathrm{EFMV}} \psi_{\lambda} = \la \psi_{\lambda}$ for \eqref{eigenfunction} near the singular points $z=z_0,z_1,z_2$. We have determined above that all multiplicities are linear combinations 
of the quantum numbers $(\Dg_i;l_i;\ell_i)$ and the dimension $d$, with coefficients in
$\frac{1}{2}\Zs$. Therefore, depending on whether $d$ is odd or even, the vertex operator is algebraically integrable when the quantum numbers $[-\Dg_i;l_i;\ell_i]$ that define the 
representation at each point are either integers or half-integers. This setup is equivalent to placing unitary irreducible representations of the compact real form $\mathrm{SO}_{\Rs}(d+2)$ at 
each point (or the double cover thereof). It would be interesting to explore the generalization of this result to non-integer conformal weights. 
\smallskip 

\noindent 
\textit{CMS multiplicities for all vertex systems:} As a final comment we want to rewrite the
relations between the CMS multiplicities and the weight and spin quantum numbers in terms of 
the parameters that appeared in our discussion of the generalized Weyl algebra, see previous 
section. Recall the parameters of the generalized Weyl algebra,
\begin{equation}
    \ag := \frac{d-3}{2}+\ell_1+\ell_2\,, \qquad \nu_{1} = l_1-\ell_1-\ell_2\,, \qquad \nu_2 = l_2-\ell_1-\ell_2\,.
\end{equation}
To determine a universal formula for the CMS multiplicities of all 1-dimensional vertex 
systems, we use the four extra parameters
\begin{equation}
    \bg := \ell_1-\ell_2+\frac{d-5}{2}\,, \qquad \cg_i := -\mathrm{i}\left(\Dg_i- \frac{d}{2}\right)\,. 
\end{equation}
The parameters $\gamma_i$ had appeared in our construction of the Hamiltonian already, see
eq.\ \eqref{new_params}. Only the parameter $\beta$ is new. Of course, the map from spin quantum 
numbers $(l_i;\ell_i)$ to $\ag,\bg,\nu_1,\nu_2$ can be inverted as
\begin{align}
   & l_1 = \nu_1+\ag+\frac{3-d}{2}\,, \qquad  l_2 = \nu_2+\ag+\frac{3-d}{2}\,, \\
    &\ell_1 = \frac{\ag+\bg+4-d}{2}\,,\qquad \ell_2 = \frac{\ag-\bg-1}{2}\,.
\end{align}
If we insert these formulas into the expressions for multiplicities we listed above, 
these become completely universal to all 1-dimensional vertex systems, i.e.\ they no longer depend on type of the 
vertex (I, II, or III). Explicitly one finds 
\begin{equation}
    k = \nu_1-\nu_2-\frac{1}{2}\,, \qquad \mathrm{or} \qquad k= \nu_2-\nu_1-\frac{1}{2}\,, 
\end{equation}
and
\begin{align}
    m_{1,0} &= -\frac{1}{2}-(\nu_1+\nu_2)-\mathrm{i}\cg_3 -2\ag -2\bg\,,  \\
    m_{2,0}&= \frac{3}{2} +(\nu_1+\nu_2) -\mathrm{i}\cg_3-2\ag+2\bg\,, \\
    m_{3,0} &= \frac{5}{2} +(\nu_1+\nu_2) + \mathrm{i} \cg_3 +2\ag+ 2\mathrm{i}(\cg_1-\cg_2)\,, \\
    m_{4,0}&= \frac{5}{2} +(\nu_1+\nu_2) + \mathrm{i}\cg_3+2\ag -2\mathrm{i}(\cg_1-\cg_2)\,, \\
    m_{1,1} &= \frac{3}{2} -(\nu_1+\nu_2) + \mathrm{i} \cg_3-4\ag\,, \\
    m_{2,1} &= -\frac{1}{2} -(\nu_1+\nu_2) + \mathrm{i}\cg_3\,, \\
    m_{3,1} &= \frac{5}{2} +(\nu_1+\nu_2) -\mathrm{i}\cg_3 + 2\ag + 2\mathrm{i}(\cg_1+\cg_2)\,, \\
    m_{4,1} &= \frac{5}{2} +(\nu_1+\nu_2) -\mathrm{i}\cg_3+2\ag -2\mathrm{i}(\cg_1+\cg_2)\,.
\end{align}
In particular, the MST$_2$-STT-scalar (type II) case in all $d\geq 3$ is obtained by 
imposing the additional relation $\bg = \ag - 1$, equivalent to $\ell_2=0$.

\section{Conclusion and Outlook}

In this work, we have constructed the fourth order differential equation that 
characterizes all vertices of the three types listed in eq.\ \eqref{eq:list}, each of which appear in OPE diagrams of scalar $N$-point functions in $d$-dimensional conformal 
field theory. These three cases cover all vertices that
contribute one degree of freedom, and that are therefore fully characterized 
by one single differential equation. This equation has been worked out in 
section~\ref{sect:ConstructionOperator}. It was then reformulated very elegantly in section~\ref{sect:GeneralizedWeyl} as an 
eigenvalue equation for a special Hamiltonian that resides in a 
generalized Weyl algebra. Finally, we mapped the operator to the Hamiltonian 
of the lemniscatic elliptic CMS model of \cite{etingof2011107} in section~\ref{sect:MappingElliptic}.  

As we have pointed out before, comb channel OPE diagrams in $d=3$ and $d=4$ 
dimensions contain vertices with at most one degree of freedom, so the 
theory we developed here exhausts all those vertices. Obviously, it would be 
very interesting to extend the analysis we carried out here to vertices with 
a higher number of degrees of freedom, such as e.g. the central vertex of 
the snowflake channel for a $N=6$-point function in $d=3$, or the (comb 
channel) vertices of type $(2,2,0)$ in $d > 4$, to name just two examples. 
It is certainly possible to compute these operators in the same way as we 
did in section~\ref{sect:ConstructionOperator}. However, in these more general cases, one will have to work with multiple independent cross ratios and higher order Gaudin Hamiltonians. 
We believe that the elegant reformulation in terms of generalized Weyl 
algebras outlined here for the single variable case can be extended 
to the multi-variable context. Whether the associated Hamiltonians can be 
mapped to a CMS type system remains an interesting question for future 
research. 

As we explained in the introduction, the vertex operators studied here 
appear within the full set of differential equations that characterize a 
scalar $N$-point function by taking OPE limits on all the three branches attached to a given vertex of the OPE diagram. The limiting 
procedure eliminates all but one of the cross ratios. Within the original 
multi-variable system, the vertex differential operators act on all of the 
cross ratios, whereas the single variable operators we have constructed here 
emerge only after taking the OPE limit. An example of the full  
vertex operator for the central vertex of a scalar 5-point function was 
constructed in \cite{Buric:2021ywo}, and we have checked that the 
single variable operator described above for vertices of type I indeed arises 
in the OPE limits. For the other two cases in our list \eqref{eq:list}, 
the simplest realization as OPE limits involves scalar $N$-point 
functions with $N=6$, $d \geq 4$ and $N=7$, $d=4$, respectively. Even 
though the full vertex operators in these cases have not been constructed, 
we were able to obtain explicit expressions in the OPE limit. 

In addition to vertices, OPE diagrams contain another important element, 
namely the links that represent the exchange of intermediate primaries. 
One can prepare the differential operators at an individual link much 
in the same way as for the vertices by taking OPE limits at all branches 
surrounding said link. The resulting differential operators are expected to 
include the Casimir operators for spinning $4$-point blocks. We have verified this 
in a few examples and shall present a more general theory in a forthcoming paper. 

While the solution theory for Casimir equations of spinning $4$-point functions 
is sufficiently well understood, at least for applications to the conformal 
bootstrap, see e.g.\  \cite{Costa:2011dw,Iliesiu:2015akf,Echeverri:2016dun,Karateev:2017jgd,
Schomerus:2016epl,Schomerus:2017eny}, very little is known so far about the vertex 
systems. For the single variable vertices we analyzed in this paper we will address 
the solution theory in future research. In this context, it may be interesting to note 
that crystallographic elliptic CMS models have made a recent equally 
unexpected appearance in the context of Seiberg-Witten theory for 
$\mathcal{N}=2$ supersymmetric gauge theory \cite{Argyres:2021iws}. 
For the lemniscatic model we described above, a gauge theory 
realization has also been announced. 

Once the solution theory for the `local' vertex systems is 
well under control, one can also hope to build up `global' solutions 
for the full problem of scalar $N$-point functions. We have started 
to explore this in the light-cone limit, where the link (Casimir) operators  
simplify drastically, with very promising first results. In fact, 
it seems likely that multi-point blocks in the light-cone limit can 
be accessed with little additional effort. Such blocks have 
featured recently in \cite{Vieira:2020xfx}. Outside the light-cone limit, 
the powerful tools of integrable systems, and in particular the method of 
separation of variables \cite{Sklyanin:1987ih,Sklyanin:1995bm}, may 
become relevant. Applications of this method to higher dimensional 
conformal field theories would require its extension to higher rank
algebras: this was initiated in~\cite{Sklyanin:1992sm,Smirnov:2001},
and has been the subject of many further developments in the past few
years~\cite{Gromov:2016itr,Maillet:2018bim,Ryan:2018fyo}
(see~\cite{Ryan:2020rfk,Maillet:2020ykb,Derkachov:2020zvv,Gromov:2020fwh,
Cavaglia:2021mft} 
for some of the most recent works and references therein). While there are 
still some steps left to carry out before our new approach can fully 
materialize into computational algorithms for multi-point conformal 
blocks, a clear path is emerging. 
\bigskip 

\noindent 
\textbf{Acknowledgements:} We are grateful to Gleb Arutyunov, Luke Corcoran, 
Pavel Etingof, Aleix Gimenez-Grau, Mikhail Isachenkov, Apratim Kaviraj, 
Madalena Lemos, Pedro Liendo, Junchen Rong, Joerg Teschner and Benoît 
Vicedo for useful discussions. This project received funding from the 
German Research Foundation DFG under Germany’s Excellence Strategy -- EXC 
2121 Quantum Universe -- 390833306 and from the European Union’s Horizon 
2020 research and innovation programme under the MSC grant agreement 
No.764850 “SAGEX”.

\appendix

\section{Map from \texorpdfstring{$\mathfrak{so}_{\Cs}(6)$}{so(6;C)} embedding space to \texorpdfstring{$\mathfrak{sl}_{\Cs}(4)$}{sl(4;C)} twistors}
\label{app:twistors_from_emb_space}
We use indices $A,B,C=0,\dots,5$ to label an orthonormal basis in the fundamental representation of $\mathfrak{so}_{\Cs}(6)$, and $a,b,c=1,2,3,4$ to label a basis in the fundamental representation of $\mathfrak{sl}_{\Cs}(4)$. We saw that irreducible representations of $\mathfrak{so}_{\Cs}(6)$ are sections of a line bundle over the space of isotropic flags in $\Cs^6$
\begin{equation}
    \mathrm{Span}(X) \subset  \mathrm{Span}(X,Z) \subset  \mathrm{Span}(X,Z,W) = \mathrm{Span}(X,Z,W)^{\perp} \subset \mathrm{Span}(X,Z)^{\perp} \subset \mathrm{Span}(X)^{\perp} \subset \Cs^6,
\end{equation}
where $\mathbb{V}^{\perp}$ is the orthogonal complement of the vector subspace $\mathbb{V} \subset \Cs^{6}$  with respect to the 6-dimensional metric  $(\eta_{AB})$. These sections are equivalent to certain functions $F(X,Z,W)$ of three vectors in $\Cs^6$ that are null and mutually orthogonal with respect to the Minkowski metric,
\begin{equation}
    X^2 = Z^2 = W^2 = X\cdot Z = X\cdot W = Z\cdot W = 0\,.
    \label{null_orthogonal}
\end{equation}
Said functions must be homogeneous of fixed multi-degree, and invariant under the gauge transformations that preserve the isotropic flag,
\begin{equation}
F(X, Z+\bg_{10} X,W+\bg_{20} X + \bg_{21} Z) = F(X,Z,W)\,. 
\end{equation}
Depending on the choice of real form of $\mathfrak{so}_{\Cs}(6)$ (or equivalently the signature of $(\eta_{AB})$), as well as the choice of representation, one must either apply reality conditions on some of the $X,Z,W$ or impose that $F$ is holomorphic in some of the $X,Z,W$ variables. For the reflection positive and integer spin representations of CFT$_4$, $X$ is real and $F$ is holomorphic in $Z,W \in \Cs^6$. In this case, the space of vectors $(X,Z,W) \in \Rs^6 \times  (\Cs^6)^2$ satisfying \eqref{null_orthogonal} is informally known as embedding space. The gauge constraints can be explicitly solved by a change of variables 
\begin{equation}
    C^{(0)}_A := X_A, \qquad C^{(1)}_{AB} := (X\wedge Z)_{AB}, \qquad C^{(2)}_{ABC} := (X\wedge Z \wedge W)_{ABC}, 
    \label{C_tensors}
\end{equation}
such that $ F(X,Z,W) =  F'(C^{(0)},C^{(1)},C^{(2)})$ for some function $F'$. 

Similarly, irreducible representations of $\mathfrak{sl}_{\Cs}(4)$ are sections of a line bundle over the space of flags in $\Cs^4$,
\begin{equation}
    \mathrm{Span}(Y_1) \subset  \mathrm{Span}(Y_1,Y_2) \subset  \mathrm{Span}(Y_1,Y_2,Y_3) \subset \Cs^4\,.
\end{equation}
This is equivalent to functions $\Psi(Y_1,Y_2,Y_3)$ of three (arbitrary) vectors in $\Cs^4$. Said functions must also be homogeneous of fixed multi-degree, and invariant under the gauge transformations that preserve the flag,
\begin{equation}
\Psi(Y_1, Y_2+c_{21} Y_1,W+c_{31} Y_1 + c_{32} Y_2) = \Psi(Y_1,Y_2,Y_3)\,. 
\end{equation}
Once again, the gauge constraints can be explicitly solved by the change of variables to gauge-invariant tensors 
\begin{equation}
S_a := Y_{1a}, \qquad X_{ab} := (Y_1\wedge Y_2)_{ab}, \qquad  \bar{S}_{abc} := (Y_1 \wedge Y_2\wedge Y_3)_{abc}\,,
\label{flag_to_twistor}
\end{equation}
such that $\Psi(Y_1,Y_2,Y_3) = \Psi'(S,X,\bar{S})$ for some function $\Psi'$. The gauge invariant, anti-symmetric, and $\mathfrak{su}(2,2)$-covariant tensors $(S,X,\bar{S})$ are known as \emph{twistor} variables in the physics literature. Similarly to the previous case, reality conditions on $Y_1,Y_2,Y_3$ or holomorphicity conditions on $\Psi$ are required to realize irreducible representations of real forms of $\mathfrak{sl}_{\Cs}(4)$. In particular, the reflection positive and half-integer spin representations of CFT$_4$ are realized by imposing that $X_{ab}$ is real and $\Psi'$ is a holomorphic function of $S$ and $\bar{S}$. This follows from the fact that $\mathrm{SU}(2,2)$ is the double cover of the Lorentzian conformal group $\mathrm{SO}(2,4)$.

More generally, as $\mathfrak{so}_{\Cs}(6) \cong \mathfrak{sl}_{\Cs}(4)$, representations of the latter can be mapped to representations of the former. As a result, there exists a map from homogeneous, gauge-invariant functions on $\mathfrak{so}_{\Cs}(6)$ embedding space to homogeneous functions of the twistor variables \eqref{flag_to_twistor}. This translates into a map from the gauge-invariant tensors \eqref{C_tensors} in $\Cs^6$ to the gauge-invariant tensors \eqref{flag_to_twistor} in $\Cs^4$. To determine explicit expressions, we make use of the chiral $\Gamma$-matrices $\Gamma^A_{ab}$ defined for example in \cite[Appendix~B]{SimmonsDuffin:2012uy}. If $\tensor{M}{^A_B}\in \mathfrak{so}_{\Cs}(6)$, then there exists $\tensor{L}{_a^b} \in \mathfrak{sl}_{\Cs}(4)$ such that 
\begin{equation}
    \tensor{M}{^A_B} \Gamma^B_{ab} = \tensor{L}{_a^c}\, \Gamma^A_{cb} + \tensor{L}{_b^d}\, \Gamma^A_{ad}\,. 
\end{equation}
These $\Gamma$-matrices are anti-symmetric, such that we can define their duals with respect to the $4$-dimensional $\epsilon$-tensor,
\begin{equation}
\bar{\Gamma}^{Aab} := \frac{1}{2} \epsilon^{abcd} \Gamma^A_{ab}\,. 
\end{equation}
The fundamental identities of the $\Gamma$-matrices can also be found in \cite[Appendix~B]{SimmonsDuffin:2012uy}. The Clifford relations are
\begin{equation}
    \bar{\Gamma}^{A ab} \Gamma^{B}_{bc} +  \bar{\Gamma}^{B ab} \Gamma^{A}_{bc}  = -2 \eta^{AB} \dg_c^a\,,
\end{equation}
while the contraction identity is
\begin{equation}
\eta_{AB} \Gamma^A_{ab}\Gamma^B_{cd}=2 \epsilon_{abcd}\,.
\end{equation}
The map from gauge-invariant tensors in $\mathfrak{so}(1,5)$ embedding space to twistor variables is then given by
\begin{align}
& C^{(0)}_A = X_A = \frac{1}{4} X_{ab} \bar{\Gamma}_A^{ab}\,, \\
& C^{(1)}_{AB} = (X\wedge Z)_{AB} = \frac{1}{\sqrt{2}} \bar{S}^a \Gamma_{Aab} \bar{\Gamma}_B^{bc} S_c\,, \\
& C^{(2)}_{ABC} = (X\wedge Z \wedge W)_{ABC} =\frac{1}{2 \sqrt{2}}  S_a \bar{\Gamma}_A^{ab} \Gamma_{Bbc} \bar{\Gamma}_B^{cd} S_d\,,  \\
& \bar{C}^{(2)}_{ABC} =(X\wedge Z \wedge \bar{W})_{ABC} =\frac{1}{2 \sqrt{2}}  \bar{S}^a \Gamma_{Aab} \bar{\Gamma}_{B}^{bc} \Gamma_{Ccd} \bar{S}^d\,. 
\end{align}
Here, we defined the dual tensors
\begin{equation}
\bar{S}^a := \frac{1}{3!}\epsilon^{abcd} \bar{S}_{bcd}, \qquad \bar{X}^{ab} := \frac{1}{2} \epsilon^{abcd} X_{cd}. 
\end{equation}
We can now summarize our nomenclature for various spinning representations of maximal spin depth in $d=4$:
\begin{itemize}
    \item we call a self-dual (respectively anti-self-dual) representation any function on embedding space that is a homogeneous polynomial of order $\ell \in \Zs_+$ in $W$ (respectively a polynomial of order $-\ell \in \Zs_+$ in $\overline{W}$). In twistor space, we see that these representations correspond to homogeneous polynomials of order $j \in 2 \Zs_+$ in $S$ and $\bar{\jmath} \in 2 \Zs_+$ with $j>\bar{\jmath}$ (respectively $j<\bar{\jmath}$). 
    \item We call a chiral (respectively anti-chiral) representation any function on twistor space that is a polynomial of order $j \in \Zs_+$ in $S$ (respectively $\bar{\jmath} \in \Zs_+$ in $\bar{S}$) with $\bar{\jmath} = 0$ (respectively $j=0$). In cases where $j$ (respectively $\bar{\jmath}$) is an even integer, these coincide with the self-dual (respectively anti-self-dual) parts of $\mathfrak{so}(4)$ representations with rectangular Young tableaux of height $h_1=\dots=h_{l}=2$ and length $l=\ell = j/2$ (respectively $l=-\ell=\bar{\jmath}/2$). 
\end{itemize}

With this map, it is easy to directly relate the MST$_2$-MST$_2$-scalar tensor 
structures in $\mathfrak{so}(1,5)$ embedding space with those of twistor space:
\begin{align}
X_i \cdot X_j & = \frac{1}{4} \bar{X}_i^{ab} X_{jab} = - \frac{1}{4} \mathrm{tr}\, \bar{X}_i X_j, \\
 X_{jk} V_{i,jk} &= \bar{S}_i X_j \bar{X}_k S_i \\
 (\mK_{\bar{\imath}j})^2 &= (\bar{S}_i S_j)^2 \label{k_to_I} \\
 \mho_{ij,k}^2 &= (S_i \bar{X}_k S_j)^2 \label{mho_to_K} \\
\bar{\mho}_{ij,k}^2 &= (\bar{S}_i X_k \bar{S}_j)^2. \label{mhob_to_Kb}
\end{align}
It is important to note that the squared tensor
structures on the left hand side of \eqref{k_to_I}, \eqref{mho_to_K}, \eqref{mhob_to_Kb} are also perfect
squares of $\mathfrak{so}_{\Cs}(6)$ embedding space variables. This means that we can compute 3-point functions of any
half-integer spin fields in our formalism.

\section{Comments on scalar products and unitarity}
\label{app:scalar_product}
\subsection{Integral formula for the \texorpdfstring{$\mathrm{SO}(d+2)$}{SO(d+2)}-invariant scalar product}
\label{ssec:BargmannTodorov}
Consider an arbitrary finite-dimensional and irreducible representation of $\mathfrak{so}(d+2)$, labeled by a Young tableau with row lengths $-\Dg \geq l_1 \geq \dots \geq l_L$, $L:= \mathrm{rank}(\mathfrak{so}(d))$. The latter can be represented as a tensor on $\Cs^{d+2}$,
\begin{equation}
    F_{A^{(0)}_1\dots A^{(0)}_{-\Dg}A^{(1)}_1\dots A^{(1)}_{l_1} \dots A^{(L)}_1 \dots A^{(L)}_{l_L}} := F_{\{ A\}_{L+1}}, \,\,\, A_{i}^{(j)} = 1,\dots, d+2\,,
\end{equation}
satisfying the same (anti)-symmetry and tracelessness conditions as in subsection \ref{subsec:MSTinembspace}. By contracting the first family of indices with a null polarization vector $X \in \Cs^{d+2}, X^2=0$,
\begin{equation}
    F_{\Dg,\{A\}_L}(X) :=   X^{{A^{(0)}_1}} \dots X^{A^{(0)}_{-\Dg}} F_{A^{(0)}_1\dots A^{(0)}_{-\Dg}A^{(1)}_1\dots A^{(1)}_{l_1} \dots A^{(L)}_1 \dots A^{(L)}_{l_L}}\,,
\end{equation}
these $[-\Dg,l_1,\dots,l_L]$ tensors are equivalent to transverse $[l_1,\dots,l_L]$ tensor-valued homogeneous functions on the complex light-cone in $\Cs^{d+2}$: 
\begin{equation}
    F_{\Dg,\{A\}_L}(\la_0 X) = \la_0^{-\Dg} F_{\Dg,\{A\}_L}(X)\,, \qquad X^{A^{(j)}_i} F_{\Dg,\{A\}_L}(X) = 0\,,
\end{equation}
with the gauge equivalence relation 
\begin{equation}
 F_{\Dg,\{A\}_L}(X) \sim  F_{\Dg,\{A\}_L}(X) + C_{\{A\}_L\backslash \{A_i^{(j)}\}} (X)\, X_{A_i^{(j)}}\,, 
\end{equation}
valid for any transverse $[l_1,\dots,l_j-1,\dots,l_L]$- tensor valued function $X\mapsto \vct{C}(X)$ of homogeneity $-\Delta-1$.
The $\mathrm{SO}(d+2)$-invariant scalar product of two irreducible tensors is given by the contraction of indices,
\begin{equation}
    \langle F,G \rangle = \bar{F}_{\{A\}_{L+1}} \dg^{\{AA'\}_{L+1}} G_{\{A'\}_{L+1}}\,, \qquad \dg^{\{BB'\}_{L+1}} := \prod_{j=0}^{L} \prod_{i=1}^{l_j} \dg^{B_i^{(j)} {B'}_i^{(j)}} \,.
\end{equation}
A result of Bargmann and Todorov (c.f. \cite{Bargmann_1977}, Proposition 4.1) recasts this scalar product as an integral over $\Cs^{d+2}$:
\begin{equation}
      \langle F_{\Dg}, G_{\Dg} \rangle = \int_{\Cs^{d+2}} \dd^{d+2} X \dg(X^2)   \dd^{d+2} \bar{X} \dg(\bar{X}^2)  \rho_{d+2}(\bar{X}\cdot X) \overline{F_{\Dg}^{\{A\}_L}}(\bar{X}) \, W_{\{AB\}_L}^{\bar{X},X} G_{\Dg}^{\{B\}_L}(X)\,. 
      \label{BT_integral}
\end{equation}
The measure is given by 
\begin{equation}
\rho_{d+2}(t):= \frac{8}{\pi^{d+1}\Gamma(d/2)} t^{-(d-2)/4} K_{(d-2)/2}(2\sqrt{t})\,,
\end{equation}
where $K_{\epsilon}(s)$ is the modified Bessel function of the second kind, while to ensure gauge invariance, the indices are contracted with the tensor
\begin{equation}
    W_{\{AB\}_L}^{\bar{X},X} := \prod_{j=0}^L \prod_{i=1}^{l_j} W_{A_i^{(j)} B_i^{(j)}}^{\bar{X},X}\,, \qquad W_{AB}^{\bar{X},X} := \dg_{AB}-\frac{X_A \bar{X}_B}{\bar{X}\cdot X}\,, 
\end{equation}
satisfying
\begin{equation}
    \bar{X}^A W_{AB}^{\bar{X},X} = 0 = W_{AB}^{\bar{X},X} X^B\,. 
\end{equation}
While there are other integral formulas for $\mathrm{SO}(d+2)$-invariant scalar products (e.g. in \cite{SimmonsDuffin:2012uy}), the Bargmann-Todorov scalar product ensures that the adjoint of the $X_A$ operator is the so-called Thomas-Todorov operator,
\begin{equation}
X_A^{\dagger} = D_A = \left(X^B \frac{\ds}{\ds X^B} + \frac{d}{2}\right) \frac{\ds}{\ds X^A} - \frac{1}{2} X_A \frac{\ds^2}{\ds X^B \ds X_B}\,.   
\end{equation}

We would now like to reduce the integral formula \eqref{BT_integral} to the Poincaré patch, 
\begin{equation}
X^A(X^+,x) = X^+ \chi^A(x)\,, \qquad X^+ := X^{d+1}+\mathrm{i} X^{d+2}\,, \qquad \chi(x) := \left(\frac{x^2-1}{2}, \mathrm{i}\frac{x^2+1}{2}, -x^a \right)\,. 
\end{equation}
First, the measure is modified as
\begin{equation}
\dd^{d+2} X = \frac{\mathrm{i}}{2} \dd(X^2) \dd X^+(X^+)^{d-1} \dd^d x\,, \qquad \dd^{d+2} \bar{X} = -\frac{\mathrm{i}}{2} \dd(\bar{X}^2) \dd \bar{X}^+(\bar{X}^+)^{d-1} \frac{\dd^d \tilde{x}}{\tilde{x}^{2d}}\,,
\end{equation}
where $\tilde{x}^a := -\bar{x}^a/\bar{x}^2$ is the generalized antipode of $x^a$ on $\Cs^d \cup \{\infty\}$. Next, the scalar product in the measure is expressible as
\begin{equation}
     \bar{X} \cdot X = \frac{1}{2} \abs{X^+}^2   \frac{(x-\tilde{x})^2}{\tilde{x}^2}\,. 
\end{equation}
Finally, the fields can be rewritten as
\begin{equation}
G_{\Dg}^{\{B\}_L}(X) = (X^+)^{-\Dg} g_{\Dg}^{\{b\}_L} (x) e_{\{b\}_L}^{\{B\}_L}(x) \,, \qquad e_{b}^B(x) := \frac{\ds \chi^A}{\ds x^{b}} (x)\,, 
\end{equation}
and 
\begin{equation}
\overline{F_{\Dg}^{\{A\}_L}}(\bar{X}) = (\bar{X}^+)^{-\Dg} \overline{f_{\Dg}^{\{a\}_L}} \left( \frac{\tilde{x}}{\tilde{x}^2}\right) e_{\{a\}_L}^{\{A\}_L}\left( \frac{\tilde{x}}{\tilde{x}^2}\right) \sim  (\bar{X}^+)^{-\Dg}\overline{f_{\Dg}^{\{c\}_L}} \left( \frac{\tilde{x}}{\tilde{x}^2}\right) I_{\{c\}_L}^{\{a\}_L}(\tilde{x})\, e_{\{a\}_L}^{\{A\}_L}(\tilde{x})\,,
\end{equation}
where "$\sim$" denotes gauge equivalence, $I^a_b(x) := \dg^a_b - 2 x^a x_b/x^2$, and the $\Cs^{d+2}$ contraction of the $e$-tensors with the $W$-tensors  yields
\begin{equation}
e_a^A(\tilde{x}) W_{AB}^{\bar{X},X} e_b^B(x) = I_{ab}(\tilde{x}-x)\,.     
\end{equation}
Putting all of these changes of variables together yields
\begin{align*}
\langle F_{\Dg}, G_{\Dg} \rangle = \frac{1}{4} \int_{\Cs \times \Cs^d} & \dd^2 X^+ \abs{X^+}^{2(d-\Dg-1)} \dd^d x \frac{\dd^d \tilde{x}}{\tilde{x}^{2d}} \, \rho_{d+2}\left( \frac{\abs{X^+}^2 (x-\tilde{x})^2}{2 \tilde{x}^2} \right) \\
& \overline{f_{\Dg}^{\{c\}_L}} \left( \frac{\tilde{x}}{\tilde{x}^2}\right) I_{\{c\}_L}^{\{a\}_L}(\tilde{x})\, I_{\{ab\}_L}(x-\tilde{x}) g_{\Dg}^{\{b\}_L} (x)\,.
\end{align*}
The integral over $X^+$ is factorized by the change of variables 
\begin{equation}
t := \bar{X}\cdot X = \frac{\abs{X^+}^2 (x-\tilde{x})^2}{2 \tilde{x}^2}\,, \qquad \dd(\abs{X^+}^2) = \frac{2 \tilde{x}^2}{(x-\tilde{x})^2} \dd t\,,
\end{equation}
and the adjoint naturally appears as
\begin{equation}
(f_{\Dg}^{\dagger})^{\{a\}_L} (x) := x^{-2\Dg} I_{\{b\}_L}^{\{a\}_L}(x) f_{\Dg}^{\{b\}_L}\left(\frac{x}{x^2}\right), 
\end{equation}
such that 
\begin{align*}
\langle F_{\Dg}, G_{\Dg} \rangle = \left\{ \frac{2^{d-\Dg} \pi }{4} \int_0^{\infty} \dd t\, t^{d-\Dg-1} \rho_{d+2}(t) \right\} \int_{\Cs^d} \frac{\dd^d x\, \dd^d \tilde{x}}{(x-\tilde{x})^{2(d-\Dg)}} (f_{\Dg}^{\dagger})^{\{a\}_L} (\tilde{x}) I_{\{ab\}_L}(\tilde{x}-x) g_{\Dg}^{\{b\}_L} (x)\,.
\end{align*}
The $t$-integral is computed using \cite[Eq.~10.43.19]{dlmf}, 
\begin{equation}
\int_0^{\infty} \dd s \, s^{\mu-1} K_{\nu}(s) = 2^{\mu-1} \Gamma\left(\frac{\mu-\nu}{2} \right) \Gamma\left(\frac{\mu+\nu}{2} \right), \,\,\, \forall \, \mathrm{Re}(\mu) > \abs{\mathrm{Re}(\nu)}\,.
\end{equation}
and the final result can be concisely written as
\begin{equation}
\langle F_{\Dg}, G_{\Dg} \rangle = \mathcal{N}^{(d+2)}_{-\Dg} \int_{\Cs^d} \dd^d x \, \dd^d\tilde{x}\, (f_{\Dg}^{\dagger})^{\{a\}_L} (\tilde{x}) \frac{I_{\{ab \}_L}(\tilde{x}-x)}{(\tilde{x}-x)^{2(d-\Dg)}} g_{\Dg}^{\{b\}_L}(x)\,, 
\label{SP_shadowDelta}
\end{equation}
where the normalization is
\begin{equation}
\,\,\, \mathcal{N}^{(d+2)}_{-\Dg}:=\frac{2^{d-\Dg}}{\pi^d} \frac{\Gamma(d+2-2\Dg)\Gamma(d+1-\Dg)}{\Gamma(d/2)}\,.
\end{equation}

\subsection{Generalizations to different real forms of \texorpdfstring{$\mathfrak{so}_{\Cs}(d+2)$}{so(d+2;C)}}
\label{ssec:real_forms}
In contrast with standard tensor contraction, formula \eqref{SP_shadowDelta} has the advantage of generalizing to infinite-dimensional unitary representations of $\mathrm{SO}(1,d+1)$ and $\mathrm{SO}(2,d)$ by changing the domain of integration of $x$ and re-applying the same algebraic manipulations of section \ref{ssec:BargmannTodorov}. We distinguish three cases:

\paragraph{$(2,d)$:} if $X \mapsto F_{\Dg,\{A\}_L}(X)$, $\Dg \in \Rs_{\geq 0}$ denotes a reflection-positive representation of $\mathrm{SO}(1,d+1)$ (i.e. a unitary representation of $\mathrm{SO}(2,d)$), then we replace \eqref{SP_shadowDelta} by an integral over two independent variables $x^a,\tilde{x}^a \in \Rs^{1,d-1}$ in $d$-dimensional Minkowski space. In this context, the scalar product is conventionally recast as 
   \begin{equation}
\langle F_{\Dg}, G_{\Dg} \rangle = \mathcal{N}^{(d+2)}_{-\Dg} \int_{\Rs^{1,d-1}} \dd^d x \, \mathbf{S}_{-\Dg}^{(2,d)}[f_{\Dg}^{\dagger}]^{\{a\}_L} (\tilde{x}) \dg_{\{ab \}_L} g_{\Dg}^{\{b\}_L}(x)\,, 
\label{SP_lorentzian}
\end{equation}
   where $\mathbf{S}_{-\Dg}^{(2,d)}$ is the well-known (non-normalized) shadow transform,
   \begin{equation}
\mathbf{S}^{(2,d)}_{-\Dg}[f]^{\{a\}_L} (x) := \int_{\Rs^{1,d-1}}\dd^d \tilde{x}\, \frac{I_{\{ab\}_L}(x-\tilde{x})}{(x-\tilde{x})^{2(d-\Dg)}} f_{\Dg}^{\{b\}_L} (\tilde{x})\,,
\end{equation}
   consisting of a non-local integral transform mapping a primary of conformal dimension $\Dg$ to a primary of conformal dimension $d-\Dg$.
    
\paragraph{$(1,d+1)$:} If $X \mapsto F_{\Dg,\{A\}_L}(X)$, $\Dg \in \frac{d}{2} + \mathrm{i} \Rs_+$ denotes principal series representations of $\mathrm{SO}(1,d+1)$, then we replace \eqref{SP_shadowDelta} by an integral over two independent variables $x^a,\tilde{x}^a \in \Rs^{d}$ in $d$-dimensional Euclidean space. Once again, we can re-express the analytic continuation of \eqref{SP_shadowDelta} in terms of the Euclidean shadow transform
   \begin{equation}
\mathbf{S}^{(1,d+1)}_{-\Dg}[f]^{\{a\}_L} (x) := \int_{\Rs^{d}}\dd^d \tilde{x}\, \frac{I_{\{ab\}_L}(x-\tilde{x})}{(x-\tilde{x})^{2(d-\Dg)}} f_{\Dg}^{\{b\}_L} (\tilde{x}). 
\end{equation}
This integral transform is known as a Knapp-Stein intertwining operator \cite{Dobrev:1977qv} between the principal series representation of weight $\Delta$, and the principal series representation of weight $\bar{\Delta} = d-\Delta$. On the other hand, if $f_{\Delta}$ transforms in a principal series representation of weight $\Delta$, then $\overline{f_{\Delta}}$ will transform in a principal series representation of weight $\bar{\Delta}$. We can thus obtain the scalar product between principal series representations by replacing the shadow transform with complex conjugation, i.e.  
    \begin{equation}
        \langle F_{\Dg}, G_{\Dg}\rangle = \mathcal{N}_{-\Dg}^{(d+2)} c_{\Dg}^{(d+2)} \int_{\Rs^d} \dd^d x\, \overline{f_{\Delta}^{\{a\}_L}(x)}\, \dg_{\{ab\}_L}\, g_{\Dg}^{\{b\}_L}(x), 
        \end{equation}
        where $c_{\Delta}^{(d+2)}$ is some normalization.
        
\paragraph{$(0,d+2)$:} Back to unitary representations of the compact real form, where $-\Dg \in \Zs_{\geq 0}$, we can make the change of coordinates from antipode $\tilde{x}$ back complex conjugate $\bar{x}$ and rewrite \eqref{SP_shadowDelta} as
     \begin{equation}
         \langle F_{\Dg},G_{\Dg}\rangle = \mathcal{N}_{-\Dg}^{(d+2)} \int_{\Cs^d} \dd^d x \, \dd^d\bar{x}\, \frac{\bar{f}^{\{a\}_L}(\bar{x}) I_{\{ab\}_L}(x+x^2\bar{x}) I_{\{c\}_L}^{\{b\}_L}(x) g^{\{c\}_L}(x)}{(1+2x\cdot \bar{x}+x^2\bar{x}^2)^{d-\Delta}}
         \label{SP_compact}
     \end{equation}
     It is known (see e.g. \cite[Eq.~2.48]{delduc1985classical}), that the denominator in \eqref{SP_compact} appears in the Kähler potential
     \begin{equation}
         \mathcal{K}(x,\bar{x}) := \log(1+2x\cdot\bar{x}+x^2\bar{x}^2)
     \end{equation}
     of the Grassmannian 
     \begin{equation}
         \mathrm{Gr}(2,d+2) := \mathrm{SO}_{\Rs}(d+2)/H_{\Rs}, \qquad H_{\Rs}:=\mathrm{SO}_{\Rs}(2) \times \mathrm{SO}_{\Rs}(d),
         \label{grass_K}
     \end{equation}
     which is not only a Kähler manifold, but also a Hermitian symmetric space. A more succinct formula for this scalar product is then given by
        \begin{equation}
         \langle F_{\Dg},G_{\Dg}\rangle = \mathcal{N}_{-\Dg}^{(d+2)} \int_{\mathrm{Gr}(2,d+2)} (\ds\bar{\ds} \mathcal{K})^{\wedge d} \,e^{\mathcal{K} \Delta} \bar{f}_{\Dg}^{\{a\}_L}(\bar{x}) I_{\{ab\}_L}(\ds_{\bar{x}} \mathcal{K}) I_{\{c\}_L}^{\{b\}_L}(x) g_{\Dg}^{\{c\}_L}(x),
         \label{SP_Kaehler}
     \end{equation}
     where $\ds,\bar{\ds}$ denote the holomorphic and anti-holomorphic exterior derivatives on the complex manifold $\mathrm{Gr}(2,d+2)$. The above formula does not lend itself well to direct computation. Instead, for practical purposes, we will introduce a compact analogue of the shadow transform, defined as
     \begin{equation}
         \dd^d x \, \mathbf{S}^{(0,d+2)}_{-\Dg}[f^{\dagger}]_{\{c\}_L}(x) := \int_{\bar{x}} (\ds\bar{\ds} \mathcal{K})^{\wedge d} \,e^{\mathcal{K} \Delta} \bar{f}^{\{a\}_L}(\bar{x}) I_{\{ab\}_L}(\ds_{\bar{x}} \mathcal{K})I^{\{b\}_L}_{\{c\}_L}(x),
         \label{S_holo}
     \end{equation}
     where $\int_{\bar{x}}$ denotes integration of anti-holomorphic top forms in the Dolbeault cohomology of $\mathrm{Gr}(2,d+2)$, such that $\dd^n x\,\mathbf{S}^{(0,d+2)}_{-\Dg}[f^{\dagger}]^{\{c\}_L}(x)$ is a holomorphic top form. We will now determine the image of the finite-dimensional representation $[-\Dg,l_1,\dots,l_L]$ of $\mathrm{SO}_{\Rs}(d+2)$ under the compact shadow transform.

     First, if we denote the action of the conformal generators on a $[l_1,\dots,l_L]$ tensor-valued field $\vct{f}_{\Dg}(x)$ as $\mathcal{T}_{\ag}^{(\Dg)}(x^a,\ds_a)$, then the highest and lowest weight vectors of the representation $[-\Dg,l_1,\dots,l_L]$ with respect to the first Cartan element are given by
     \begin{align*}
     & \vct{K}^{(\Dg)}_{\mu} \vct{f}_{\Dg} = 0 \iff \vct{f}_{\Dg}(x) = x^{-2\Dg} \vct{I}(x) \vct{v}\,, \qquad \forall \vct{v} \in [l_1,\dots ,l_L]\,, \\
     & \vct{P}^{(\Dg)}_{\mu} \vct{f}_{\Dg} = 0 \iff \vct{f}_{\Dg}(x) = \vct{v}\,, \qquad \qquad \qquad \,\, \forall \vct{v} \in [l_1,\dots ,l_L]\,.
     \end{align*}
     Inserting both of these expressions for $f_{\Delta}^{\{a\}_L}$ in \eqref{S_holo}, it is easy to deduce how $\dd^d x\, \mathbf{S}_{-\Dg}^{(0,d+2)}[f^{\dagger}](x)$ transforms under the left action of $H_{\Rs}$ on $\mathrm{Gr}(2,d+2)$ (the latter are defined in \eqref{grass_K}). Combining these $H_{\Rs}$-covariance properties with the holomorphicity constraint is enough to determine, up to normalization, the image of the highest and lowest weights under the compact shadow transform:
     \begin{equation*}
     \mathbf{S}^{(0,d+2)}_{-\Dg}[ \vct{v}^{\dagger}](x) \propto (x^2)^{-\frac{d}{2}} \bar{\vct{v}}^{\mathsf{T}} \vct{I}(x)\,, \qquad  \mathbf{S}^{(0,d+2)}_{-\Dg}[\left( x^{-2\Dg} \vct{I}(x) \vct{v} \right)^{\dagger} ](x) \propto (x^2)^{-\frac{d}{2}+\Dg} \bar{\vct{v}}^{\mathsf{T}}\,.
     \end{equation*}
     Next, we know that all other states in the representation $[-\Dg,l_1,\dots,l_L]$ are given by the repeated action of either $\vct{P}^{(\Dg)}_{\mu}$ or $\vct{K}^{(\Dg)}_{\mu}$ on either the highest weight vectors $x^{-2\Dg} \vct{I}(x) \vct{v}$ or the lowest weight vectors $\vct{v}$ respectively, until the opposite weights are reached. Integrating by parts the action of conformal generators in \eqref{S_holo}, we can deduce that the compact shadow transform continues to satisfy the interwining property,
     \begin{align*}
         \mathbf{S}^{(0,d+2)}_{-\Dg}[(\vct{K}^{(\Dg)}_{\mu} \vct{f}_{\Dg})^{\dagger}](x) &= - \mathbf{S}^{(d+2)}_{-\Dg}[\vct{f}_{\Dg}^{\dagger}](x) \vct{P}_{\mu}^{(d-\Dg)}, \\        \mathbf{S}^{(d+2)}_{-\Dg}[(\vct{P}^{(\Dg)}_{\mu} \vct{f}_{\Dg})^{\dagger}](x) &= - \mathbf{S}^{(0,d+2)}_{-\Dg}[\vct{f}_{\Dg}^{\dagger}](x) \vct{K}_{\mu}^{(d-\Dg)}.
     \end{align*}
     This allows us to compute the compact shadow transform of any vector in $[-\Dg,l_1,\dots,l_L]$ from the shadow transform of the highest/lowest weights determined above. The compact shadow transform thus maps the space of $[l_1,\dots,l_L]$-tensor valued polynomials in $x$ with $x^a \ds_a = \mathrm{diag}(-d,\dots,-d+2\Dg)$ to a finite dimensional space of $[l_1,\dots,l_L]$-tensor valued rational functions of $x$ with $x^a \ds_a = \mathrm{diag}(-d,\dots,-d+2\Dg)$. Both spaces are meromorphic functions in $x^a$, and they are paired together with scalar product 
     \begin{equation}
         \langle F_{\Dg}, G_{\Dg}\rangle = \mathcal{N}_{-\Dg}^{(d+2)}\int_{S^1 \times S^{d-2}} \dd^d x\, \mathbf{S}^{(0,d+2)}_{-\Dg}[f^{\dagger}]_{\{a\}_L}(x)\, g_{\Dg}^{\{a\}_L}(x).
     \end{equation}
     Note that the integration domain, $S^1 \times S^{d-2}$, is isomorphic to the $H_{\Rs}$-orbit of any non-zero point in $\mathrm{Gr}(2,d+2)$.        
    
    \subsection{Iterated integration over Poincaré patches}
    \label{ssec:nested_poincare}
    For each of the $L$ remaining Cartan generators in $\mathfrak{so}_{\Cs}(d+2)$, we can repeat the procedure in section \ref{ssec:BargmannTodorov} to write the scalar product of the representation $[-\Dg,l_1,\dots,l_L]$ as iterated integrals over coordinates $x,y_1,\dots,y_{\nu_{\mathrm{max}}}$, $\nu_{\mathrm{max}} \leq L$. Having discussed the analytic continuations of the integral formula for all relevant real forms of $\mathfrak{so}_{\Cs}(d+2)$, we will make the representative choice 
    \begin{equation}
    (x,y_1,\dots,y_{\nu_{\mathrm{max}}}) \in \Rs^d \times \Cs^{d-2} \times \dots \times \Cs^{d-2 \nu_{\mathrm{max}}},
    \label{x,y,...,y}
    \end{equation}
    which are local coordinates on the flag manifolds
\begin{equation} 
\mathrm{SO}(1,d+1)/(\mathrm{SO}(1,1) \times \mathrm{SO}(d))\ltimes \mathbb{R}^d 
\quad , \quad 
\mathrm{SO}_\bC(d+2-2\nu)/(\mathrm{SO}_\bC(2) \times \mathrm{SO}_\bC(d-2\nu))\ltimes 
\mathbb{C}^{d-2\nu} \,,
\end{equation}  
for $\nu = 1, \dots, \nu_{\mathrm{max}}$. At the STT level ($\nu_{\mathrm{max}}=1$) we introduce null polarization vectors in $\Cs^d$ and write
    \begin{equation}
        f_{\Dg,l_1,\{a\}_{L-1}}(x,z_1) := f_{\Dg,a_1^{(1)}\dots a_{l_1}^{(1)} a_{1}^{(2)} \dots a_{l_L}^{(L)}}(x) z_1^{a_1^{(1)}}  \dots z_1^{a_{l_1}^{(1)}}. 
    \end{equation}
    Then, applying the reduction to the $\Cs^{d-2}$ Poincaré patch of $z \in \Cs^{d}$,
\begin{equation}
z^a(z^+,y_1) = z^+ \psi_1^A(y_1)\,, \qquad z^+ := z^{d-1}+\mathrm{i} z^{d}, \qquad \psi_1(x) := \left(\frac{y_1^2-1}{2}, \mathrm{i}\frac{y_1^2+1}{2}, -y_1^{\ag} \right) ,
\end{equation}
    such that
    \begin{equation}
    f_{\Dg,l_1,\{a\}_{L-1}}(x,z_1) = (z_1^+)^{l_1} f_{\Dg,l_1,\{\ag\}_{L-1}}(x,y_1) e^{\{a\}_{L-1}}_{\{\ag\}_{L-1}}(y_1)\,, \qquad e^a_{\ag}(y_1) := \frac{\ds \psi_1^a}{\ds y_1^{\ag}}(y_1)\,,
    \end{equation}
    yields
    \begin{align}
        \langle F_{\Dg}, G_{\Dg} \rangle = \mathcal{N}^{(d+2)}_{-\Dg} \mathcal{N}^{(d)}_{l_1} \int_{\Rs^d \times (S^1\times S^{d-3})}& \dd^d x\, \dd^{d-2} y_1 \nonumber \\
        & \mathbf{S}^{(0,d)}_{l_1} \circ \mathbf{S}^{(1,d+1)}_{-\Dg}[f_{\Dg,l_1}^{\dagger}]_{\{\ag\}_{L-1}} (x,y_1) \, g_{\Dg}^{\{\ag\}_{L-1}}(x,y_1)\,.
        \label{full_isospin_integral}
    \end{align}
    The above choice of domain then corresponds to a principal series representation of $\mathrm{SO}(1,d+1)$. Another choice relevant for CFT corresponds to representations where $l_1 \in \frac{2-d}{2} + i \Rs$ is on the principal series of the $\mathrm{SO}(1,d-1)$ subgroup of $\mathrm{SO}(2,d)$. Such representations are known as \emph{light-ray operators} in the CFT literature \cite{kravchuk2018light}. The scalar product for such representations can then be obtained by replacing $\mathbf{S}_{l_1}^{(0,d)} \rightarrow \mathbf{S}_{l_1}^{(1,d-1)}$, $S^1\times S^{d-3} \rightarrow \Rs^{d-2}$ in \eqref{full_isospin_integral}. 
    
   For the remaining spins $l_2,\dots,l_L$, the only physically relevant reality condition is $l_2,\dots l_L \in \Zs_{\geq 0}$ and $y_2,\dots,y_L$ as in \eqref{x,y,...,y}. Omitting the domain of integration over $\Rs^d \times \prod_{\nu=1}^L (S^1\times S^{d-2\nu-1})$, the integral formula of the scalar product at maximal depth is given by
\begin{equation}
    \langle F_{\Dg}, G_{\Dg} \rangle = \mathcal{N}^{(d+2)}_{[-\Dg,\{l_{\nu}\}]} \int \dd^d x\prod_{\nu=1}^L \dd^{d-2\nu} y_{\nu}\, \mathbf{S}_{[-\Dg,\{l_{\nu}\}]} [f_{\Dg,\{l_{\nu}\}}^{\dagger}](x,\{y_{\nu}\}) \, g_{\Dg,\{l_{\nu}\}} (x,\{y_{\nu}\})\,,
\end{equation}
where
\begin{equation}
\mathcal{N}^{(d+2)}_{[-\Dg,\{l_{\nu}\}]} := \mathcal{N}^{(d+2)}_{-\Dg} \mathcal{N}^{(d)}_{l_1} \dots     \mathcal{N}^{(d+2-2L)}_{l_L}, \,\,\ \mathbf{S}_{[-\Dg,\{l_{\nu}\}]} := \mathbf{S}^{(0,d+2-2L)}_{l_L} \circ \dots \circ \mathbf{S}^{(0,d)}_{l_1} \circ \mathbf{S}^{(1,d+1)}_{-\Dg}.
\end{equation}
    
\subsection{Application to the scalar products of \texorpdfstring{$3$}{3}-point vertex systems}
$3$-point functions are invariants in the tensor product of three irreducible unitary representations of $\mathrm{SO}(2,d)$, from which they naturally inherit a $\mathrm{SO}(2,d)$-invariant scalar product. In the notation of section~\ref{sect:reviewsummary}, this takes the form
\begin{align*}
    \langle t, t' \rangle_{\mathrm{H}} = \prod_{i=1}^3 \mathcal{N}_{[-\Dg_i;l_i;\ell_i]} \int & \dd^dx\dd^{d-2} y_{1i} \dd^{d-4} y_{2i}  \\
    & \bigotimes_{i=1}^3 \mathbf{S}^{(d+2)}_{[-\Dg_i;l_i;\ell_i]} \left[(\Omega t)^{\dagger}\right] \,  \Omega t',
\end{align*}
where we have suppressed the dependence of $\Omega$ and $t$ on the quantum numbers $[-\Dg_i;l_i;\ell_i]$ for convenience. The subscript "H" in $\langle -,-\rangle_{\mathrm{H}}$ stands for "Haar", because this formula for the scalar product can be understood as descending from the scalar product on $G\times G \times G$ defined by the Haar measure. In the conformal partial wave literature, it has now become common practice to analytically continue the conformal weights $\Dg$ from the physical region $\Dg \in \Rs_{\geq 0}$ to the domain of the $\mathrm{SO}(1,d+1)$ principal series representations, $\Dg \in \frac{d}{2}+ \mathrm{i} \Rs$. For similar reasons, it will be useful to analytically continue the STT spins from the physical region $l_i \in \Zs_{\geq 0}$ to the domain of the $\mathrm{SO}(1,d-1)$ principal series representations, $l_i \in \frac{2-d}{2}+\mathrm{i} \Rs$. In these regions, we can replace the shadow transforms in the scalar product with complex conjugation. In particular, for the STT-STT-scalar system (vertices of type I), this allows us to write the scalar product as
\begin{equation}
    \langle t, t' \rangle_{\mathrm{H}} = \mathcal{N}_{[-\Dg_i;l_i]} \int_{(\Rs^d)^3 \times (\Rs^{d-2})^2}  \dd^d x_1\,\dd^d x_2\,\dd^dx_3\,\dd^{d-2}y_1\,\dd^{d-2}y_2 \, (\omega^{\dagger} \omega) \, t^{\dagger}t'\,,
    \label{basicSP}
\end{equation}
where we defined
\begin{equation}
    \Omega^{(\Dg_i;l_i)}(X_i;Z_i) =: \prod_{i=1}^3 (X_{i}^+)^{-\Dg_i} \prod_{i=1}^2 (z_i^+)^{l_i}  \omega^{(\Dg_i;l_i)}(x_i;y_i)\,,
\end{equation}
and
\begin{equation}
    f_{\Dg,l}^{\dagger}(x,y) = x^{-2 \bar{\Dg}} y^{2\bar{l}} \,\overline{f_{\Dg,l}}\,(x/x^2, I(x) \cdot y/y^2 )\,.
\end{equation}
Note that $t^{\dagger} = \bar{t}$ because of the conformal invariance of $\mathcal{X}$. To reduce $\langle -,-\rangle_{\mathrm{H}}$ to an integral over cross-ratio space, we need to factorize the global $\mathrm{SO}(1,d+1)$ symmetry acting on the integration variables $x_1,x_2,x_3 \in \Rs^{d}$ and $y_1,y_2 \in S^1 \times S^{d-3}$. This can be achieved either by applying the Faddeev-Popov method to a given conformal frame (passive picture), or by directly reducing the integration variables to the cross-ratio via conformally covariant changes of variables (active picture). We will adopt the latter approach. 

First, let us decompose the prefactor $\om^{(\Dg_i;l_i)}$ as 
\begin{align}
& \om^{(\Dg_i;l_i)}(x_i,y_i):= \om^{(\Dg_i)}_{\mathrm{sc}}(x_i) \om^{(l_i)}_{\mathrm{sp}}(x_i,y_i)\,, \\
& \om^{(\Dg_i)}_{\mathrm{sc}}(x_i) := \abs{X_{23;1}}^{\Dg_1} \abs{X_{31;2}}^{\Dg_2} \abs{X_{12;3}}^{\Dg_3}, \\
& \om^{(l_i)}_{\mathrm{sp}}(x_i,y_i) = \left(N_{23;1} \cdot \psi_1 \right)^{l_1} \left(N_{23;1} \cdot I(x_{12}) \psi_2 \right)^{l_2},
\end{align}
where
\begin{equation}
X_{ij;k} := x_{ik}/x_{ik}^2 - x_{jk}/x_{jk}^2\,, \qquad X_{ij;k}^2 = \frac{x_{ij}^2}{x_{ik}^2 x_{jk}^2}\,, \qquad N_{ij;k} := \frac{X_{ij;k}}{\abs{X_{ij;k}}}\,, 
\end{equation}
and $\psi_i = \psi_1(y_i) = \left(\frac{y_i^2-1}{2}, \mathrm{i} \frac{y_i^2+1}{2}, -y_i^{\ag} \right)$ is the projective null vector in $\Cs^{d-2}$ defined above. Now, because $\bar{\Dg}_i = d-\Dg_i$ on the principal series, the prefactors in the integral simplify to
\begin{align}
&\om^{(\Dg_i)}_{\mathrm{sc}} (\om^{(\Dg_i)}_{\mathrm{sc}})^{\dagger} = \om_{\mathrm{sc}}^{(\Dg_i)} \om_{\mathrm{sc}}^{(d-\Dg_i)} = \abs{x_{12}}^{-d} \abs{x_{23}}^{-d} \abs{x_{31}}^{-d}, \\
& \om^{(l_i)}_{\mathrm{sp}} (\om^{(l_i)}_{\mathrm{sp}})^{\dagger} = \om^{(l_i)}_{\mathrm{sp}} \om^{(2-d-l_i)}_{\mathrm{sp}} = (N_{23;1} \cdot \psi_1)^{2-d} (N_{23;1} \cdot I(x_{12}) \psi_2)^{2-d}.
\end{align}
In the second line, we used the transformation properties of $(x_i,y_i)$ under simultaneous $\mathrm{SO}(2,d)$ and $\mathrm{SO}(1,d-1)$ inversion,
\begin{align*}
&(x_i,y_i) \mapsto (x_i/x_i^2, -I(x_i) \cdot y_i/y_i^2)\,, \\
& I^a_b(x_{ij}) \mapsto I^a_c(x_i) I^c_d(x_{ij}) I^d_b(x_j)\,, \\
& N_{ij;k}^a \mapsto I^a_b(x_k) N_{ij;k}^b\,, \\
& \psi_i^a \mapsto -y_i^2 I^a_b(x_i) \psi_i^b\,.
\end{align*}
We can then re-express the measure coming from the $x_i \in \Rs^{d}$ as
\begin{align*}
\prod_{i=1}^3 \dd^d x_i \, (\om_{\mathrm{sc}}  \om^{\dagger}_{\mathrm{sc}}) =& \frac{\dd^d x_1 \dd^d x_2 \dd^d x_3}{\abs{x_{12}}^d \abs{x_{23}}^d\abs{x_{31}}^d} \\
=& \dd^d x_1 \dd^d x_{21}^{-1} \frac{\dd^d X_{23;1}}{\abs{X_{23;1}}^d} \\
=&  \dd^d x_1 \dd^d x_{21}^{-1} \dd \log \abs{X_{23;1}} \dd^{d-1} N_{23;1}\,, 
\end{align*}
where $\dd^{d-1} N$ denotes the measure of the $(d-1)$-sphere in $\Rs^d$. Recall that in $x^a,y^{\ag}$-variables, the cross-ratio takes the form
\begin{equation}
\mathcal{X} = \frac{1}{2} \frac{\psi_1 \cdot I(x_{21}^{-1}) \psi_2}{(N_{23;1} \cdot \psi_1)(N_{23;1} \cdot I(x_{21}^{-1}) \psi_2)}\,. 
\end{equation}
Since $I(x)$ is independent of $\abs{x}$, the non-compact moduli of the $x_i$-integrals fully factorize to
\begin{equation}
\int_{\Rs^d \times \Rs^d \times \Rs_+} \dd^d x_1 \dd^d x_{21}^{-1} \dd \log \abs{X_{23;1}} =\frac{\abs{\mathrm{SO}(1,d+1)}}{\abs{\mathrm{SO}(d)}}\,, 
\end{equation}
where $\abs{G}$ denotes the volume of $G$. 
Here, we understand $\dd^d x_1 \dd^d x_{21}^{-1} \dd \log \abs{X_{23;1}} $ as integrals over translations, SCTs, and dilations respectively. Next, we change isospin variables to 
\begin{align*}
\dd^{d-2} y_1 \dd^{d-2}y_2 \, (\om_{\mathrm{sp}}  \om^{\dagger}_{\mathrm{sp}}) = \dd^{d-2} y_1  \dd^{d-2} y_{2;1} \,  (N_{23;1} \cdot \psi_1)^{2-d} (N_{23;1} \cdot \psi_{2;1})^{2-d},
\end{align*}
where $y_{2;1} := I(x_{21}^{-1}) \cdot y_2$ and $\psi_{2;1}^a := \psi^a_1(y_{2;1})$, such that 
\begin{equation}
\mathcal{X} = \frac{1}{2} \frac{\psi_1 \cdot \psi_{2;1}}{(N_{23;1} \cdot \psi_1) (N_{23;1} \cdot \psi_{2;1})}\,.
\end{equation}
As we have now eliminated all ambiguity, let us denote $N^a_{23;1} \equiv N^a \in S^{d-1}$. We can always find a rotation matrix $\La_N \in \mathrm{SO}(d)$ such that $N^a = \tensor{(\La_N)}{^a_b} \dg^b_1$, where if $\psi(y)=\left(\frac{y^2-1}{2}, \mathrm{i} \frac{y^2+1}{2},-y^{\ag}  \right)$, then $e_1 := (0,0,\dg^{\ag}_1)$. We can thereby absorb any appearance of $N$ in the integrand as the action of $\La_N^{-1}$ on $\psi_1$, $\psi_{2;1}$, i.e.
\begin{align*}
 (N_{23;1} \cdot \psi_1)^{2-d} (N_{23;1} \cdot \psi_{2;1})^{2-d} &= (e_1 \cdot \La_N^{-1} \psi_1 )^{2-d} (e_1 \cdot \La_N^{-1} \psi_{2;1} )^{2-d} , \\
\mathcal{X} &= \frac{1}{2} \frac{(\La_N^{-1} \psi_1) \cdot (\La_N^{-1} \psi_{2;1})}{ (e_1 \cdot \La_N^{-1} \psi_1 ) (e_1 \cdot \La_N^{-1} \psi_{2;1})}\,.
\end{align*}
Now, recall that $\psi^a(y) = \frac{z^a}{z^+}$, for some null vector $z \in \Cs^d$, so that 
\begin{equation}
\La_N^{-1} \cdot  \psi(y) = \frac{(\La_N^{-1} z)^+}{z^+} \psi (\La_N^{-1} \cdot y)\,.
\end{equation}
As we know from CFT, linearity of the $\La$-action in $\Cs^d$ embedding space ensures that $\frac{(\La_N^{-1} z)^+}{z^+} $ is a function of $y$, and it is determined by the Jacobian of the action of $\La_{N}^{-1}$ on $y \in \Cs^{d-2}$, 
\begin{equation}
 \frac{(\La_N^{-1} z)^+}{z^+} = \abs{\det \frac{\ds(\La_N^{-1} \cdot y)}{\ds y}}^{\frac{1}{d-2}}.
\end{equation}
It is precisely for this reason that we can write 
\begin{equation}
\dd^{d-2} y \,  \left(\frac{(\La_N^{-1} z)^+}{z^+}\right)^{2-d} = \dd^{d-2} y \, \abs{\det \frac{\ds(\La_N^{-1} \cdot y)}{\ds y}} = \dd^{d-2}(\La_N^{-1} \cdot y)\,.
\end{equation}
If we define $y_1':= \La_N^{-1} \cdot y_1$, $y_{2}':= \La_N^{-1} \cdot y_2$, and
\begin{equation}
    e_1 \cdot y'_i := {y'_i}^{\parallel}, \qquad y_i -{y'_i}^{\parallel} \, e_1 := y_i^{\perp}, 
\end{equation}
this yields 
\begin{equation}
\dd^{d-2} y_1 \dd^{d-2}y_2 \, (\om_{\mathrm{sp}}  \om^{\dagger}_{\mathrm{sp}}) = \frac{\dd^{d-2} y_1' \dd^{d-2} y_2'}{(y_1'^{\parallel})^{d-2} (y_2'^{\parallel})^{d-2}}\,, \qquad \mathcal{X} =- \frac{(y_1'-y_2')^{\ag} \dg_{\ag\bg} (y_1'-y_2')^{\bg}}{4(y_1'^{\parallel}) (y_2'^{\parallel})}\,.
\end{equation}
Having eliminated the dependence on $N = N_{23;1}$, we can then factorize the integral over $S^{d-1}$ as
\begin{equation}
\int_{S^{d-1}} \dd^{d-1} N = \abs{S^{d-1}}= \frac{\abs{\mathrm{SO}(d)}}{\abs{\mathrm{SO}(d-1)}}\,.
\end{equation}
To reach this point, we have exclusively made changes of variables given by conformal transformations. In summary, we have simplified the scalar product to 
\begin{align}
\langle t, t'\rangle =& \prod_{i=1}^3 \mathcal{N}^{(d+2)}_{[-\Dg_i;l_i]} \,  \frac{\abs{\mathrm{SO}(1,d+1)}}{\abs{\mathrm{SO}(d-1)}} \\
& \int_{(\Rs^{d-2})^2} \frac{\dd^{d-2} y_1 \dd^{d-2} y_2}{(y_1^{\parallel})^{d-2} (y_2^{\parallel})^{d-2}} \bar{t} \left(\mathcal{X}(y_i) \right) t'\left(\mathcal{X}(y_i)  \right).
\label{conf_pair_bCFT}
\end{align}
 From the Faddeev-Popov perspective, this expression is equivalent to the partial gauge fixing
\begin{equation}
(x_1^{\star a},x_2^{\star a},x_3^{\star a}) = (0, \dg^a_1, \infty)\,, 
\end{equation}
with a residual $\mathrm{SO}(1,d-2)$ symmetry on the $a=2,\dots , d$ plane remaining. To get a better intuition of the kinematics at hand, it is worth noting that \eqref{conf_pair_bCFT} is equivalent to the conformal invariant pairing of scalar 2-point functions in $(d-2)$-dimensional Euclidean boundary CFT, with boundary at $y^{\parallel} = 0$ (see section \ref{ssec:bCFT} for further details). To obtain a scalar product with discrete orthonormal basis, we will restrict the domain of $(y_1,y_2)$ from $(\Rs_+ \times \Rs^{d-3})^2$, to a $\mathrm{SO}(1,d-2)$-invariant submanifold for which the image of $(y_1,y_2) \mapsto \mathcal{X}(y_1,y_2)$ is a compact domain. To find this restriction, we can write the cross-ratio in manifestly $\mathrm{SO}(1,d-2)$-invariant form as
\begin{equation}
s = 1-2 \mathcal{X} = \frac{ (y_2-\mathcal{P}y_1)^2+(y_2-y_1)^2}{ (y_2-\mathcal{P}y_1)^2-(y_2-y_1)^2},
\end{equation}
where $\mathcal{P}y := - y^{\parallel} e_1 + y^{\perp}$ maps a point in half-space to its mirror image. Since $\mathcal{P}$ commutes with conformal transformations that preserve the boundary, both $(y_2-y_1)^2$ and $(y_2-\mathcal{P}y_1)^2$ are $\mathrm{SO}(1,d-2)$-invariant. Thus, the restriction of these conformal invariants to
\begin{equation}
(y_2-y_1)^2 \leq 0, \quad (y_2-\mathcal{P}y_1)^2 \geq 0 \Longrightarrow  \abs{s} \leq 1,
\label{regge_domain}
\end{equation}
defines an invariant submanifold where $s$ is bounded. Note that this requires a continuation to the $(d-2)$-dimensional Minkowski metric, with respect to which $y_2$ and $y_1$ are then timelike separated. To proceed further, we act with $\mathrm{SO}(1,d-2)$ transformations on $y_i$:
\begin{equation}
    y_i \rightarrow y_i -y_1^{\perp} \rightarrow \frac{y_i-y_1^{\perp}}{y_1^{\parallel}} \equiv y_i^{\star}, 
\end{equation}
which maps a generic pair of isospins to the gauge
\begin{equation}
    y_1^{\star \ag} \!= \dg^{\ag}_1, \qquad y_2^{\star \ag} \!= \frac{y^{\ag}_2-(y_1^{\perp})^{\ag}}{y_1^{\parallel}} \in \Rs^{d-2}.
    \label{gauge_phi_theta_n}
\end{equation}
On the domain \eqref{regge_domain}, $y_2^{\star}$ is a spacelike vector and can be parameterized as \begin{equation}
y_2^{\star} = e^{\psi} \left( \cosh \phi \, e_1 + \sinh\phi\, n^{\perp} \right),
\end{equation}
where $n^{\perp} \in \mathcal{H}^{n-1}$ is the two-sheeted hyperboloid of unit timelike vectors in $\Rs^{1,n-1}$. With these variables, we can write the cross-ratio as
\begin{equation}
    s = 1-2\mathcal{X} = \frac{\cosh\psi}{\cosh \phi}, \qquad  1 \leq \cosh^2 \psi \leq  \cosh^2 \phi.
    \end{equation}
    
Now, it's easy to transform the measure to 
\begin{align*}
    \frac{\dd^{d-2} y_1 \dd^{d-2} y_2}{(y_1^{\parallel})^{d-2} (y_2^{\parallel})^{d-2}} =& \frac{\dd^{d-2} y_1 \dd y_2^{\parallel} \dd^{d-3} y_{21}^{\perp}}{(y_1^{\parallel})^{d-2} (y_2^{\parallel})^{d-2}} \\
    =& \frac{\dd^{d-2} y_1}{(y_1^{\parallel})^{d-2}} \frac{ \dd( y_2^{\parallel}/y_1^{\parallel}) \dd^{d-3} (y_{21}^{\perp}/y_1^{\parallel})}{ (y_2^{\parallel}/y_1^{\parallel})^{d-2}} \\
    =& \frac{\dd^{d-2} y_1}{(y_1^{\parallel})^{d-2}} \frac{\dd^{d-2} y_2^{\star}}{(y_2^{\star \parallel})^{d-2}}.
\end{align*}
This factorizes the $y_1$ measure, leading to another normalization that corresponds to the volume of half-space,
\begin{equation*}
    \mathcal{V}_{\mathrm{hs}} = \int_0^{\infty} \dd x \, x^{2-d} \int_{\Rs^{d-3}} \dd^{d-3} y.
\end{equation*}
Next, applying the parameterization \eqref{gauge_phi_theta_n} to $y_2^{\star}$, we get
\begin{align*}
   \frac{\dd^{d-2} y_2^{\star}}{(y_2^{\star \parallel})^{d-2}} =& \frac{\dd(e^{\psi}) (e^{\psi})^{d-3} \dd\phi \sinh^{d-4}\phi \, \dd^{d-4} n^{\perp}}{(e^{\phi})^{d-2} \cosh^{d-2} \phi} \\
   =&  \dd \psi  \dd(\cosh\phi)\cosh^{2-d}\phi\,\sinh^{d-5}\phi\, \dd^{d-4}n^{\perp} \\
   =& \frac{\dd(s \cosh \phi)}{\sqrt{s^2\cosh^2\phi-1}} \frac{\dd(\cosh^2\phi)}{2 \cosh\phi} (\cosh^2\phi)^{\frac{2-d}{2}} (\cosh^2\phi-1)^{\frac{d-5}{2}} \dd^{d-4}n^{\perp} \\
   =& \frac{1}{2} \dd^{d-4} n^{\perp} \, \dd t\, t^{\ag-\frac{3}{2}} (1-t)^{-\frac{1}{2}} \, \dd s \, (1-s^2)^{\ag-\frac{1}{2}}\,, \qquad t:= \frac{\cosh^2 \phi - \cosh^2\psi}{\cosh^2\phi-1} \in [0,1].
\end{align*}
The $t$-integral yields
\begin{equation}
    \int_0^1 \dd t\, t^{\ag-\frac{3}{2}} (1-t)^{-\frac{1}{2}} =  \frac{\Gamma(\ag-1/2)}{\Gamma(\ag)} \sqrt{\pi},  
\end{equation}
and the unit vector $\dd^{d-4} n^{\perp}$ integrates to an extra $\abs{H^{d-4}}$ volume. Putting everything back together, we obtain 
\begin{align}
    \langle t, t' \rangle_{\mathrm{H}} = \frac{ \sqrt{\pi}}{2}  \prod_{i=1}^4 &\mathcal{N}_{[-\Dg_i;l_i]}^{(d+2)} \,  \mathcal{V}_{\mathrm{hs}} \, \frac{\abs{\mathrm{SO}(1,d+1)} \, \abs{\mathrm{SO}(1,d-5)}}{\abs{\mathrm{SO}(d-1)} \abs{\mathrm{SO}(d-5)}} \frac{\Gamma(\ag-1/2)}{\Gamma(\ag)} \nonumber \\
    & \int_{-1}^{+1} \dd s \, (1-s^2)^{\ag-\frac{1}{2}}\, \bar{t}\left( \frac{1-s}{2}\right) t'\left(\frac{1-s}{2}\right). 
\end{align}
We have thus proven that for the STT-STT-scalar vertex, the scalar product descending from the Haar measure coincides, up to normalization, with the Gegenbauer scalar product,
\begin{equation}
    \langle t, t'\rangle_{\mathrm{H}} \propto \langle t,t'\rangle_{\ag}. 
\end{equation}
For a space of functions with the appropriate boundary conditions at $\mathcal{X}=0,1 \iff s = -1,+1$, the hermiticity constraint of the 2-STT vertex operator \eqref{H_basis} relative to this scalar product is then given by
\begin{align}
H^{(d;\Dg_i;l_i)}(\mathcal{X},\ds_{\mathcal{X}}) =\,& h_0^{(d;d-\Dg_i;2-d-l_i)}(\mathcal{X})\, + \nonumber \\
&\sum_{q=1}^4 (-)^q [\mathcal{X}(1-\mathcal{X})]^{\frac{1}{2}-\alpha}\ds_{\mathcal{X}}^q [\mathcal{X}(1-\mathcal{X})]^{\alpha+1-\frac{3}{2}} h_q^{(d;d-\Dg_i;2-d-l_i)}(\mathcal{X})\,, 
\end{align}
where $\ag = \frac{d-3}{2}$ in this case. In principle, one could obtain the generalization of this scalar product to  MST$_2$-STT-scalar, and  MST$_2$-MST$_2$-scalar in $d=4$ by direct computation, using the scalar product on arbitrary spinning primaries that was derived in the previous section. However, given the full expression of the vertex Hamiltonian, we can instead make a simple self-adjointness ansatz for a modified Gegenbauer scalar product,
\begin{align}
H^{(d;\Dg_i;l_i;\ell_i)}(\mathcal{X},\ds_{\mathcal{X}}) \stackrel{!}{=}\,& h_0^{(d;d-\Dg_i;2-d-l_i;\ell_i)}(\mathcal{X})\, + \nonumber \\
&\sum_{q=1}^4 (-)^q [\mathcal{X}(1-\mathcal{X})]^{\frac{1}{2}-\alpha}\ds_{\mathcal{X}}^q [\mathcal{X}(1-\mathcal{X})]^{\alpha+1-\frac{3}{2}} h_q^{(d;d-\Dg_i;2-d-l_i;\ell_i)}(\mathcal{X}).
\end{align}
For the values of $(d;\Dg_i;l_i;\ell_i)$ where the vertex system is 1-dimensional, this ansatz is valid if and only if $\ag = \ell_1+\ell_2 + \frac{d-3}{2}$.

\subsection{Relationship with boundary and projective space in \texorpdfstring{CFT$_{d-2}$}{CFT in (d-2)}:}
\label{ssec:bCFT}
Consider once again two STTs and one scalar field in $d$ dimensions, and define
\begin{equation}
    \oo_{\Dg_i,l_i}(X_i,Z_i) := (X_i^+)^{-\Dg_i} \Phi_{\Dg_i,l_i}(x_i,z_i)\,,
\end{equation}
as well as 
\begin{equation}
    \Omega_{\mathrm{sc}}(x_1,x_2,x_3) := \abs{X_{23;1}}^{-\Dg_1} \abs{X_{31;2}}^{-\Dg_2} \abs{X^a_{23;1}}^{-\Dg_3}\,, \qquad N^a := \frac{X_{23;1}}{\abs{X_{23;1}}}\,. 
\end{equation}
Then a general 3-point function of such representations is given by
\begin{equation}
\Omega_{\mathrm{sc}}^{-1} \langle \Phi_{\Dg_1,l_1}(x_1,z_1) \Phi_{\Dg_2,l_2}(x_2,z_2) \Phi_{\Dg_3}(x_3) \rangle = (N\cdot z_1)^{l_1} (N\cdot z_{2;1})^{l_2}\, t (\mathcal{X})\,,
\label{bCFT}
\end{equation}
where
\begin{equation}
    \mathcal{X} = \frac{z_1 \cdot z_{2;1}}{2 (N\cdot z_1) (N\cdot z_{2;1})}\,.
\end{equation}
The right hand side of eq.\ \eqref{bCFT} takes the exact same form as the 2-point function of two scalars with conformal dimensions $(-l_1,-l_2)$ in $(d-2)$-dimensional boundary CFT with $\mathcal{X}= -\xi$ when $N^2 = 1$, or $(d-2)$-dimensional projective space CFT with $\mathcal{X}=\eta$ when $N^2 = -1$. Using this kinematic equivalence, we could have directly obtained the Gegenbauer formula for the scalar product from \cite[Eq.~4.10]{mazavc2019analytic}, by continuing $\xi=-\mathcal{X} \in [0,\infty)$ to $\xi \in [-1,0]$. However, in this correspondence, the CFT$_{d-2}$ embedding space vectors $P \in \Rs^{1,d-1}$ are replaced with complex null vectors $z \in \Cs^d$, and there is no reality condition to distinguish projective space kinematics and bCFT kinematics in the case of finite dimensional representations of the rotation/Lorentz group. That being said, in Lorentzian signature, we expect our setup to coincide with the kinematics of Euclidean projective space CFT after analytically continuing to principal series representations of the $\mathrm{SO}(1,d-1)$ subgroup.

\section{The \texorpdfstring{$d$}{d}-deformation of the \texorpdfstring{MST$_2$-MST$_2$}{MST2-MST2}-scalar vertex operator}
\label{app:d_def}
\subsection{Comparison with one-dimensional vertex systems}
For all combinations of $(d;\Dg_i;l_i;\ell_i)$ that yield one-dimensional vertex systems, the Hamiltonian can be written as 
\begin{equation}
    H^{(d;\Dg_i;l_i;\ell_i)} = \tilde{H}^{(\cg_i;\nu_i;\ag;\bg)} + \Delta \tilde{E}^{(\cg_i;\nu_i;\ag;\bg;d)}\,, 
    \label{tildeH}
\end{equation}
where $(\Dg_i;l_i;\ell_i) \leftrightarrow (\cg_i;\nu_i;\ag;\bg)$ is the $d$-dependent bijection of seven parameters defined in subsection \ref{ssec:CMS_multiplicities}, and $\Delta \tilde{E}^{(\cg_i;\nu_i;\ag;\bg;d)}$ is a constant energy shift determined by 
\begin{equation}
    \Delta \tilde{E} - E_{\mathrm{EFMV}} =L_{\mathrm{EFMV}}-\tilde{H},
\end{equation}
with $E_{\mathrm{EFMV}}^{(\Dg_i;l_i;\ell_i;d)}$ given in \ref{ssec:ECMS}. Even for the two-dimensional vertex systems $d>4,\ell_1,\ell_2\neq 0$, we can obtain a $d$-dependent, MST$_2$-MST$_2$-scalar Hamiltonian for a one-dimensional system by restricting to the $\mathcal{Y}=0$ plane:
\begin{equation}
    H^{(d>4;\Dg_i;l_i;\ell_1,\ell_2)}(\mathcal{X},\ds_{\mathcal{X}}) := H^{(d>4;\Dg_i;l_i;\ell_1,\ell_2)}(\mathcal{X},\mathcal{Y}=0,\ds_{\mathcal{X}},\ds_\mathcal{Y}=0)\,.
\end{equation}
This $d$-deformation of the  MST$_2$-MST$_2$-scalar  operator is qualitatively different from the $d=4$ or $\ell_2=0$ cases for several reasons:
\begin{itemize}
    \item[1)] First, while we can still write the whole operator $H^{(d>4;\Dg_i;l_i;\ell_1,\ell_2)}$ in \eqref{tildeHd} as an elliptic CMS Hamiltonian, two of its multiplicities will no longer be linear in the quantum numbers --- instead
   \begin{align*}
m_{1,0} = \frac{7-d}{2}-(l_1+l_2)-\Dg_3-2\, \sqrt{\left(\ell_1+\frac{d-4}{2}\right)^2 + 2\ell_2\left(\frac{d-4}{2}-\ell_1\right)+\ell_2^2}\,, \\
m_{2,0} =  \frac{7-d}{2}-(l_1+l_2)-\Dg_3+2\, \sqrt{\left(\ell_1+\frac{d-4}{2}\right)^2 + 2\ell_2\left(\frac{d-4}{2}-\ell_1\right)+\ell_2^2}\,, 
\end{align*}
and the remaining multiplicities are 
\begin{align*}
& k(d;\Dg_i;l_i;\ell_1,\ell_2) = k(d;\Dg_i;l_i;\ell_1,0)\,, \\
& m_{i,\nu}(d;\Dg_i;l_i;\ell_1,\ell_2) = m_{i,\nu}(d;\Dg_i;l_i;\ell_1+\ell_2,0)\,,\qquad (i,\nu) \in \{3,4\} \times \{0\} \cup \{1,2,3,4\}\times \{1\}\,.
\end{align*}

\item[2)] Second, there is no choice of $\ag$ such that $H^{(d>4;\Dg_i;l_i;\ell_1,\ell_2)}$ is hermitian with respect to the Gegenbauer scalar product $\langle -,- \rangle_{\ag}$, nor any scalar product with a measure of the form $\mathcal{X}^{a}(1-\mathcal{X})^b$, $\mathcal{X} \in [0,1]$. 

\item[3)] It goes hand in hand with reason (1) that $H^{(d>4;\Dg_i;l_i;\ell_1,\ell_2\neq 0)}$ will now exhibit an explicit dependence on dimension after the reparametrization $(\Dg_i;l_i;\ell_i) \leftrightarrow (\cg_i;\nu_i;\ag;\bg)$, i.e.
\begin{equation}
       H^{(d>4;\Dg_i;l_i;\ell_1,\ell_2\neq 0)} = \tilde{H}^{(d>4;\cg_i;\nu_i;\ag;\bg)}. 
\end{equation}
In fact, the generalization of \eqref{tildeH} to $H^{(d>4;\Dg_i;l_i;\ell_1,\ell_2\neq 0)}$ is given by 
\begin{equation}
    H^{(d;\Dg_i;l_i;\ell_i)} = \tilde{H}^{(\cg_i;\nu_i;\ag;\bg)} + (d-4) (\ag-\bg-1) H_{\mathrm{def}}^{(\cg_i;\nu_i;\ag;\bg)}(\mathcal{X},\ds_\mathcal{X}) + \Delta \tilde{E}^{(\cg_i;\nu_i;\ag;\bg;d)}\,, 
    \label{tildeHd}
\end{equation}
where
\begin{align*}
    H_{\mathrm{def}}^{(\cg_i;\nu_i;\ag;\bg)}(\mathcal{X},\ds_\mathcal{X}) =& 4 \mathcal{X}(1-\mathcal{X})^2\ds_\mathcal{X}^2\\
    &+ 2(1-\mathcal{X})\left[ (\nu_1+\nu_2-1)\mathcal{X} -(\nu_1+\nu_2+1) -2\ag -2\mathrm{i}\cg_3 \right]\ds_\mathcal{X} + 4 \nu_1\nu_2 \mathcal{X} \\
    & +\frac{1}{4\mathcal{X}} \left( 4 \left(\gamma _1-\gamma _2\right)^2+\left(2 \alpha +2 \mathrm{i} \gamma _3+2 \nu _1+2 \nu _2+3\right)^2\right),
\end{align*}
is the $d \neq 4$ deformation, and $\Delta \tilde{E}^{(\cg_i;\nu_i;\ag;\bg;d>4)}$ is also a constant obtained from 
\begin{equation}
   \Delta \tilde{E} - E_{\mathrm{EFMV}} =L_{\mathrm{EFMV}}-\tilde{H}-(d-4)(\ag-\bg-1) H_{\mathrm{def}}.
\end{equation}
\end{itemize}

\subsection{The constant shift for the CMS Operator}
\label{ssec:ECMS}
In section \ref{sect:MappingElliptic} (more specifically Eq. \eqref{h_to_g}), we determined the Hamiltonian of all one-dimensional vertex systems in terms of the CMS operator $L_{\mathrm{EFMV}}$ up to a constant shift $E_{\mathrm{EFMV}}$. This was generalized in the previous section to $H^{(d\neq 4;\Dg_i;l_i;\ell_i \neq 0)}(\mathcal{X},\mathcal{Y} = 0, \ds_\mathcal{X},\ds_\mathcal{Y}=0)$ with generalized CMS multiplities that are no longer linear in the dimension and quantum numbers. In all of these cases, the constant shift in the Hamiltonian is given by
\begin{equation}
    H^{(d;\Dg_i;l_i;\ell_i)}(\mathcal{X},\ds_\mathcal{X}) = L_{\mathrm{EFMV}}(\mathcal{X},\ds_\mathcal{X}) + E_{\mathrm{EFMV}}^{(d;\Dg_i;l_i;\ell_i)}.
    \end{equation}
    To write out $E_{\mathrm{EFMV}}$ explicitly, we make use of the previous change of variables to $(\cg_i;\nu_i;\ag,\bg)$ and expand the dimension around $d=4$, $\bg=0$, and $\ag=0$, i.e. 
\begin{align*}
d:= 4+2\varepsilon, \,\,\, E_{\mathrm{EFMV}} := \sum_{m = 0}^4 \varepsilon^m E_{\mathrm{EFMV}}^{(m)}, \qquad E_{\mathrm{EFMV}}^{(m)} := \sum_{n=0}^3\bg^n E_{\mathrm{EFMV}}^{(m,n)} , \qquad E_{\mathrm{EFMV}}^{(m,n)} := \sum_{p=0}^4\ag^p E_{\mathrm{EFMV}}^{(m,n,p)}.
\end{align*}
The simplest coefficients are at the highest order in each of the three expansion parameters,
\begin{align*}
& E_{\mathrm{EFMV}}^{(4)} = - 128/3, \qquad E_{\mathrm{EFMV}}^{(3)} = \frac{64(\bg-\ag) - 1856}{3}, \qquad E_{\mathrm{EFMV}}^{(2,2)} =32, \\
&   E_{\mathrm{EFMV}}^{(1,3)} =E_{\mathrm{EFMV}}^{(0,0,4)} = -16, \qquad E_{\mathrm{EFMV}}^{(0,3)} = 8(1-2\ag). 
\end{align*}
The next highest order terms are also relatively simple, 
\begin{align*}
& E_{\mathrm{EFMV}}^{(2,1)} = \frac{64(\ag-3\mathrm{i}\cg_3) + 896}{3}, \qquad E_{\mathrm{EFMV}}^{(1,2)} = 16(2\mathrm{i}\cg_3+\ag-2), \qquad E_{\mathrm{EFMV}}^{(1,0,2)} = 16 (15 + 6 \nu_1-2\nu_2), \\
& E_{\mathrm{EFMV}}^{(0,1,2)}  = 8(1+8\nu_1-8\nu_2), \qquad E_{\mathrm{EFMV}}^{(0,0,3)} = 80 \nu_2 - 176 \nu_1-16\mathrm{i}\cg_3 - 64.
\end{align*}
We can then write all terms at $\mathcal{O}(\varepsilon,\beta)$ as
\begin{align*}
E_{\mathrm{EFMV}}^{(1,1)} =& 56 \nu_1(\nu_1-\frac{12}{7}\ag -1) + 8 \nu_2(\nu_2+4\ag+1) + 16\nu_1\nu_2 + 8 ( 4\cg_1^2-4\cg_2^2+\cg_3^2)    \\
&-16 \mathrm{i}\cg_3(\nu_1+\nu_2-2\ag)+\frac{688}{3} \ag -16 \ag^2 +\frac{230}{3}.
\end{align*}
The remaining terms that fit on one line are 
\begin{align*}
& E_{\mathrm{EFMV}}^{(0,1,1)} =  -16 \nu_1(2\nu_1-1) +16\nu_2(2\nu_2+1) +16\mathrm{i}\cg_3 + 32(\cg_1^2-\cg_2^2+\cg_3^2) +\frac{1120}{3}, \\
& E_{\mathrm{EFMV}}^{(0,1,0)}  = 28\nu_1(\nu_1+1)-4\nu_2(\nu_2+1) -8\nu_1\nu_2 +8 \mathrm{i}\cg_3(\nu_1+\nu_2 + \frac{3}{2})  -4 \cg_3^2 + 16(\cg_2^2-\cg_1^2) -\frac{347}{3} \\
 & E_{\mathrm{EFMV}}^{(0,0,2)} =  60 \nu_2(\nu_2+\frac{31}{15}) -124 \nu_1( \nu_1+1)+88\nu_1\nu_2 +4\mathrm{i}\cg_3(1-2\nu_1-2\nu_2) +60 \cg_3^2 - 32(\cg_1^2+\cg_2^2) +183.
\end{align*}
Finally, we have
\begin{align*}
 E_{\mathrm{EFMV}}^{(1,0,1)} =& 56\nu_1^2 -8 \nu_2^2 -16 \nu_1\nu_2 - \frac{1208}{3} \nu_2 - \frac{1928}{3} \nu_1\\
&+16 \mathrm{i}\cg_3(\nu_1+\nu_2+ \frac{3}{2}) -8\cg_3^2 + 32(\cg_2^2-\cg_1^2) - \frac{278}{3}, \\
 E_{\mathrm{EFMV}}^{(1,0,0)} =& \frac{856}{3} \nu_1(\nu_1+1) - \frac{616}{3} \nu_2(\nu_2+1) +16\nu_1\nu_2 \\
&+16 \mathrm{i}\cg_3(\nu_1+\nu_2-\frac{3}{2}) -\frac{616}{3} \cg_3^2 + \frac{544}{3}\cg_2^2 - \frac{928}{3} \cg_1^2 - \frac{4874}{3}, \\
 E_{\mathrm{EFMV}}^{(0,2)} =&    16\ag\left( -\ag + \nu_1+\nu_2+\mathrm{i}\cg_3+2 \right) +12\nu_1(\nu_1+1) +12\nu_2(\nu_2+1) + 8\nu_1\nu_2\\
& +4\mathrm{i}\cg_3(2\nu_1+2\nu_2+3) + 59,
\end{align*}
along with 
\begin{align*}
    E_{\mathrm{EFMV}}^{(0,0,1)}  =& 16\nu_2^3 -240 \nu_1^3+52 \nu_2^2-364 \nu_1^2  \\
    &+  48 \nu_1\nu_2(\nu_1+\nu_2 +104)+\frac{2168}{3} \nu_1 - \frac{1352}{3} \nu_2+  4\mathrm{i}\cg_3(3+2\nu_1+2\nu_2) \\
&  + 4 \cg_3^2(13 + 28\nu_1-4\nu_2) + \cg_1^2(32 \nu_1-96\nu_2) -32\cg_2^2(1+3\nu_1-\nu_2) + 64 \mathrm{i}\cg_1\cg_2\cg_3 + \frac{899}{3},
\end{align*}
and 

\begin{align*}
E_{\mathrm{EFMV}}^{(0,0,0)} =& 4 \nu_2^4 - 60\nu_1^4  + 8 \nu_2^3 -120 \nu_1^3 + 24 \nu_1^2 \nu_2^2 +\frac{1084}{3} \nu_1^2- \frac{724}{3} \nu_2(\nu_2+1)  + 28\nu_1\nu_2  + \frac{1276}{3} \nu_1 \\
& + 16 \mathrm{i}\cg_1\cg_2\cg_3(1+2\nu_1+2\nu_2)-2\cg_3^4 - 64 \cg_1^2(\cg_1^2-\cg_2^2)  \\
&- \cg_3^2 \left( \frac{751}{3} + 56 \cg_1^2 - 8 \cg_2^2-52\nu_1^2 - 56\nu_1+8\nu_1 - 8 \nu_1\nu_2  + 12\nu_2^2 \right)   \\
& -\cg_1^2 \left( \frac{1330}{3}+56\nu_2^2 -8 \nu_1^2  +40 \nu_2-24\nu_1-16\nu_1\nu_2    \right)  \\
&+ \cg_2^2\left( \frac{766}{3} +8\nu_2^2-56\nu_1^2 +24\nu_2-40\nu_1+16\nu_1\nu_2  \right).
\end{align*}



\end{document}